\pdfoutput=1

\RequirePackage{fix-cm} 
\documentclass[oneside,phd]{snuthesis}

\ifpdf
	\usepackage[pdftex]{graphicx}
\else
	\usepackage[dvipdfmx]{graphicx}
\fi

\usepackage{epsfig}
\usepackage{wrapfig}
\usepackage{graphicx}
\usepackage{xcolor}
\usepackage{setspace}
\usepackage{url}
\usepackage{graphicx}
\usepackage{listings}
\usepackage{footnote}
\usepackage{longtable}
\usepackage{algorithmic}
\usepackage{mathtools}
\usepackage{booktabs}
\usepackage{makecell}
\usepackage{soul}
\usepackage{diagbox}
\usepackage{tabularx}
\usepackage{hyperref}
\usepackage{cleveref}
\usepackage{subfig}
\usepackage{multicol}
\usepackage{multirow}
\usepackage{tipa}
\usepackage[comma,authoryear,round]{natbib}
\usepackage[flushleft]{threeparttable}
\usepackage{scalerel,amssymb}

\newcolumntype{C}[1]{>{\centering\arraybackslash}p{#1}}

\title{\bf Voice Biomarker Analysis and Automated Severity Classification of Dysarthric Speech in a Multilingual Context}
\title*{다언어 환경에서의 마비말장애 음성 바이오마커 분석 및 자동 중증도 분류}

\academicko{Linguistics}
\departmenten{Graduate School of Linguistics}
\departmentko{언어학과}

\author{Eunjung Yeo}
\author*{여 은 정} 

\studentnumber{2021-36540}

\advisor{Minhwa Chung}
\advisor*{Minhwa Chung} 

\graddate{August 2024}

\submissiondate{July 2024}

\approvaldate{July 2024}

\committeemembers%
{이 호 영}%
{정 민 화}%
{James Whang}
{김 회 린}%
{박 형 민}%

\begin{document}
\include{snutocstyle} 

\pagenumbering{Roman}
\makefrontcover
\makeapproval

\newcommand{\fix}[1]{\textcolor{red}{#1}}

\cleardoublepage
\pagenumbering{roman}

\keyword{dysarthric speech, multilingual speech analysis, automatic dysarthria severity classification, automatic dysarthria pronunciation assessment}
\begin{abstract}
Dysarthria, a motor speech disorder, severely impacts voice quality, pronunciation, and prosody, leading to diminished speech intelligibility and reduced quality of life. Accurate assessment is crucial for effective treatment, but traditional perceptual assessments are limited by their subjectivity and resource intensity. To mitigate the limitations, automatic dysarthric speech assessment methods have been proposed to support clinicians on their decision-making. While these methods have shown promising results, most research has focused on monolingual environments. However, multilingual approaches are necessary to address the global burden of dysarthria and ensure equitable access to accurate diagnosis. This thesis proposes a novel multilingual dysarthria severity classification method, by analyzing three languages: English, Korean, and Tamil.

Each language has its unique phonemic and prosodic systems, necessitating that multilingual assessment systems consider these characteristics to achieve optimal classification performance. Therefore, we first conduct multilingual analyses using two different types of features: paralinguistic features to analyze different speech dimensions (phonation, articulation, prosody), and Goodness of Pronunciation (GoP) for phoneme-level pronunciation errors. We hypothesize and validate that differences in language structures result in different optimal feature sets and the impact of each feature on speech intelligibility.

Based on these findings, we propose a set of clinical-knowledge-driven features designed to comprehensively capture dysarthric speech characteristics, where the feature list encompasses voice quality-, pronunciation-, and prosody-related features. For each language, the feature set is validated through two-step validation, which includes statistical analysis and clinical analysis. Multilingual analyses are performed based on the validated feature set, to distinguish between language-universal feature set (intersection feature set of the three languages) and language-specific features (unique features of the language). Furthermore, we perform classification experiments using language-universal features. Our results indicate that classification performance is suboptimal when relying solely on language-universal features, highlighting the importance of incorporating language-specific characteristics.

Building on these insights, we propose a multilingual dysarthria severity classification methodology using eXtreme Gradient Boosting (XGBoost). Our approach uniquely integrates both language-universal and language-specific features to enhance model robustness and optimize classification performances. This approach significantly outperforms traditional multilingual classification methods that employ language-universal features only, achieving a 7.33\% relative improvement.

In conclusion, this thesis establishes a new paradigm for multilingual dysarthria assessment using voice biomarkers. Our findings demonstrate that both language-universal and language-specific features are essential for accurate severity classification. This verifies the potential of our approach, which considers linguistic characteristics, to significantly advance the performance of multilingual dysarthria severity classification. We anticipate that this method will contribute to providing more inclusive, efficient, and precise tools for evaluating dysarthria across diverse linguistic populations.

\end{abstract}

\tableofcontents
\listoffigures
\listoftables

\cleardoublepage
\pagenumbering{arabic}

\chapter{Introduction}

\section{Motivation}
\subsection{What is Dysarthria?}\label{ssec:1-dysarthria}
Dysarthria, a prevalent motor speech disorder arising from neuromuscular damage, disrupts multiple dimensions of speech production, including phonation, respiration, resonance, articulation, and prosody \citep{enderby1980frenchay, enderby2013disorders}. Research indicates that dysarthria affects a significant proportion of individuals with various neurological conditions, including Parkinson's Disease, multiple sclerosis, cerebral palsy, and amyotrophic lateral sclerosis (ALS) \citep{muller2001progression, perez2014prevalence, danesh2013clinical, yorkston2001communication, nordberg2013speech, traynor2000clinical}. The resulting reduction in speech intelligibility – the degree to which a speaker's intended message is understood by a listener – hinders effective communication and may significantly diminish one's quality of life. Consequently, accurate and reliable assessment of speech intelligibility is paramount for monitoring disease progression, evaluating therapeutic interventions, and ultimately improving patient outcomes.

\subsection{Dysarthria Assessment: Perceptual assessment to Automatic assessment}\label{ssec:1-assessment}

The current gold standard for dysarthria assessment relies on the perceptual judgment of trained speech-language pathologists (SLPs) using standardized frameworks such as the Mayo Clinic System for Differential Diagnosis of Dysarthria \citep{simmons1997use, duffy2012motor} and the Frenchay Dysarthria Assessment (FDA, FDA-2) \citep{enderby1980frenchay, enderby2013disorders}. The Mayo Clinic System comprehensively evaluates 38 dimensions across five aspects of speech production, while FDA-2 focuses on eight key domains (see Tables \ref{tab:mayo}, \ref{tab:fda}). These frameworks offer a structured approach, prioritizing speech intelligibility as the primary target for clinical intervention.
\begin{table}[ht]
\centering
\caption{Classification of Dysarthrias in Mayo Clinical Rating System}
\resizebox{\textwidth}{!}{
\begin{tabular}{C{3cm}|C{4cm}|C{8cm}|C{3cm}}
\toprule
\textbf{Types} & \textbf{Localization} & \textbf{Primary Deficit; Auditory Signs} & \textbf{Characteristic Disease(s)} \\ \midrule
Flaccid & Lower motor neuron & Weakness; Hypernasality, breathiness & ALS \\ \midrule
Spastic & Upper motor neuron & Spasticity; Misarticulation, slow speech rate, harsh/strained voice & ALS, CP, MS \\ \midrule
Ataxic & Cerebellum & Incoordination; Monostress phoneme prolongation, slow speech rate & MS, MSA \\ \midrule
Hyperkinetic & Basal ganglia & Rigidity; Monopitch, monoloudness, variable speech rate & HD \\ \midrule
Hypokinetic & Basal ganglia & Involuntary movements; Rapid rate, reduced loudness, monopitch, monoloudness & MSA, PD \\ \midrule
Mixed & Multiple motor systems & More than one & ALS \\  \bottomrule
\end{tabular}
}
\label{tab:mayo}
\begin{spacing}{1.0}
\small
* ALS (Amyotrophic Lateral Sclerosis), CP (Cerebral Palsy), MS (Multiple Sclerosis), MSA (Multiple System Atrophy), HD (Huntington's Disease)
\end{spacing}
\end{table}

\begin{table}[ht]
\centering
\caption{Rating categories in FDA-2.}
\resizebox{\textwidth}{!}{
\begin{tabular}{C{3cm}|C{12cm}}
\toprule
\textbf{Rating section} &  \textbf{Description}\\ \midrule
Reflexes & Ratings for cough, swallow, and dribble/drool \\\midrule
Respiration & Ratings at rest and in speech \\\midrule

Lips & Ratings for at rest, spread, seal, alternate, and in speech \\\midrule
Palate & Ratings for fluids, maintenance, and in speech \\\midrule
Laryngeal & Ratings for time, pitch, volume, and in speech \\\midrule
Tongue & Ratings for at rest, protrusion, elevation, lateral, alternate, an in speech \\\midrule
Intelligibility & Ratings for words, sentences, and conversation \\\midrule
Influencing Factors & Includes hearing, sight, teeth, language, mood, posture, rate (words per minute), and sensation\\
\bottomrule
\end{tabular}
}
\label{tab:fda}
\end{table}

However, reliance on subjective judgment introduces concerns regarding inter-rater and intra-rater reliability \citep{zyski1987identification, kearns1988interobserver, zeplin1996reliability}. Additionally, these evaluations can be time-consuming and labor-intensive, hindering their widespread clinical application.

Recent advancements in spoken language processing have led to the development of automatic methods that promise enhanced objectivity, and efficiency. The ultimate goal of these techniques is to support clinicians by providing quantifiable measures for evaluating dysarthric speech, ultimately striving to achieve assessments comparable to those of professional clinicians. Further details on the previous studies are presented in \Cref{chap:related}.





\subsection{The Need for Multilingual Studies on Dysarthric Speech}\label{ssec:1-nonEnglish}
Despite the promising results of automatic assessment methods, the vast majority of research to date has focused on monolingual environments. This limited scope neglects the needs of the diverse linguistic populations affected by dysarthria worldwide. To address the global burden of dysarthria and ensure equitable access to accurate diagnosis and care, it is imperative to develop assessment tools that can be effectively applied in multilingual settings, specifically catering to underrepresented languages. With multilingual settings, we can anticipate performance enhancements in low-resource or less-studied languages, leveraged by high-resource languages.

Before developing a multilingual dysarthria assessment, it is crucial to understand the characteristics of dysarthria in each analyzed language. While speech motor control and its associated disorders may appear consistent across languages \citep{miller2014introduction}, and it might initially seem reasonable to assume that speech motor activities—such as breathing, vocal fold vibrations, and tongue and lip movements—are identical for speakers of any language, these physical activities are tailored to meet the specific demands of each language \citep{kim2024does}. These differences can be categorized into two aspects: phonemic differences and prosodic differences.

Phonemic differences refer to the varying phoneme inventories across languages, where different values are used to distinguish phonetic or phonemic contrasts. For instance, different voice onset time (VOT) values are employed to differentiate categories in each language \citep{lisker1964cross}, with English having two distinctions (fortis vs lenis), Korean having three (plain vs tense vs aspirated), and Tamil having none. Consequently, speakers with dysarthria who experience similar difficulties in controlling vocal fold vibration and oral constriction might exhibit different stop consonant errors depending on their language. Therefore, the impact of acoustic features is expected to differ across languages. For example, \citet{kim2017cross} found that the abnormal distribution of VOT for stop consonants is a potential language-specific factor affecting intelligibility in Korean dysarthric speech compared to English dysarthric speech. While this primarily relates to plosives, this can be extended to other acoustic features of phones/phonemes, such as centralization for vowels, friction duration for fricatives, degree of nasality for nasals, and phonation contrasts for both plosives and fricatives.

Prosodic differences, which pertain to the rhythmic structure of languages \citep{liss2013crosslinguistic, kim2017cross}, play a crucial role in how easily listeners understand speech. These rhythmic patterns vary significantly across languages and can greatly influence intelligibility. For instance, in stress-timed languages like English \citep{arvaniti2012usefulness}, stress is essential for conveying meaning. In these languages, stressed syllables are typically longer than unstressed ones. Consequently, distortions in segment duration can significantly impact intelligibility, as the length and prominence of stressed syllables are crucial for conveying messages. Conversely, syllable-timed languages, such as Korean \citep{mok2008korean}, rely on relatively equal duration for successive segments to maintain intelligibility. This consistent syllable timing creates a more uniform rhythm, which is vital for listeners to predict and process speech accurately. Disruptions to this rhythmic regularity can lead to intelligibility problems, as listeners rely on the steady pace of syllable production. Mora-timed languages, such as Tamil \citep{nespor201149}, add another layer of complexity. These languages depend on mora timing, where moras are units of sound that are even shorter than syllables. Similar to syllable-timed languages, maintaining consistent mora timing is crucial for intelligibility in mora-timed languages. To sum up, the same speech characteristics may affect speech intelligibility differently, based on the language's rhythm.

In conclusion, to develop effective multilingual dysarthria assessment tools, it is essential to account for the unique characteristics of each language. This understanding will ensure that the tools are robust and sensitive to the specific features of dysarthric speech in diverse linguistic contexts, ultimately resulting in more effective multilingual dysarthria assessment tools. Hence, this paper focuses on three aspects: multilingual analyses of speech characteristics across languages, the proposal of clinical-knowledge-driven features expected to comprehensively reflect dysarthric speech characteristics, and multilingual severity classification.

\section{Research Questions and Thesis Contributions}\label{sec:1-questions}
This thesis delves into the realm of multilingual dysarthria assessment, focusing on three languages with distinct characteristics: English, Korean, and Tamil.

We commence by conducting a comprehensive multilingual analysis to identify the shared and language-specific features that characterize dysarthric speech across these languages. Based on the insights learned from the analyses, we propose a clinically-driven feature set specifically designed to comprehensively capture dysarthric speech characteristics across different languages. Furthermore, we introduce a novel multilingual dysarthria severity classification method that incorporates language-specific characteristics to enhance assessment accuracy.

To guide our research efforts and ensure a targeted approach, we established the following key research questions:
\begin{itemize}
\item \textbf{[Multilingual analysis]} What are the language-universal and language-specific features that characterize degraded speech intelligibility (dysarthria severity level) in English, Korean, and Tamil dysarthric speech?
\item \textbf{[Clinical-knowledge-driven features]} How can we design a clinically-driven feature set that effectively captures degraded speech intelligibility (dysarthria severity levels) across three languages?
\item \textbf{[Multilingual dysarthria severity classification]} How can we develop a robust multilingual dysarthria severity classification model that leverages both shared and language-specific features to improve assessment accuracy across English, Korean, and Tamil?
\end{itemize}

By addressing these critical questions, this thesis strives to contribute to a deeper understanding of dysarthric speech and its manifestation across languages.  Furthermore, we aim to develop novel assessment tools that are not only accurate but also inclusive of diverse linguistic populations.  Our findings have the potential to inform the development of tailored interventions, ultimately improving the quality of life for individuals with dysarthria worldwide.

\section{Thesis Structure}
\label{sec:organization}
The remainder of this thesis is structured as follows. Chapter 2 offers a comprehensive review of existing research on automatic dysarthria severity classification. Chapter 3 presents the datasets utilized in our analyses and details the rationale behind their selection. Chapter 4 conducts a multilingual analysis of speech dimensions using paralinguistic features for dysarthria severity classification, while Chapter 5 focuses on phoneme pronunciation across three languages by proposing an enhanced Goodness of Pronunciation metric. Based on the findings of the previous chapters, Chapter 6 introduces a clinical-knowledge-driven feature set designed to capture the comprehensive aspects of dysarthric speech and validates its efficacy in dysarthria severity classification. Chapter 7 proposes a novel method for multilingual dysarthria severity classification. Finally, Chapter 8 concludes the thesis by summarizing key findings and contributions and offering recommendations for future research directions.

The work presented in \Cref{chapter:pronunciation} presents our previous study, which proposes a novel methodology for multilingual pronunciation assessment of dysarthric speech, employing Goodness of Pronunciation with Uncertainty Quantification leveraged by a cross-lingual pre-trained self-supervised learning (SSL) model\footnote{\cite{yeogop2023} \textbf{Yeo, E. J.*}, Choi, K.*, Kim, S.,  \&  Chung, M. (2023) Speech Intelligibility Assessment of Dysarthric Speech by using Goodness of Pronunciation with Uncertainty Quantification. \textit{In Proc. Interspeech}, pp. 166-170.}.
\Cref{chapter:analysis} and \Cref{chapter:classification} significantly extends upon our previously published research. Specifically, \Cref{chapter:analysis} builds upon our preliminary multilingual analysis of clinical-knowledge-based features\footnote{\cite{yeo2022multilingual} \textbf{Yeo, E. J.}, Kim, S., \& Chung, M. (2022). Multilingual analysis of intelligibility classification using English, Korean, and Tamil dysarthric speech datasets. \textit{In Proc. O-COCOSDA}, pp. 1-6.}. \Cref{chapter:classification} introduces a novel method for multilingual dysarthria severity classification that employs both language-universal and language-specific features\footnote{\cite{yeo2022cross} \textbf{Yeo, E. J.}, Choi, K., Kim, S.,  \&  Chung, M. (2022). Cross-lingual dysarthria severity classification for English, Korean, and Tamil. \textit{In Proc. APSIPA ASC}. pp. 566-574.}.

\chapter{Review of Related studies}\label{chap:related}


\section{An overview of automatic dysarthric speech assessment}\label{sec:2-monolingual}
The field of automatic dysarthric speech assessment has witnessed significant advancements, driven by the need for objective, efficient, and scalable tools to assess dysarthric speech. This section provides an overview of two prominent approaches in this domain: machine learning and deep learning, highlighting key studies and their contributions and limitations.

\subsection{Machine learning approach}
\begin{figure*}
  \centering
\includegraphics[width=0.9\textwidth]{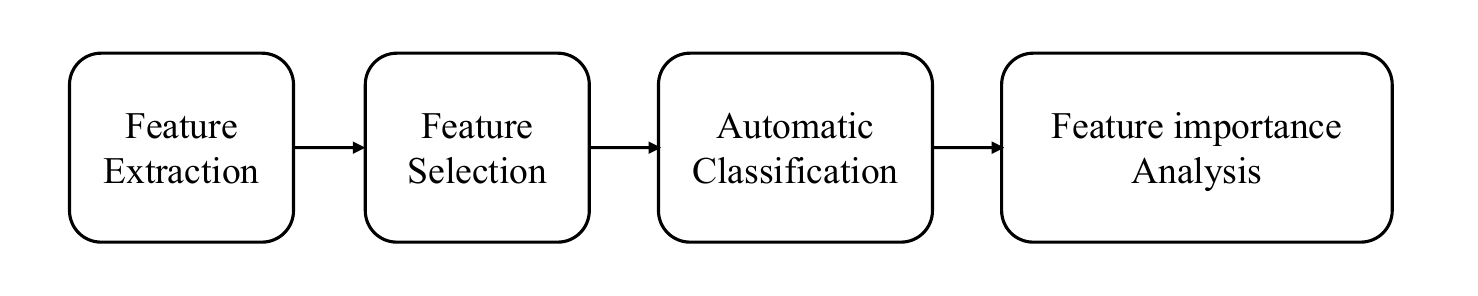}
  \caption{Workflow of machine learning approach.}
  \label{fig:ml-approach} 
\end{figure*}

The machine learning approach to automatic dysarthric speech assessment, as illustrated in Figure \ref{fig:ml-approach}, generally consists of four steps: feature extraction, feature selection, automatic classification, and feature importance analysis. In the feature extraction step, a list of potential voice biomarkers is extracted from the speech signal. Next, in the feature selection step, a subset of features is automatically chosen based on certain criteria, as not all potential features may be relevant to the automatic classification task. The automatic classification step involves training a machine learning classifier, such as Support Vector Machine (SVM), Random Forest (RF), or eXtreme Gradient Boosting (XGBoost), to predict the target variable (e.g., dysarthria severity level or speech intelligibility score) based on the selected features. Finally, feature importance analysis is conducted to determine which features have the greatest impact on the assessment outcome. 

A critical aspect of the machine learning approach is the careful selection of features that can effectively capture the unique characteristics of dysarthric speech. These features can be broadly categorized into two types: paralinguistic features and clinical-knowledge-driven features. \textbf{Paralinguistic features} focus on quantifiable acoustic properties of speech itself. The openSMILE toolkit \citep{eyben2010opensmile} provides standardized feature sets for general acoustic analysis, including ComParE 2016 \citep{schuller2016interspeech}, GeMAPS \citep{eyben2015geneva}, and eGeMAPS \citep{eyben2015geneva}. Among these, eGeMAPS has emerged as the most widely used feature set in dysarthria research, due to its comprehensive representation of various acoustic properties and its proven ability to distinguish between dysarthric and healthy speech. For instance, \citet{xue2019acoustic} reported that eGeMAPS features could effectively discriminate between the two groups, achieving multiple R-squared scores up to 0.6823 for the dysarthria detection task. Similarly, \citet{vanbemmel23_interspeech} demonstrated the efficacy of eGeMAPS for both dysarthria detection and speech intelligibility score regression, achieving 98.06\% accuracy for classification and 9.89 Root Mean Squared Error (RMSE) and R² of 0.65 for regression. This consistent performance has led to the adoption of eGeMAPS as a baseline feature set in numerous dysarthric speech assessment studies \citep{narendra2021automatic, pesenti2022effect, yeo2023automatic, javanmardi2024pre}.
Another paralinguistic feature set is the DisVoice feature set \citep{DisVoice}, which is a paralinguistic feature set for disorder speech analysis. Its advantage comes from designing feature sets to reflect the speech dimensions, which are glottal, phonation, articulation, and prosody. \citet{joshy2023dysarthria} found its efficacy in dysarthria severity classification, achieving up to 85.65\% classification accuracy in the English dataset. They also found that articulation and prosody dimensions are particularly efficient in English dysarthria severity classification.

While paralinguistic features offer a valuable starting point, they may not fully capture the nuanced acoustic and articulatory characteristics of dysarthric speech that are relevant to clinical diagnosis and treatment planning. To address this, many studies have focused on investigating \textbf{clinical-knowledge-driven features}, specifically tailored to the pathophysiological aspects of dysarthria. For example, \citet{narendra2018dysarthric} extracted glottal source parameters to quantify the impaired phonatory control in dysarthric speech patients, achieving 93.52\%, 94.29\%, and 91.38\% accuracy for non-words, words, and sentence materials, respectively, when combined with the openSMILE toolkit. Similarly, \citet{kadi2013discriminative} integrated voice quality and prosodic features, demonstrating the effectiveness of this approach with a 93\% accuracy in classifying dysarthria severity levels.  In the realm of pronunciation assessment, \citet{hernandez2019acoustic} focused on acoustic features of fricatives, finding that fricative duration and spectral moments can effectively distinguish dysarthric speech from healthy speech with an accuracy of 82\%. Furthermore, \citet{yeo2021automatic} utilized phoneme-level pronunciation features for dysarthria severity classification, achieving 77.38\% F1-score with the SVM classifier, while \citet{kim2012automatic} used pronunciation error patterns (correct, substitution, deletion) as input features for intelligibility prediction, achieving 8.14 RMSE. Moreover, to comprehensively reflect the multifaceted nature of dysarthria, researchers often strategically combine features from different speech dimensions, such as voice quality, articulation/pronunciation/phonetic quality, and prosody \citep{kim2012combination, vasquez2018towards}. Experiments have consistently demonstrated that integrating features across multiple dimensions leads to enhanced performance in dysarthria assessment tasks, supporting clinical findings that speech intelligibility of dysarthric speakers are affected by multiple speech dimensions.

In addition to feature extraction, feature selection techniques have been employed to refine the feature set further. Feature selection is not only valuable for enhancing classification performance but also for increasing computational efficiency and interpretability by identifying the most salient voice biomarkers. For instance, \citet{vanbemmel23_interspeech} employed LASSO regression as a feature selection method to mitigate overfitting, while \citet{yeo2021automatic} incorporated both Recursive Feature Elimination (RFE) and Extra Trees Classifier (ETC), demonstrating improved classification accuracy after feature selection. \citet{hernandez2020prosody} presented a comprehensive experimental results, employing four feature selection methods, which includes filter, LASSO, RFE, and ETC method for English and Korean dysarthria severity classification.

Furthermore, analysis of feature importance scores has been used to gain deeper insights into the specific speech characteristics that drive automatic assessment. This can be achieved through methods like coefficient analysis of linear regression models \citep{wei23_slate} or the internal scoring mechanisms of tree-based classifiers \citep{yeo2021automatic, yeo2022cross}. Additionally, SHAP (SHapley Additive exPlanations) values \citep{lundberg2017unified}, a model-agnostic method, have been employed to further elucidate feature importance \citep{kovac2021multilingual, zhang2023artificial}. These analyses reveal that different features have varying impacts on the assessment task, therefore can provide clinicians with a more nuanced understanding of the underlying factors contributing to dysarthria. Ultimately, this knowledge can inform more targeted and effective diagnosis and treatment planning.

\subsection{Deep learning approach}
Deep learning has emerged as a promising avenue for automatic dysarthric speech assessment, leveraging neural networks' ability to learn complex representations directly from raw speech data, such as waveforms or spectrograms. This eliminates the need for explicit feature engineering and potentially reveals subtle patterns indicative of dysarthria that might be missed by hand-crafted features through feature engineering.

Researchers have explored various neural network architectures for this purpose, including distance-based convolutional neural networks (CNNs) for capturing local spectral patterns \citep{janbakhshi2021automatic}, LSTM-based models for modeling temporal dependencies in speech signals \citep{mayle2019diagnosing, bhat2020automatic}, and hybrid CNN-RNN models for combining both spatial and temporal information \citep{shih2022dysarthria, ye2022hybrid}. While these models have shown promise, their widespread application in dysarthria assessment was initially limited by the need for large annotated datasets to achieve optimal performance. Consequently, research in this area often focused on the binary task of dysarthria detection, leaving more nuanced assessments, like severity classification, relatively unexplored.

However, recent advancements in self-supervised learning (SSL) have transformed the landscape of automatic dysarthria assessment. SSL models, such as wav2vec 2.0 and its variants \citep{baevski2020wav2vec, babu22_interspeech}, HuBERT \citep{Hsu2021HuBERTSS}, and WavLM \citep{Chen2021WavLMLS}, are pre-trained on vast amounts of unlabeled speech data, allowing them to learn generalizable speech representations. These pre-trained models can then be fine-tuned for specific tasks using smaller amounts of labeled data, addressing the issue of data scarcity that previously hindered the application of deep learning for dysarthric speech assessments. Leveraging a large amount of healthy training data, SSL models achieve general acoustic information, which can be efficiently fine-tuned with small amounts of dysarthric speech recordings. This paradigm shift has led to state-of-the-art performance in various downstream tasks for pathological speech, including automatic speech recognition (ASR) \citep{hernandez2022cross, violeta2022investigating} and automatic speech assessments \citep{jiang2021towards, bayerl2022detecting, getman2022wav2vec2, bayerl2022detecting, favaro2023interpretable}. 

 \begin{figure*}
  \centering
\includegraphics[width=0.8\textwidth]{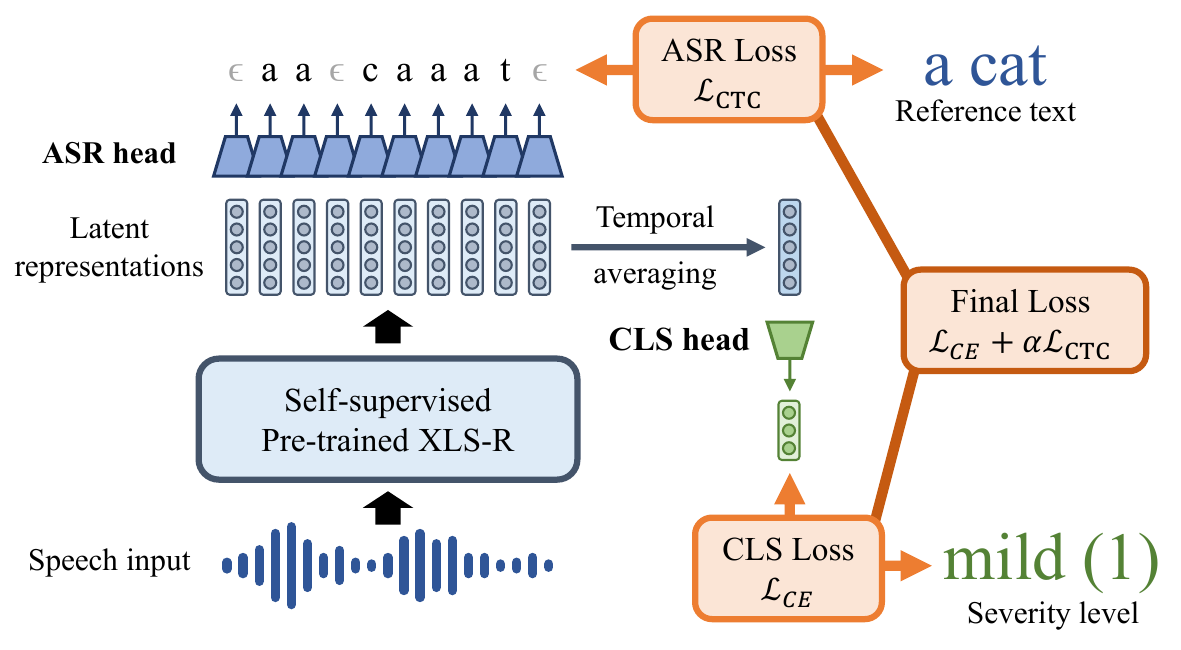}
  \caption{Example of deep learning approach. The method leverages SSL models fine-tuned with Multi Task Learning \citep{yeo2023automatic}.}
  \label{fig:DL-overview} 
\end{figure*}

To further enhance the performance and generalization capabilities of deep learning models, researchers have explored multi-task learning (MTL). This technique involves training a model on multiple related tasks simultaneously, allowing the model to leverage shared knowledge across tasks and potentially improve performance on each individual task. \citet{bayerl2022detecting} employed MTL setting stuttering detection as a main task and gender classification as an auxiliary task. \citet{yeo2023automatic} combined dysarthria severity classification with ASR as an auxiliary task (see Figure \ref{fig:DL-overview}). Both showed the effectiveness of the MTL framework compared to the Single Task Learning (STL) framework, which only trains main tasks, excluding the auxiliary task. \citet{joshy2023dysarthria} proposed a framework that employs both multi-head attention and multi-task learning for dysarthria severity classification. Experiments have shown that the effectiveness of the MTL framework achieves better performances compared to the Single Taks Learning (STL) framework, which trains main tasks only, excluding the the auxiliary task.

While self-supervised learning (SSL) has propelled deep learning models to state-of-the-art performance in automatic dysarthria assessment, consistently outperforming traditional hand-crafted feature-based approaches \citep{favaro2023interpretable, yeo2023automatic}, a significant challenge remains: the inherent difficulty in interpreting the decision-making process of these complex models. In the clinical context, interpretability is paramount for establishing trust with clinicians and patients, understanding the rationale behind assessment outcomes, and facilitating informed treatment decisions based on a clear understanding of the underlying factors contributing to dysarthria.

Recent studies have focused on developing interpretable neural networks for automated dysarthric speech assessment \citep{korzekwa19_interspeech, tu2017interpretable, xu2023dysarthria}. In \citet{korzekwa19_interspeech}, end-to-end deep learning models, including CNNs, RNNs, and transformers, are utilized to evaluate dysarthric speech intelligibility. These models are trained directly on raw speech signals, enabling them to learn representations that capture speech patterns relevant to intelligibility assessment. To address the black-box nature of neural networks, the authors employ attention mechanisms and visualization techniques to identify which parts of the speech signal the model focuses on during the assessment. \citet{tu2017interpretable} employs intermediate representations that can be mapped to perceptual speech attributes such as nasality, voice quality, articulatory precision, and prosody. Networks are trained on datasets annotated with these perceptual attributes, allowing them to learn correlations between speech signal features and perceptual labels. 
Similarly, \citet{xu2023dysarthria} presents a model with a bottleneck layer trained to jointly learn a severity classification label and four clinically interpretable features: articulation precision, consonant-vowel transition precision, hypernasality, and voice quality. This approach ensures that the model not only predicts the severity of dysarthria but also provides insights into specific speech characteristics that influence the overall assessment.

In other aspects of interpretable assessments, certain studies have concentrated on decoding misarticulation characteristics, a prominent feature of dysarthric speech across languages and a key factor affecting speech intelligibility \citep{yeo2022multilingual, de2002intelligibility}. For instance, a framework was proposed to measure the level of phonetic impairment using the activations of hidden neurons \citep{abderrazek2022validation, abderrazek2022interpreting}. While this method provides an overall assessment of phonetic characteristics in utterances, it falls short of offering phoneme-level evaluations, which are crucial for identifying specific phonemes that need pronunciation training in clinical settings.

A common phoneme-level speech assessment approach involves using parallel neural networks (NNs) trained on parallel datasets. These NNs are trained with the same set of utterances recorded by both healthy speakers and patients to distinguish whether each phoneme in the utterances is from healthy or disordered speech \citep{miller2020assessing, quintas2022automatic}. However, acquiring parallel datasets is challenging, especially for disordered speech, and this approach often limits the analysis to predefined speech materials, which may not reflect the natural speech patterns used in everyday communication.
Another conventional approach for phoneme-level pronunciation evaluation is the Goodness of Pronunciation (GoP) method \citep{witt2000phone}. GoP measures the similarity between produced and canonical phoneme pronunciations and has two significant advantages in automatic speech assessments. While GoP has frequently been used to assess non-native (L2) speech pronunciation, applications on speech disorders have been under-studied \citep{pellegrini2014goodness, Fontan2015PredictingDS}. We propose a novel GoP metric for dysarthric speech assessment in \Cref{chapter:pronunciation}.

\section{An overview of multilingual analysis on dysarthric speech}\label{ssec:1-multilingual}
A few studies have analyzed voice biomarkers from a multilingual cohort of participants. \cite{whitehill2010studies} identified that the most affected speech dimensions in Cantonese-speaking and English-speaking individuals with dysarthria associated with PD include reduced pitch, loudness variation, and imprecise consonants. \cite{rusz2021speech} performed a speech analysis on Czech, English, German, French, and Italian speakers in the early stages of PD, involving 448 participants. They found that monopitch, longer pauses, and imprecise consonants significantly distinguished subjects from healthy controls and PD patients. \cite{kovac2021multilingual} focused on identifying acoustic features whose effectiveness in dysarthria detection is independent of the speaker's language. They recorded seven monologues and sustained phonation tasks from 59 PD patients and age- and gender-matched healthy controls speaking Czech or American English. Their statistical test revealed that the number of interword pauses per minute has the best discrimination power between the two groups. \cite{kovac2022exploring} conducted a multilingual analysis using speech recordings from 214 Czech-speaking subjects, 29 American English-speaking subjects, 115 Israeli-speaking subjects, 100 Colombian Spanish-speaking subjects, and 48 Italian-speaking subjects. The biomarkers assessing the prominence of the second formant, monopitch, and the number of pauses during text reading showed the highest effectiveness in dysarthria detection. \cite{favaro2023multilingual} analyzed a total of 24 voice biomarkers from American English, Italian, Castilian Spanish, Colombian Spanish, German, and Czech. Their statistical analysis revealed that the best acoustic biomarkers across languages were F0 standard deviation, duration of pause, the percentage of pauses, duration of silence, and speech rhythm standard deviation. To sum up, multilingual studies on dysarthria assessment have focused on identifying features that show the same pattern across languages. Such features are referred to in various terminologies by studies, including language-agnostic, language-independent, and language-universal features. It is important to consider such terminology narrowly, meaning that the features show the same trend for the languages analyzed in the study.

While multilingual analyses have predominantly focused on identifying language-universal biomarkers, some studies have also highlighted the language-specific aspects of voice biomarkers. For instance, \cite{kim2017cross} examined the differences in acoustic vowel space (AVS) and voice onset time (VOT) contrast scores between Korean- and American English-speaking subjects with PD. Their statistical test revealed that the abnormal distribution of VOT could be a potential language-specific contributor to intelligibility in dysarthria. Furthermore, different multiple regression models were found in American English and Korean, even with the same predictor variables and with speakers matched on speech intelligibility across languages. \cite{hernandez2020prosody} suggested that different prosodic feature lists are required depending on the language (American English and Korean). \cite{yeo2022multilingual} presented that while pronunciation features were found to be language-universal, voice quality features and prosodic features tend to present different patterns across English, Korean, and Tamil.

These collective findings highlight the complex interplay between language-universal and language-specific factors in dysarthric speech. A comprehensive understanding of both is paramount for the field of speech pathology. Research into language-universal features can inform the development of standardized assessment and diagnostic tools with broad applicability, improving efficiency and cross-cultural utility. Concurrently, investigating language-specific features is essential for creating finely-tuned tools that accurately assess and diagnose dysarthria within specific linguistic contexts. This approach has the potential to significantly enhance diagnostic accuracy, enabling more targeted and effective interventions for individuals with dysarthria across diverse linguistic backgrounds.

\section{An overview of multilingual automatic assessments on dysarthric speech}\label{ssec:1-automatic}
While automatic dysarthric speech assessment has demonstrated promising results in monolingual settings, its application to diverse linguistic populations remains an active area of research. This section explores existing studies on multilingual approaches to dysarthria assessment, highlighting the challenges encountered, strategies employed, and future directions for the field.

Multilingual assessments typically involve training machine learning models on combined data from multiple languages, with the goal of identifying language-universal acoustic voice biomarkers that generalize across linguistic contexts. While promising in its potential for efficient and scalable assessment tools, this approach has limitations. \citet{orozco2016automatic} observed varying prediction accuracy across different language combinations (60-99\%), compared to higher accuracy within single language groups (85-99\%). Similarly, \citet{kovac2021multilingual}, using logistic regression, reported a decline in classification accuracy when transitioning from monolingual to multilingual assessment. This suggests that relying solely on universal features may not adequately capture the nuanced manifestations of dysarthria across diverse languages. These findings underscore the need to consider both language-universal and language-specific features in multilingual dysarthria assessment. While universal features may capture fundamental aspects of dysarthria that transcend language barriers, incorporating language-specific features can improve the sensitivity and specificity of assessment tools. 




The current landscape of multilingual dysarthria assessment underscores the need for continued research in this domain. While promising strides have been made in identifying language-universal features, the under-exploration of language-specific features and the challenges in cross-language generalization remain significant obstacles. We anticipate this thesis to make significant strides in this aspect, ultimately improving the diagnosis and treatment of individuals with dysarthria worldwide. Ultimately, the goal of multilingual dysarthria assessment is to create inclusive, efficient, and precise tools that can accurately evaluate dysarthric speech across diverse linguistic populations.



\chapter{Dysarthric Speech Datasets}\label{chap:datasets}


\section{An overview of Dysarthric speech datasets}
Table \ref{tab:dysarthric_db} provides a comprehensive overview of seven publicly available dysarthric speech datasets described in the literature. For each dataset, the table outlines key details such as language, underlying disease type, the number of speakers and utterances, type of speech material, and public availability status.

\begin{table}[htbp]
\centering
\caption{Dysarthric Speech Datasets}
\label{tab:dysarthric_db}
\resizebox{\textwidth}{!}{%
\begin{tabular}{@{} p{2cm} p{3cm} p{2cm} p{3cm} p{3cm} p{7cm} p{2cm}@{}}
\toprule
Dataset & Language & Disease Type & \# of Speakers & \# of Utterances & Material Type & Availability \\
\midrule
UA Speech & American English & CP & 19 dysarthria & 10,279 dysarthria & \begin{tabular}[c]{@{}l@{}}Prompt words (10 digits, 26 radio \\ alphabet letters, command words, \\ common words, uncommon words)\end{tabular} & O \\
\midrule
Nemours & American English & CP or head trauma & 11 dysarthria & 814 recordings & Nonsense sentences (the X is Y the Z) & X \\
\midrule
\multirow{2}{*}{TORGO}  & American English & CP, (ALS) & 7 healthy, & 5,980 healthy,  & Words, sentences, describing a photo & O \\
& & &  8 dysarthria  & 2,762 dysarthria & \\
\midrule
\multirow{2}{*}{EasyCall} & Italian & PD, HD, & 24 healthy, & 396-462 healthy,  & Command words, non-commands & O \\
& &ALS  &  31 dysarthria & 18-414 dysarthria & \\
\midrule
\multirow{2}{*}{QoLT} & Korean & CP & 10 healthy,& 100 healthy, & Sentences & O \\
& & &   70 dysarthria  & 700 dysarthria & \\
\midrule
\multirow{2}{*}{SSNCE}  & Tamil & CP & 10 healthy, & 3,650 healthy, & Words, sentences & O 
\\
& &  &  30 dysarthria &10,950 dysarthria  & \\
\midrule
\multirow{2}{*}{PC-GITA} & Colombian  & PD & 50 healthy, & - & Vowels, DDK, words, sentences, & X \\
&Spanish &  &   50 dysarthria   &   & reading text, monologue &\\

\bottomrule
\end{tabular}%
}
\end{table}

For the present study, we selected datasets based on two criteria: (1) public availability, ensuring accessibility and reproducibility of our research, and (2) the inclusion of sentence-level speech materials. Sentence-level recordings were preferred over word-level data due to their richer acoustic information, particularly regarding prosodic features, which are crucial for our comprehensive evaluation of acoustic features across various speech dimensions, including voice quality, pronunciation, and prosody. 
Based on these criteria, three publicly available datasets were chosen for analysis: TORGO (English), QoLT (Korean), and SSNCE (Tamil). Table \ref{tab:dataset} provides a summary of the number of speakers and utterances within each of these datasets. Moreover, different etiology of dysarthric speech can result in very different acoustic characteristics \citep{simmons1997use, duffy2012motor}. Note that all three datasets include speech recordings from patients diagnosed with Cerebral Palsy (CP), except for one male mild dysarthria speaker diagnosed with Amyotrophic Lateral Sclerosis (ALS) in the TORGO English dataset. 

\section{TORGO English dataset}
The TORGO dataset \citep{rudzicz2012torgo} encompasses speech samples from 15 participants: 7 healthy controls (4 male, 3 female) and 8 individuals with dysarthria (5 male, 3 female). Dysarthria severity was initially classified based on Frenchay assessment scores \citep{enderby1980frenchay}, yielding categories of mild (2 speakers), mild-to-moderate (1 speaker), moderate-to-severe (1 speaker), and severe (4 speakers). To ensure balanced sample sizes across languages and enhance comparability, categories with single speakers were merged, resulting in a moderate category with two speakers. For analysis, a total of 570 utterances were utilized (156 healthy, 414 dysarthric).

\section{QoLT Korean dataset}
The QoLT dataset \citep{choi2012dysarthric} comprises Korean dysarthric speech from 10 healthy controls (5 male, 5 female) and 70 individuals with Cerebral Palsy (45 male, 25 female). Five speech pathologists rated speech intelligibility on a 5-point Likert scale, classifying dysarthria severity into mild (25 speakers), mild-to-moderate (26 speakers), moderate-to-severe (12 speakers), and severe (7 speakers). For consistency across languages, mild-to-moderate and moderate-to-severe groups were combined, resulting in 40 moderate speakers. The corpus contains five unique sentences repeated twice per speaker, yielding 800 total utterances (100 healthy, 700 dysarthric).

\section{SSNCE Tamil dataset}
The SSNCE dataset \citep{ta2016dysarthric} is a publicly available collection of Tamil dysarthric speech from 10 healthy controls (5 male, 5 female) and 20 individuals with Cerebral Palsy (13 male, 7 female). Two speech pathologists assessed speech intelligibility on a 7-point Likert scale (0: healthy, 1-2: mild, 3-4: moderate, 5-6: severe), resulting in 7 mild, 10 moderate, and 3 severe dysarthric speakers. The dataset contains 262 unique phrases per speaker, representing both common and uncommon utterances. A total of 7860 sentences (2620 healthy, 5240 dysarthric) were analyzed in this study.

\begin{table}[h]
\caption{
The number of speakers and utterances
}
\label{tab:dataset}

\centering
\resizebox{0.85\textwidth}{!}
{
\begin{tabular}{c|c|c|c|c|c|c|c|c}
\hline
\multirow{2}{*}{Language} & \multicolumn{2}{c|}{healthy} &\multicolumn{2}{c|}{mild} & \multicolumn{2}{c|}{moderate} & \multicolumn{2}{c}{severe} \\
\cline{2-9}
& spk & utt & spk & utt & spk & utt & spk & utt \\
\hline
English & 7 & 156 & 3 & 220 & 2 & 88 & 3 & 106 \\
\hline
Korean & 10& 100 & 25 & 250 & 38 & 380 & 7 & 70 \\
\hline
Tamil & 10 & 2620 & 7 & 1834 & 10 & 2620 & 3 & 786 \\
\hline
\end{tabular}
}
\end{table}
\chapter{Multilingual Speech Dimension Analysis for Dysarthric Speech Assessment Using Paralinguistic Features}\label{chap:paralinguistics}

\section{Background and Research questions}
While the Mayo Clinic System and FDA-2 have been translated and adapted for use in various languages, concerns remain regarding their cross-linguistic validity and cultural sensitivity \citep{liss2013crosslinguistic, kim2024does}. These frameworks were primarily developed based on observations of individuals speaking American English, raising questions about the applicability of the classification criteria to dysarthria manifestations in speakers of different native languages \citep{kim2017cross, kim2024does}. To ensure accurate assessment, it is crucial to understand both the similarities and differences in dysarthric speech manifestations across languages.

As previously mentioned in Chapter \ref{chap:related}, the efficacy of paralinguistic features in automatic dysarthria assessment has been validated through previous studies. While it is crucial to consider a wide range of features, not all features hold equal significance. Moreover, in high-dimensional datasets, some features may be highly correlated, which leads to multicollinearity issues that can compromise the accuracy and reliability of the feature importance analysis \citep{ayesha2020overview}. To address this, it is essential to eliminate highly correlated, redundant features. Various studies have explored effective techniques for feature selection. For instance, \cite{fewzee2012dimensionality} compared dimension reduction techniques, including Elastic Net regression and Principal Component Analysis (PCA), in emotional speech recognition tasks. Similarly, \cite{wei23_slate} investigated the use of LASSO regression, Elastic Net regression, and Redundant Variable Removal (RVR) to mitigate multicollinearity problems in the context of speech intelligibility assessment for L2 speech. Additionally, \cite{vanbemmel23_interspeech} performed LASSO analysis before regression and classification experiments, to mitigate multicollinearity and avoid overfitting.

Building on these approaches, we introduce a two-step feature selection method (TSFS) that employs three conventional approaches for solving multicollinearity (LASSO, Elastic Net, and Hierarchical Clustering), followed by traditional feature selection methods (filter, wrapper, and embedded methods). This method aims to identify the optimal subset of features for dysarthria severity classification in each language while avoiding multicollinearity and overfitting. The ultimate objective of this research is to comprehensively understand the speech dimensions affected by dysarthria through a detailed analysis of paralinguistic features, leveraging our new TSFS method. Furthermore, through multilingual analysis, we aim to identify both universal and language-specific aspects of speech intelligibility in dysarthric speech. This approach will help in developing more accurate and robust assessment tools tailored to different linguistic backgrounds. Our investigation is guided by the following research questions:

\begin{itemize}
\item How effective is the proposed Two-Step Feature Selection method?
\item What are the most informative paralinguistic features for dysarthria severity assessment in English, Korean, and Tamil?
\item How are the importance of speech dimensions similar and different across English, Korean, and Tamil?
\end{itemize}

\section{Paralinguistic features}
Acoustic features were extracted using the DisVoice Python library \citep{vasquez2018towards}. DisVoice is an open-source acoustic feature extraction toolkit designed to comprehensively capture the characteristics of speech disorders across diverse dimensions: glottal, phonation, articulation, and prosody features. Notably, DisVoice excels in its consideration of speech disorder attributes, facilitating easier interpretation compared to other publicly available acoustic feature sets, such as ComParE 2016 \citep{schuller16_interspeech} and eGeMAPS \citep{eyben2015geneva}, which are primarily designed for general voice analysis.

In our analysis, we excluded glottal features, as these are primarily tailored for extraction from sustained vowel recordings rather than sentence recordings. Consequently, our analysis encompasses 619 acoustic features derived from 46 low-level descriptors (LLDs), reflecting various aspects of phonation, articulation, and prosody. To the best of our knowledge, this represents the most comprehensive and extensive feature set analyzed in the context of multilingual studies for dysarthric speech assessments. The following sections briefly introduce the features extracted for the analysis. Moreover, we describe several hypotheses on which speech dimensions are expected to have greater importance on dysarthria severity level estimation for each English, Korean, and Tamil.

\subsection{Phonation features}
Phonation features capture irregular patterns in vocal fold vibrations and are derived from voiced segments where the vocal folds vibrate. These features can be extracted from both sustained vowels and continuous speech. Common phonation impairments are analyzed using stability-related features such as jitter and shimmer, which represent variations in the fundamental period and amplitude of the voice signal, respectively. Pitch perturbation quotient (PPQ) measures long-term perturbations of the fundamental period, while amplitude perturbation quotient (APQ) captures long-term variations in peak-to-peak amplitude. Additionally, the first and second derivatives of the fundamental frequency (F0) and the signal's energy content are included. APQ and PPQ are calculated using 100 ms frames with an 80 ms time shift, while other phonation features are computed using 20 ms frames with a 15 ms time shift. Four statistical functionals (mean, standard deviation, skewness, and kurtosis) are calculated per feature, resulting in a 28-dimensional feature vector for each utterance.

\subsection{Articulation features}
Articulation LLDs include spectral features of onset transitions (unvoiced to voiced segments), spectral features of offset transitions (voiced to unvoiced segments), and formants. The spectral features are designed to model the difficulties dysarthric patients face in initiating and stopping vocal fold movements. Transitions are detected by identifying the presence of the fundamental frequency. To capture these transitions, 40 ms chunks are taken from both sides of each border. The energy content during these transitions is modeled using 22 frequency bands based on the Bark scale, along with 12 MFCCs and their first and second derivatives, resulting in a total of 116 spectral features.

Formant features capture two key abilities of the speaker: maintaining the tongue in a specific position during phonation and producing vowels accurately. Six features are extracted in total, including the first and second formants and their respective first and second derivatives.

\subsection{Prosody features}
Prosody LLDs are further categorized into features related to fundamental frequency (F0), energy, and duration. F0-based features encompass measures like the F0-contour, tilt, and mean squared error (MSE) of a linear estimation of F0 for each voiced segment. Energy-based features include analyses of the energy contour for voiced and unvoiced segments, tilt, MSE, and energy measures for the first and last segments. Duration-based features focus on temporal aspects, including voiced rate, duration of voiced/unvoiced segments and pauses, and various duration ratios. These features collectively provide a comprehensive representation of the timing and rhythm of speech.

\subsection{Hypothesis}\label{ssec:hypothesis}
As discussed in \Cref{ssec:1-nonEnglish}, different phonemic and prosodic systems across languages may result in language-specific characteristics for speech intelligibility degradation. While phonation features and articulation features are expected to be related to phonemic structure, prosodic features are expected to depend on prosodic structure. For example, in terms of \textit{phonation} features, English is expected to place greater importance on these features compared to Korean and Tamil. This is because voicing contrasts are used in English consonants \citep{wright2004review, raphael2021acoustic}, whereas neither Korean nor Tamil utilize these contrasts to the same extent.

In terms of \textit{articulation features}, spectral onset features are expected to be more significant than formants and spectral offsets in all three languages because the cognitive load for perception is greatest at the onset of syllables \citep{mattys2011effects}. Regarding \textit{formants}, vowel space density is known to significantly impact speech intelligibility \citep{kim2017cross}. The compression of the vowel space, frequently observed in dysarthric speech \citep{kim2011acoustic}, may have a lesser effect on intelligibility in Korean, as the loss of contrast among vowels is presumed to be more substantial in the denser vowel space of English and Tamil. Consequently, it is expected that formant features will have greater importance in English and Tamil compared to Korean, where Korean has 7 monophthongs while English and Tamil have over 10.

Lastly, in terms of \textit{prosodic features}, the impact on speech intelligibility will vary according to the rhythmic structure of each language. In English, a stress-timed rhythm emphasizes the duration and stress of syllables, making distortions in these areas particularly detrimental to intelligibility. Therefore, prosodic features related to F0 and energy are expected to have greater importance in English because variations in pitch (F0) and loudness (energy) are crucial for marking stress patterns, which are essential for conveying meaning. Conversely, Korean, with its syllable-timed rhythm, requires consistent syllable duration, and disruptions can lead to intelligibility issues. Thus, duration-related features are expected to be more important in Korean. Tamil, as a mora-timed language, relies on consistent mora timing, making variations in these shorter units crucial for maintaining intelligibility. Therefore, duration-related features will also be crucial in Tamil, with a focus on maintaining consistent mora timing.


\section{Proposed Analysis Method}\label{sec:method}
\subsection{Two-step Feature Selection Method (TFSM)}
While a wide range of features was considered, not all features are equally important. In high-dimensional data, some features may be highly correlated, leading to multicollinearity issues that can affect the analysis results. Therefore, it is crucial to remove highly correlated, redundant features. Disvoice features, in particular, are expected to exhibit high multicollinearity due to their inclusion of statistical values of LLDs. Consequently, we implement a Two-step Feature Selection Method (TFSM). First, we apply feature selection techniques that are known to mitigate multicollinearity between features. Following this, we employ traditional feature selection methods, including filter, wrapper, and embedded methods, to refine the feature set further.

\subsubsection{Step 1. Mitigating Multicollinearity}
Multicollinearity occurs when two or more features in a dataset are highly correlated, resulting in redundancy and potential issues during model training and feature importance analysis. To address this, we implement several techniques to mitigate multicollinearity: Lasso, Elastic Net, and Clustering. 

\subsubsection{1-1. Lasso}
Lasso is a regularization technique that adds a penalty term proportional to the absolute value of the magnitude of coefficients to the loss function. This penalty encourages sparsity, driving the coefficients of less important or redundant features towards zero, effectively performing feature selection. By eliminating these less informative features, Lasso can reduce multicollinearity and enhance model interpretability.

\subsubsection{1-2. Elastic Net}
Elastic Net is a hybrid regularization technique that combines the penalties of both Lasso (L1) and Ridge (L2) regression.  This method balances feature selection (L1) with coefficient shrinkage (L2), making it particularly well-suited for situations where predictors are highly correlated. The Elastic Net penalty is controlled by two parameters: alpha, which determines the overall strength of the penalty, and L1 ratio, which controls the balance between the L1 and L2 penalties. By adjusting these parameters, Elastic Net can effectively mitigate multicollinearity while retaining important features and preventing excessive coefficient shrinkage.

\subsubsection{1-3. Clustering}
This method offers a distinct approach to addressing multicollinearity by grouping similar features together based on their correlation. By identifying clusters of highly correlated features, we can select a single representative feature from each cluster, effectively reducing redundancy in the feature space. This not only simplifies the model but also mitigates the adverse effects of multicollinearity on model stability and interpretation.

\subsubsection{Step 2. Feature selection}
Feature selection is a critical phase known to enhance model performance by identifying and retaining only the most relevant features for the classification task. Our objective is to determine the optimal set of acoustic features for English, Korean, and Tamil, by employing conventional feature selection techniques. Following \citet{hernandez2020dysarthria}, we explore three main types of feature selection methods: filter, wrapper, and embedded methods, utilizing the scikit-learn library \citep{scikit-learn}.

\subsubsection{2-1. Filter methods}
Filter methods assess the relevance of features based on their intrinsic properties, independent of any machine learning algorithm. These methods are computationally efficient and help in reducing the dimensionality of the data before applying any learning algorithm. In our study, we employ Kendall's $\tau$ Rank Correlation test, a non-parametric method that evaluates the association between each acoustic feature and the dysarthria severity level. This test identifies significant differences between the levels of an independent variable (acoustic feature) and a dependent variable (dysarthria severity level). Features with p-values lower than 0.05 are considered statistically significant and are selected for further analysis. This approach helps in quickly filtering out irrelevant features, ensuring that only those with a strong statistical relationship to the outcome variable are retained.

\subsubsection{2-2. Wrapper methods: Recursive Feature Elimination (RFE)}
Wrapper methods select feature subsets based on the performance of a specified learning algorithm. These methods involve training and evaluating the model multiple times to find the best subset of features, thus being more computationally intensive compared to filter methods. We utilize Recursive Feature Elimination (RFE) with Cross-Validation (RFECV) to iteratively eliminate the least significant features. RFE works by recursively fitting the model and removing the least important feature at each step. Cross-validation is integrated to validate the performance of the model on different subsets of the data, ensuring robustness. The feature set that achieves the highest classification accuracy during cross-validation is identified as the optimal subset. This method helps in selecting features that contribute the most to the model's predictive power.

\subsubsection{2-3. Embedded methods: Extra Trees Classifier (ETC)}
Embedded methods perform feature selection as part of the model training process. These methods are advantageous as they integrate feature selection and model building into a single step, often leading to more efficient and effective models. We implement the Extra Trees Classifier (ETC), an ensemble learning method that constructs multiple randomized decision trees. ETC ranks features based on their importance, which is determined by the decrease in the impurity criterion (e.g., Gini impurity or entropy) each feature brings when used in a tree. This embedded method is particularly useful for handling high-dimensional data, as it naturally selects features that enhance the model's accuracy. By incorporating feature selection within the training process, ETC ensures that the most informative features are used to build the model, leading to potentially better generalization on unseen data.

\subsection{Automatic dysarthria severity classification}
The machine learning classifiers were trained to distinguish the severity levels of each utterance into four categories: healthy (0), mild (1), moderate (2), and severe (3). 
We employed three conventional machine learning classifiers: K-Nearest Neighbor (kNN), Random Forest (RF), Support Vector Machine (SVM), Multiple Linear Perception (MLP), and extreme Gradient Boosting (XGBoost) algorithms. 

\subsubsection{k-Nearest Neighbors (kNN)}
The k-Nearest Neighbors (k-NN) algorithm is a simple, non-parametric method used for classification and regression. In k-NN classification, an object is classified by a majority vote of its neighbors, with the object being assigned to the class most common among its k nearest neighbors (where k is a positive integer, typically small). For regression, the output is the average of the values of its k nearest neighbors. The algorithm is intuitive and straightforward, but it can be computationally expensive for large datasets because it requires calculating the distance between the query and all examples in the dataset.

\subsubsection{Random Forest (RF)}
The Random Forest (RF) algorithm is an ensemble learning method primarily used for classification and regression tasks. It operates by constructing a multitude of decision trees during training time and outputting the class that is the mode of the classes (classification) or the mean prediction (regression) of the individual trees. Each tree in the forest is built from a random subset of the training data using a technique known as bootstrap aggregation, or bagging. Additionally, random subsets of features are considered for splitting nodes, which introduces diversity among the trees and helps in reducing overfitting. This method improves the robustness and generalizability of the model. Random Forest is known for its high accuracy, ability to handle large datasets with higher dimensionality, and robustness to noise and overfitting, making it a versatile and powerful tool in machine-learning applications. 

\subsubsection{Support Vector Machine (SVM)}
Support Vector Machine (SVM) is a supervised learning algorithm used for both classification and regression tasks. It works by finding the hyperplane that best divides a dataset into classes. In high-dimensional space, SVM constructs a hyperplane or set of hyperplanes in such a way that the margin between the classes is maximized. Points that are closest to the hyperplane are called support vectors, and they are the critical elements of the training set. SVMs are effective in high-dimensional spaces and are versatile because different kernel functions can be specified for the decision function.

\subsubsection{Multiple Linear Perceptron (MLP)}
A Multilayer Perceptron (MLP) is a class of feedforward artificial neural network (ANN). An MLP consists of at least three layers of nodes: an input layer, one or more hidden layers, and an output layer. Each node (except for the input nodes) is a neuron that uses a nonlinear activation function. MLPs use backpropagation for training the network. They are capable of learning non-linear models and are used for various tasks such as classification, regression, and complex function approximation.

\subsubsection{eXtreme Gradient Boosting (XGBoost)}
Extreme Gradient Boosting (XGBoost) is an optimized distributed gradient boosting library designed to be highly efficient, flexible, and portable. It implements machine learning algorithms under the Gradient Boosting framework. XGBoost provides a parallel tree boosting that solves many data science problems in a fast and accurate way. It is particularly known for its speed and performance and is used for supervised learning tasks. XGBoost excels in handling large-scale data and can be used for both classification and regression problems. It includes several advanced features like handling missing data, regularization, and tree pruning.

\subsection{Feature importance analysis}
We further conduct feature importance analysis to better understand how each paralinguistic feature contributes to the severity classification decisions. We employ SHapley Additive exPlanations (SHAP) values, a unified measure of feature importance grounded in cooperative game theory. SHAP values quantify the contribution of each feature to a machine learning model's prediction by considering its marginal contribution across all possible feature combinations. This approach provides a fair and comprehensive assessment of feature importance, accounting for both individual feature effects and interactions with other features. By utilizing SHAP values, we gain a consistent and interpretable measure of feature importance, facilitating a deeper understanding of the model's decision-making process and the underlying factors influencing its predictions.

\section{Experiments}
\subsection{Experimental Settings}

\subsubsection{Feature selection and Classification}
In our experimental setting in feature selection for multicollinearity, we employed three methods: LassoCV, ElasticNetCV, and hierarchical clustering with permutation importance. For LassoCV, we selected features with coefficients greater than 0.001. For ElasticNetCV, we used l1 ratios of [0.1, 0.3, 0.5, 0.7, 0.9], also selecting features with coefficients greater than 0.001. Both methods involved feature scaling using StandardScaler and were conducted in a Leave-One-Subject-Out (LOSO) cross-validation manner to identify common features across folds.
For hierarchical clustering, we first computed the Spearman correlation matrix of the features and converted it into a distance matrix. We then performed clustering using Ward's linkage method, which minimizes the variance within each cluster. Ward's linkage is an agglomerative hierarchical clustering technique that merges clusters based on the smallest increase in total within-cluster variance, and is known to effectively group features with similar characteristics. From each cluster, one representative feature was selected to reduce redundancy. This selection process ensured that the most informative and non-redundant features were chosen. 
As for feature selection, default settings are used, with 10-fold cross-validation. 

As for severity classification, we employ a diverse range of machine learning classifiers, including k-Nearest Neighbors (kNN), Support Vector Machine (SVM), Multi-Layer Perceptron (MLP), and XGBoost. The classifiers are trained in a LOSOCV manner and are designed to perform 4-way classification (healthy, mild, moderate, and severe). To ensure optimal performance, each classifier undergoes a meticulous grid search process to identify the most suitable hyperparameters. For kNN, the grid search encompasses optimal values of the number of neighbors (3, 5, 7, 9), weights (uniform, distance), and metric (euclidean, manhattan). For the SVM classifier, we explore a range of C values (0.1, 1, 10, 100) and $\gamma$ values (1, 0.1, 0.01, 0.001), employing a radial basis function (RBF) kernel for effective non-linear separation. In the case of the MLP classifier, we examine different hidden layer sizes (50 and 100 neurons) and weight optimization solvers (SGD and Adam), utilizing ReLU activation and a maximum of 5000 iterations. Finally, for XGBoost, we investigate 300 estimators, maximum depths of 4, 5, and 6, and learning rates (η) of 0.3, 0.4, and 0.5.

\subsubsection{Feature importance extraction}
We extract SHAP values from the experiments that demonstrated the highest classification accuracy for each language. Utilizing a LOSOCV approach, we extract SHAP values from the training folds. These values are then averaged to represent the feature importance for each feature, providing a comprehensive measure of each feature's contribution to the model's predictions. Furthermore, we aggregate the SHAP values by speech dimensions, to better understand which speech dimensions affects the classifier's decision.

\subsection{Experimental Results}
\subsubsection{Multicollinearity}
To evaluate the effectiveness of the first step of our Two-Step Feature Selection (TSFS) method in mitigating multicollinearity, we calculated the Variance Inflation Factor (VIF) before and after feature selection. A VIF value greater than 10 generally indicates significant multicollinearity. By comparing the VIF values before and after applying our feature selection methods, we can assess how well the TSFS process has reduced multicollinearity among the features.

Table \ref{tab:feature_set_sizes} presents the number of features with a Variance Inflation Factor (VIF) exceeding 10, a threshold commonly used to indicate multicollinearity, for each language and feature selection method. Notably, before feature selection, a substantial number of features in English (619) and Korean (619) exhibited VIF values above 10, suggesting a high degree of multicollinearity. In Tamil, a smaller but still considerable number of features (261) exceeded this threshold. However, following Lasso and Elastic Net feature selection, no features in either English or Korean displayed VIF values above 10, indicating a successful mitigation of multicollinearity. In Tamil, only two features continued to exhibit high VIF values after Lasso and Elastic Net, suggesting a near-complete resolution of multicollinearity. Similarly, after feature clustering, only a few features (3 in English, 1 in Korean) remained with VIF values exceeding 10, further supporting the effectiveness of these methods in addressing multicollinearity. Overall, these results suggest that the employed feature selection methods were highly effective in reducing multicollinearity, paving the way for more robust and reliable subsequent analysis and modeling.

\begin{table}[ht]
\centering
\caption{The number of features that has VIF greater than 10.}
\begin{tabular}{lccc}
\toprule
 & English & Korean & Tamil \\
\midrule
total & 619 & 619 & 261 \\
lasso & 0 & 0 & 2 \\
elastic & 0 & 0 & 2 \\
clustering & 3 & 1 & 0 \\
\bottomrule
\end{tabular}
\label{tab:feature_set_sizes}
\end{table}

\begin{table}[t]
\centering
\caption{Performance comparison of kNN, RF, SVM, MLP, and XGBoost classifiers across English, Korean, and Tamil datasets.}
\begin{tabular}{lcccccc}
\toprule
Classifiers & \multicolumn{2}{c}{English} & \multicolumn{2}{c}{Korean} & \multicolumn{2}{c}{Tamil} \\
 & Accuracy & F1-Score & Accuracy & F1-Score & Accuracy & F1-Score \\
\midrule
kNN & \textbf{51.30} & 63.02 & 61.79 & 71.83 & 51.50 & 63.40 \\
RF & 46.37 & 57.88 & 66.09 & 73.75 & 54.10 & 62.48 \\
SVM & 50.88 & \textbf{63.26} & \textbf{68.25} & \textbf{76.37} & 52.35 & 63.85 \\
MLP & 49.35 & 62.19 & 64.83 & 74.48 & 52.37 & 64.70 \\
XGBoost & 47.56 & 61.42 & 61.64 & 70.78 & \textbf{54.40} & \textbf{64.93} \\
\bottomrule
\end{tabular}
\label{tab:classifier_comparison}
\end{table}

\subsubsection{Feature selection and Classification}
To determine the optimal feature set for each language, we conducted a comprehensive analysis of classification performance using various feature sets and classifiers. As shown in Table \ref{tab:classifier_comparison}, the combination of feature selection methods and classifiers yielded distinct results for each language.

For English, the subset obtained through clustering followed by ETC feature selection demonstrated the highest accuracy (51.30\%) when combined with the kNN classifier. However, the ElasticNet feature selection method, without additional clustering, achieved the highest F1-score (63.26\%) when paired with the SVM classifier.
In contrast, for both Korean and Tamil, the two-step feature selection method consistently outperformed other combinations. For Korean, the clustering + ETC subset achieved the best results with the SVM classifier, attaining an accuracy of 68.25\% and an F1-score of 76.37\%. Similarly, for Tamil, the clustering + RFE subset yielded the highest performance with the XGBoost classifier, achieving an accuracy of 54.40\% and an F1-score of 64.93\%.

These results highlight the effectiveness of our two-step feature selection approach, particularly for Korean and Tamil, where it consistently led to the highest classification performance across all evaluated metrics. While the ElasticNet method alone achieved a slightly higher F1-score for English, the clustering + ETC combination still demonstrated the highest accuracy, suggesting its potential utility in real-world applications where accurate classification is paramount. For a detailed analysis of the classification performances of the experiments other than the best-performing ones, please refer to Appendix A. 

\begin{table}[t]
\centering
\caption{Speech Dimension Analysis Across Languages}
\begin{tabular}{lrrr}
\toprule
Speech Dimension & English & Korean & Tamil \\
\midrule
Phonation & 3 & 2 & 2 \\
Spectral Onset & 6 & 6 & 12 \\
Spectral Offset & 3 & 6 & 6 \\
Formants & 2 & 3 & 5 \\
F0 & 1 & 3 & 3 \\
Energy & 2 & 3 & 1 \\
Duration & 1 & 2 & 0 \\
\bottomrule
\end{tabular}\label{tab:select_num}
\end{table}

Table \ref{tab:select_num} presents a breakdown of the selected features by speech dimension for each language, based on the optimal feature sets identified earlier. In English, the ElasticNet subset predominantly selected features related to spectral onset (6 features), spectral offset (3 features), and phonation (3 features). For Korean, the clustering + ETC subset heavily favored spectral onset and offset features (6 features each), followed by formants, F0, and energy features (3 features each). In contrast, the Tamil clustering + RFE subset greatly emphasized spectral onset (12 features), followed by spectral offsets (6 features) and formats (5 features), with fewer features selected from the other speech dimensions. This distribution highlights the varying importance of different speech dimensions in capturing the unique characteristics of dysarthric speech across the three languages.

\subsubsection{Feature importance}
\begin{figure}[t] 
\centering
\includegraphics[width=0.8\columnwidth]{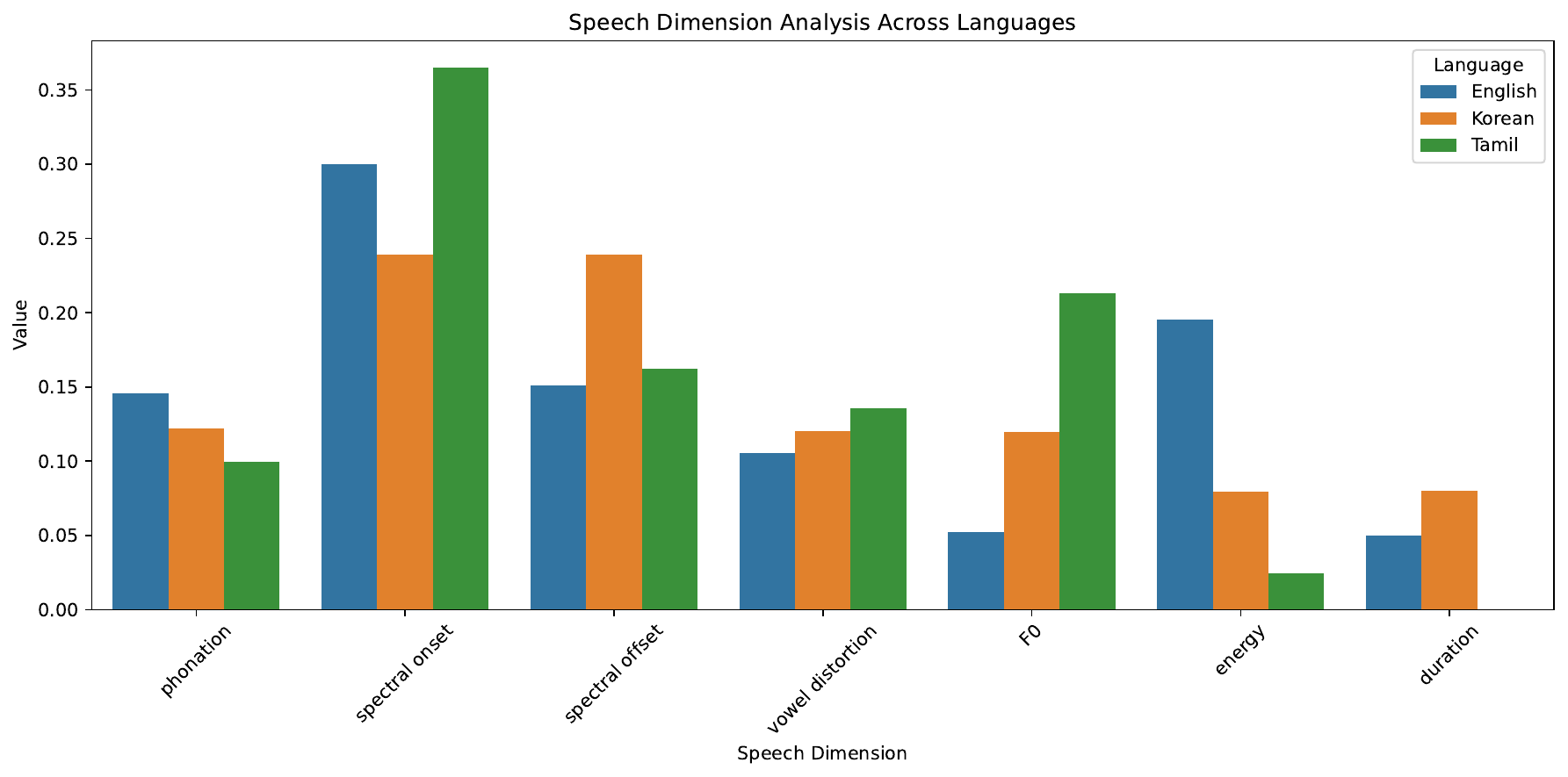}
\caption{
Feature importance scores aggregated by speech dimensions.
}
\label{figs:3-importance}
\end{figure}

The feature importance scores, aggregated by speech dimensions, reveal distinct patterns across the three languages, underscoring the necessity for a nuanced multilingual analysis in dysarthric speech assessment. While spectral features consistently play a crucial role across English, Korean, and Tamil, the relative importance of specific dimensions varies considerably. \Cref{figs:3-importance} demonstrates the feature importance scores.

In English, spectral onset features (30\%) emerge as the most influential, suggesting heightened sensitivity to initial acoustic cues in the perception of dysarthric speech. This is followed by energy (20\%), and a balanced contribution from phonation (15\%) and spectral offset (15\%) features. Formants (vowel distortion) (11\%) also play a significant role, indicating that vowel articulation difficulties are a prominent characteristic of dysarthric speech in English.

For Korean speakers, spectral offset (24\%) and spectral onset (24\%) features share the highest importance, emphasizing the importance of both the initial and final acoustic properties of speech sounds. Phonation (12\%) remains a key contributor, while formants (12\%) and F0 (12\%) also play notable roles, suggesting that both vowel articulation and pitch variations are important markers of dysarthria in Korean.

In contrast, Tamil speakers exhibited a distinct pattern, with spectral onset features being the most dominant (36\%), highlighting their heightened importance in characterizing dysarthric speech in this language. This was followed by F0 (21\%), spectral offset (16\%), and formants (14\%), aligning with the importance of spectral and articulatory features observed in English and Korean. The prominence of spectral onset in Tamil suggests that initial acoustic cues may be particularly salient in distinguishing dysarthric speech in this language.


\section{Discussion}
\subsection{Feature importance analysis}
In \Cref{ssec:hypothesis}, we outlined the hypothesis for the anticipated importance of various speech features across English, Korean, and Tamil, based on their unique phonemic and prosodic structures. The results of the feature importance scores, aggregated by speech dimensions, provide insights into the multilingual analysis of dysarthric speech assessment, largely confirming our expectations.

Firstly, we hypothesized that English would place greater importance on phonation features due to the use of voicing contrasts in English consonants \citep{wright2004review, raphael2021acoustic}, unlike Korean and Tamil. The results support this assertion, indicating that phonation features contribute 15\% to the overall importance in English. Although phonation also plays a notable role in Korean (12\%) and Tamil (10\%), the prominence in English aligns with our hypothesis.

Secondly, it was anticipated that spectral onset features would hold significant importance across all three languages due to the high cognitive load for perception at the onset of syllables \citep{mattys2011effects}. The results confirm this expectation, showing spectral onset features as the most influential in English (30\%), Korean (24\%), and Tamil (36\%). This underscores the critical role of initial acoustic cues in speech perception across different languages.

Regarding formant features, the hypothesis predicted that vowel space density would significantly impact speech intelligibility, with greater importance in English and Tamil due to their larger vowel inventories compared to Korean. The results partly support this, showing formant features with higher importance scores in the order of Tamil (14\%), Korean (12\%), and English (11\%). While the lower importance in English does not fully support our hypothesis, this can be explained by previous findings suggesting that English listeners are equipped to deal with vowel variability, likely due to dialect variation mainly carried by vowels \citep{van1996vowel}.

The hypothesis posited that prosodic features related to F0 and energy would be more significant in English, given its stress-timed rhythmic structure. The results partially support this, showing that while F0 is less highlighted in English, energy-related features still contribute significantly. Unexpectedly, Tamil showed the highest importance scores for F0, followed by Korean. For duration features, it was expected to be greater in Korean and Tamil, where maintaining the duration of segments equally is important for high speech intelligibility. This hypothesis was partially supported, with Korean showing the greatest feature importance, followed by English. For Tamil, duration features were not selected in the optimal feature set. Further analyses with finer-grained experimental design is necessary to better understand prosodic aspect of dysarthric speech in a multilingual context.

Moreover, interesting patterns aside from the hypotheses were also found. For example, consonant-related features (spectral onset, spectral offset) have greater importance scores compared to vowel-related features (formant features). This aligns with the understanding that consonants are more important in perceiving word meaning compared to vowels \citep{owren2006relative, nespor2003different}. Additionally, only Korean showed that spectral onset and spectral offset have the same amount of feature importance, which is different from English and Tamil, where the importance was skewed towards spectral onsets. This highlights a unique attribute of Korean, necessitating further analysis to understand this rationale.

In summary, the results largely align with the initial hypothesis, while certain findings, particularly for prosodic features, deviate from expectations. The significant deviation observed in prosodic features is likely due to their inherent complexity, encompassing more than just rhythmic elements. Factors such as intonation, stress patterns, and speech rate could greatly influence the importance of these features. Moreover, the extracted features do not directly reflect the linguistic characteristics, which are far more complex than what paralinguistic features can capture. Paralinguistic features are indirect reflections of linguistic differences, highlighting the need for features that more directly represent linguistic characteristics. Therefore, further investigations with features that more accurately reflect linguistic characteristics are necessary to understand the discrepancies between our hypothesis and the results.

\section{Conclusion}
This research introduces a robust multilingual approach to dysarthria severity classification, addressing limitations of existing monolingual assessment tools like the Mayo Clinic System. By leveraging paralinguistic features, particularly the DisVoice feature set, and implementing a two-step feature selection method (TSFS), we enhance the accuracy and applicability of dysarthria assessments across diverse linguistic backgrounds.

Our study validates the effectiveness of the TSFS method, which combines Lasso, Elastic Net, and Hierarchical Clustering, followed by traditional feature selection techniques to mitigate multicollinearity and select the most relevant features. This approach not only enhances the precision of feature selection but also ensures that the selected features are robust and non-redundant.

Through comprehensive analysis, we identified the most informative paralinguistic features for dysarthria severity assessment in English, Korean, and Tamil. The results indicate that while spectral features are universally important, the significance of specific speech dimensions varies across languages. We hypothesize such differences stem from differences in language structure.

Our multilingual analysis underscores the necessity of tailored approaches to dysarthria assessment, taking into account language-specific phonetic and phonological characteristics. Overall, this research contributes significantly to the field of dysarthria assessment by:

\begin{itemize}
\item Introducing an effective feature selection method for analyzing paralinguistic features in dysarthria severity classification.
\item Identifying crucial features and their importance for dysarthria severity assessment in multiple languages, enhancing the understanding of language-specific and universal aspects of dysarthric speech.
\end{itemize}

\chapter{Multilingual Pronunciation Analysis for Dysarthric Speech Assessment Using Goodness of Pronunciation} \label{chapter:pronunciation}

\begin{figure}[t] 
\centering
\includegraphics[width=0.7\columnwidth]{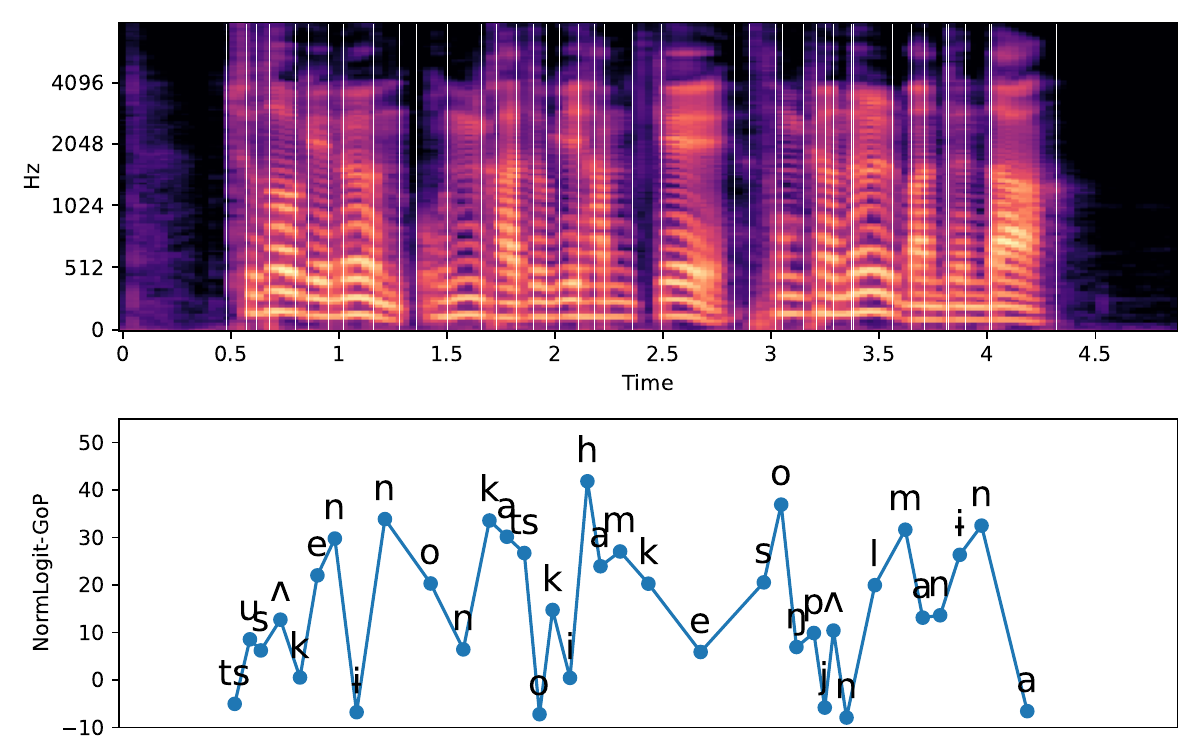}
\caption{
Example of phoneme-level assessment.
}
\label{figs:samplewise_phones}
\end{figure}


Phoneme mispronunciation is a primary characteristic of dysarthric speech \citep{darley1969differential}. As a result, prior research has aimed to automate phoneme-level speech assessment for dysarthric speech. This involves training models to assess the disorderedness of each phoneme segment by comparing its speech features to those of healthy speech phoneme segments. This approach allows for pinpointing which phonemes are mispronounced in each utterance, as illustrated in \Cref{figs:samplewise_phones}.

One prevalent method for phoneme-level speech assessment uses a parallel neural network (NN) with parallel datasets. The NN is trained with identical utterances recorded by both healthy speakers and patients, enabling it to discern whether each phoneme originates from healthy or disordered speech \citep{miller2020assessing,quintas2022automatic}. However, gathering parallel datasets, especially for disordered speech, is challenging. Additionally, this method often limits analysis to specific speech materials, which may not reflect the natural speech patterns used in daily communication.

Another traditional method for evaluating phoneme-level pronunciation is Goodness of Pronunciation (GoP) \citep{witt2000phone}. GoP measures the similarity between produced and canonical phoneme pronunciations and is typically derived from a phoneme recognizer. Unlike the parallel dataset method, GoP does not require datasets of both healthy and dysarthric speech to read the same parallel sentences. While GoP has been commonly used to assess non-native (L2) speech pronunciation, some studies have demonstrated its potential for evaluating speech disorders \citep{pellegrini2014goodness,Fontan2015PredictingDS}.

With advances in neural networks (NNs), new GoP variants that use probabilities from cutting-edge neural networks have been proposed \citep{hu2015improved,Cheng2020ASRFreePA,xu2021explore}. However, these probabilities can be misleading due to modern NNs' tendency to produce overconfident outputs, even when incorrect \citep{guo2017calibration}. This overconfidence is problematic for GoP, which relies heavily on probabilities. The issue is exacerbated with out-of-distribution (OOD) inputs, such as dysarthric speech compared to healthy speech \citep{hendrycks2017baseline}. To mitigate these problems, Uncertainty Quantification (UQ) techniques, which address OOD issues, can be employed \citep{guo2017calibration,hendrycks2017baseline}.

Another critical limitation of traditional GoP approaches is their dependence on monolingual automatic speech recognizers, which hinders their effectiveness in multilingual contexts. To overcome this, we propose a GoP method utilizing a self-supervised learning (SSL) model trained on a cross-lingual dataset for multilingual pronunciation assessment. SSL models are advantageous due to their ability to perform well on downstream tasks, such as speech and phoneme recognition, with limited data \citep{Yang2021SUPERBSP}. This method facilitates multilingual pronunciation analysis for dysarthric speech assessment.

This paper presents a comprehensive method for multilingual pronunciation analysis for dysarthric speech assessment. We enhance the traditional GoP method by integrating Uncertainty Quantification (UQ) techniques in two ways: (1) normalizing phoneme predictions and (2) modifying the scoring function. Given the significant acoustic differences between dysarthric and healthy speech \citep{wilson2000acoustic}, dysarthric speech is treated as OOD input. Consequently, we apply conventional UQ methods to improve GoP calculations for dysarthric speech assessment.

\section{Proposed Method} \label{sec:gop}
\begin{figure*}[t!] 
\centering
\includegraphics[width=1\textwidth,height=5cm]{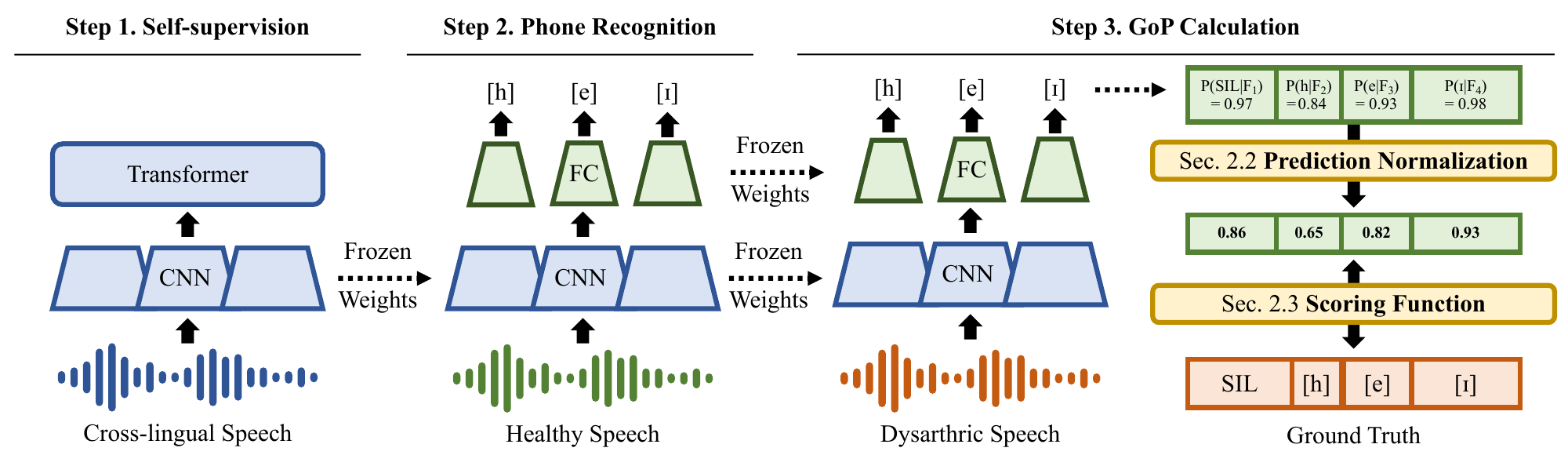}
\caption{Overview of the improved GoP with UQ methods.}
\vspace{-1em}
\label{fig:diagram}
\end{figure*}

This study utilizes various conventional Uncertainty Quantification (UQ) methods to calculate Goodness of Pronunciation (GoP) scores, as shown in \Cref{fig:diagram}.

\subsection{Prerequisite: Goodness of Pronunciation (GoP)} \label{ssec:gop-baseline}
In this subsection, we review existing GoP-based methods from a UQ perspective. Beginning with GMM-GoP, \Cref{eq:GMM} defines this method:
For a given phoneme $p$ over frames $f \in F$, with frame-wise phoneme probability and logits denoted as $P^f(p|f)$ and $L^f(p|f)$, and the total phoneme set as $Q$, \textbf{GMM-GoP} \citep{witt2000phone} is defined as the average log probability across the phoneme duration, $\mathbb{E}F[\log P^f(p|f)]$:
\begin{equation}
s\text{GMM-GoP}(p) = \frac{1}{|F|} \sum_{f \in F} \log \frac{e^{L^f(p|f)}}{ \sum_{q \in Q} e^{L^f(q|f)}}.
\label{eq:GMM}
\end{equation}
Averaging log probabilities functions as a form of temporal ensembling \citep{laine2017temporal}, a well-known UQ method, as it aggregates predictions over multiple frames into a single estimate. Moreover, directly using the probability output is a typical baseline for OOD detection \citep{hendrycks2017baseline}.

Another key baseline is \textbf{NN-GoP} \citep{hu2015improved}, which enhances GMM-GoP by leveraging deep neural networks and altering the scoring function:
\begin{align}
\Bar{P}(p|F) = \mathbb{E}F[P^f(p|f)] = \frac{1}{|F|} \sum{f \in F} P^f(p|f), \\
s_\text{NN-GoP}(p) = \log \Bar{P}(p|F) - \max_{q \in Q} \log \Bar{P}(q|F). \label{eq
}
\end{align}

Lastly, \textbf{DNN-GoP} \citep{hu2015improved} normalizes the phoneme probability using the phoneme prior \citep{hong2021disentangling}:
\begin{align}
s_\text{DNN-GoP}(p) = \Bar{P}(p|F) / P(p).
\end{align}

\subsection{Normalizing the phoneme predictions}\label{subsec:modify_pred}
There are two common techniques for calibrating the posterior prediction $P(p|F)$ by modifying its logit $L(p|F)$. The first technique involves normalization by removing the influence of the prior $P(p)$ \citep{hong2021disentangling,hu2015improved} (\textbf{Prior}), and the second technique aims to reduce peakiness through temperature scaling \citep{guo2017calibration} (\textbf{Scale}):
\begin{align}
L^f(p, f) &= L^f(p|f) - \log P(p), \
L^f_T(p|f) &= L^f(p|f) / T,
\end{align}
where $T$ is a hyperparameter, and the adjusted predictions are obtained by applying the softmax function to the modified logits.

Normalization using the prior corresponds to the concept applied in DNN-GoP, where it is generally used to separate the training distribution of the phoneme recognizer. This is crucial as majority classes often exhibit greater confidence than minority classes \citep{hong2021disentangling}. Temperature scaling is another widely used UQ baseline as it helps to avoid overly peaky posterior probability distributions.

\subsection{Modifying the scoring function}\label{subsec:modify_score}
This approach modifies the GoP score calculation by incorporating uncertainty information. Initially, we use common methods to measure data uncertainty: Entropy $H$ and Margin $M$.

\textbf{Entropy} assesses the uncertainty among a set of possible outcomes:
\begin{align}
s_\text{H}(p) = -\sum_{q \in Q} \Bar{P}(q|F) \log \Bar{P}(p|F).
\end{align}
Entropy quantifies the average information needed to determine the observed outcome. It is advantageous as it does not require ground truth labels, which can be expensive to obtain.

\textbf{Margin} represents the difference between the probability of the true class and the highest probability among other classes for a prediction:
\begin{align}
s_\text{M}(p) = \Bar{P}(p|F) - \max_{q \in Q - {p}} \Bar{P}(q|F).
\end{align}
This formula is similar to that of NN-GoP but excludes the true phoneme from $Q - {p}$.

Alternatively, \textbf{MaxLogit} \citep{hendrycks2022scaling} and \textbf{LogitMargin} make use of the logits directly. The softmax function can hide useful information in logits by normalizing their sum to one. Thus, combining the ideas of GMM-GoP (using probabilities directly) and NN-GoP (using Margin) results in:
\begin{align}
s_\text{MaxLogit}(p) &= \frac{1}{|F|} \sum_{f \in F} L^f(p|f) = \Bar{L}(p|F), \\
s_\text{LogitMargin}(p) &= \Bar{L}(p|F) - \max_{q \in Q - {p}} \Bar{L}(p|F).
\label{eq
}
\end{align}

\section{Experimental setting} \label{sec:experiment}

\subsection{Datasets} \label{ssec:corpus}
To train the acoustic model on healthy phoneme distributions, we use the Common Phone dataset \citep{klumpp2022common} and the L2-ARCTIC dataset \citep{zhao2018l2}. To evaluate the effectiveness of our proposed approach, we employ three dysarthric datasets: the UASpeech English dataset \citep{kim2008uaspeech}, the QoLT Korean dataset \citep{choi2012dysarthric}, and the SSNCE Tamil dataset \citep{ta2016dysarthric}. Each of these datasets includes dysarthric speech from individuals with Cerebral Palsy.

Our analysis concentrates on sentences from the QoLT and SSNCE dysarthric datasets, as the Common Phone and L2-ARCTIC datasets only contain sentences. For the UASpeech dataset, which exclusively comprises words, we analyze word materials. We chose the UASpeech dataset over the TORGO dataset for this task because UASpeech provides a wider variety of phonemic environments, which is essential for comprehensive phoneme pronunciation assessments.

\subsubsection{Common Phone dataset and L2-ARCTIC dataset} \label{sssec:commonPhone}
The acoustic model is trained on healthy speech using the Common Phone dataset and the L2-ARCTIC dataset due to their comprehensive phoneme coverage and detailed phonetic annotations. These datasets include most of the phonemes found in English, Korean, and Tamil.
The \textbf{Common Phone dataset} is a gender-balanced, multilingual corpus spanning six languages. It consists of over 11,000 speakers and approximately 116 hours of speech.
The \textbf{L2-ARCTIC dataset} is a speech corpus frequently used to detect mispronunciations in non-native English speakers. It contains recordings from 24 speakers with a balanced distribution of gender and native language, representing six different countries. On average, each speaker provides around 67.7 minutes of speech, with a total duration of approximately 27.1 hours.

\subsubsection{UASpeech English dysarthric datasat} \label{sssec:uaspeech}
The UASpeech dataset \citep{kim2008uaspeech} is a publicly accessible English dysarthric speech dataset that includes 15 dysarthric speakers (11 males, 4 females) and 13 age-matched healthy speakers (9 males, 4 females). The speakers are categorized according to their scores on the Frenchay Dysarthria Assessment (FDA) \citep{enderby1980frenchay}, with 5 mild speakers (score 1), 3 moderate-to-severe speakers (score 2), 3 moderate-to-severe speakers (score 3), and 4 severe speakers (score 4). Phoneme-level alignments are obtained using the Montreal Forced Aligner (MFA) \citep{mcauliffe2017montreal}.

\subsection{Experimental details} \label{ssec:detail}
In this study, we assessed GoP performance by adopting the methodology from a previous study \citep{Fontan2015PredictingDS}. Rather than evaluating the models based on their accuracy in identifying mispronunciations, our goal was to calculate the average GoP score for each utterance and examine its correlation with intelligibility scores. This approach was selected because annotating each phoneme pronunciation in dysarthric speech is both subjective and labor-intensive.

\subsubsection{Phoneme Prediction} \label{sssec:phone_pred}
In line with recent studies \citep{xu2021explore}, we extract posterior probabilities from the Wav2Vec 2.0 XLS-R model \citep{babu22_interspeech}, a cross-lingual self-supervised learning (SSL) model. SSL models are effective in downstream tasks even with limited training data due to their rich, generalizable representations learned from large amounts of unlabeled data. This allows efficient fine-tuning for specific tasks with relatively small labeled datasets, which is particularly beneficial for dysarthric speech assessment, where collecting extensive dysarthric speech recordings is challenging. Utilizing knowledge from healthy speech, SSL models have demonstrated promising results in automatic dysarthric speech assessments \citep{yeo2023automatic, stumpf2024speaker}. The XLS-R model was selected for its cross-lingual pre-training, which is especially powerful for multilingual dysarthria assessments, providing diverse phonetic and linguistic knowledge.

To reduce computational overhead and maintain phonetic characteristics inherent in convolutional features, we slightly modified the architecture by attaching a linear phone prediction head to the convolutional layer instead of the transformer layer \citep{choi2022opening}. Additionally, the loss function was simplified by removing the adaptive pooling, based on our observation that this change did not significantly impact performance.
The AdamW optimizer \citep{loshchilov2019decoupled} was used with a default learning rate of $0.001$ for three epochs. Since only the linear prediction head was trained, the final performance showed minimal sensitivity to other hyperparameters.

\subsubsection{Baselines} \label{sssec:baselines}
Three conventional experiments are conducted as baseline experiments: GMM-GoP \citep{witt2000phone}, NN-GoP \citep{hu2015improved}, and DNN-GoP. We compare these baseline experiments with our proposed experiments, which include UQ methods (\Cref{subsec:modify_pred}, \Cref{subsec:modify_score}) using the same phoneme probabilities. This comparison aims to determine if the UQ methods effectively mitigate the OOD problem in GoP calculations for dysarthric speech. To examine the correlations between GoP scores (continuous) and intelligibility scores (ordinal), we use the Kendall Rank Coefficient $\tau$. Kendall's $\tau$ measures the strength of the relationships, with a higher absolute coefficient indicating a stronger correlation between the two variables.

\begin{table}[t]
\caption{
Kendall's rank coefficient between GoP \& intelligibility severity levels.
}
\vspace{1em}
\label{tab:results}
\centering
\resizebox{0.8\textwidth}{!}
{

\begin{tabular}{c|c|c|c|c|c|c}
\toprule
Method & Norm. & Scoring Func. & English & Korean & Tamil & Average \\
\midrule
\multirow{3}{*}{Baseline} 
& None & GMM & -0.2049 & -0.5237 & -0.3571 & -0.3619 \\
& None & NN & -0.1536 & -0.4687 & -0.4003 & -0.3409 \\
& Prior & DNN-GoP & -0.1836 & -0.4237 & -0.4681 & -0.3585 \\
\midrule
\multirow{12}{*}{Proposed} 
& \multirow{4}{*}{None}  & Entropy & -0.1831 & -0.2643 & -0.3251 & -0.2575 \\
&  & Margin & -0.1628 & -0.4434 & -0.4445 & -0.3502 \\
&  & MaxLogit & -0.2164 & -0.5440 & -0.5786 & -0.4463 \\
&  & LogitMargin & -0.1732 & -0.4753 & -0.5158 & -0.3881 \\
\cline{2-7}
& \multirow{4}{*}{Scale} & Entropy & -0.1755 & -0.1974 & -0.2263 & -0.1997 \\
&  & Margin & -0.1260 & -0.4444 & -0.4210 & -0.3305 \\
&  & MaxLogit & -0.2164 & -0.5440 & -0.5786 & -0.4463 \\
&  & LogitMargin & -0.1732 & -0.4753 & -0.5158 & -0.3881 \\
\cline{2-7}
& \multirow{4}{*}{Prior} & Entropy & -0.1833 & -0.2645 & -0.3254 & -0.2577 \\
&  & Margin & -0.1630 & -0.4432 & -0.4447 & -0.3503 \\
&  & \textbf{MaxLogit} & \textbf{-0.2165} & \textbf{-0.5442} & \textbf{-0.5788} & \textbf{-0.4465} \\
&  & LogitMargin & -0.1733 & -0.4753 & -0.5160 & -0.3882 \\
\bottomrule
\end{tabular}
}
\end{table}
\section{Experimental results}
\subsection{Correlation between GoPs and intelligilbity scores}
The experimental results, summarized in \Cref{tab:results}, show Kendall’s rank coefficient between GoP and intelligibility severity levels across different methods, normalization techniques, and scoring functions for English, Korean, and Tamil. The baseline methods, including GMM, NN, and DNN-GoP, achieved average coefficients of $-0.3619$, $-0.3409$, and $-0.3585$ respectively, indicating a moderate negative correlation between GoP and intelligibility severity. Among these, the GMM method performed the worst in Korean and Tamil, while NN showed the lowest correlation in English.

In contrast, the proposed methods, which incorporate various normalization techniques (None, Scale, and Prior) and scoring functions (Entropy, Margin, MaxLogit, and LogitMargin), generally demonstrated better performance. Notably, the proposed method with the MaxLogit scoring function and prior normalization achieved the highest average coefficient of $-0.4465$, significantly outperforming the baseline methods. This improvement was especially evident in Korean and Tamil, where MaxLogit with prior normalization scored $-0.5442$ and $-0.5788$ respectively, indicating that the proposed methods are more effective in these languages.

Overall, while the baseline methods showed moderate negative correlations, the proposed methods, particularly those with prior normalization and MaxLogit scoring, significantly improved performance, indicating a stronger and more consistent negative correlation between GoP and intelligibility severity levels across multiple languages.

\section{Discussion}
\subsection{Analysis on phonemes}
\begin{figure}[t] 
\centering
\subfloat[Korean phoneme \textipa{/i/}.]{\includegraphics[width=0.43\columnwidth]{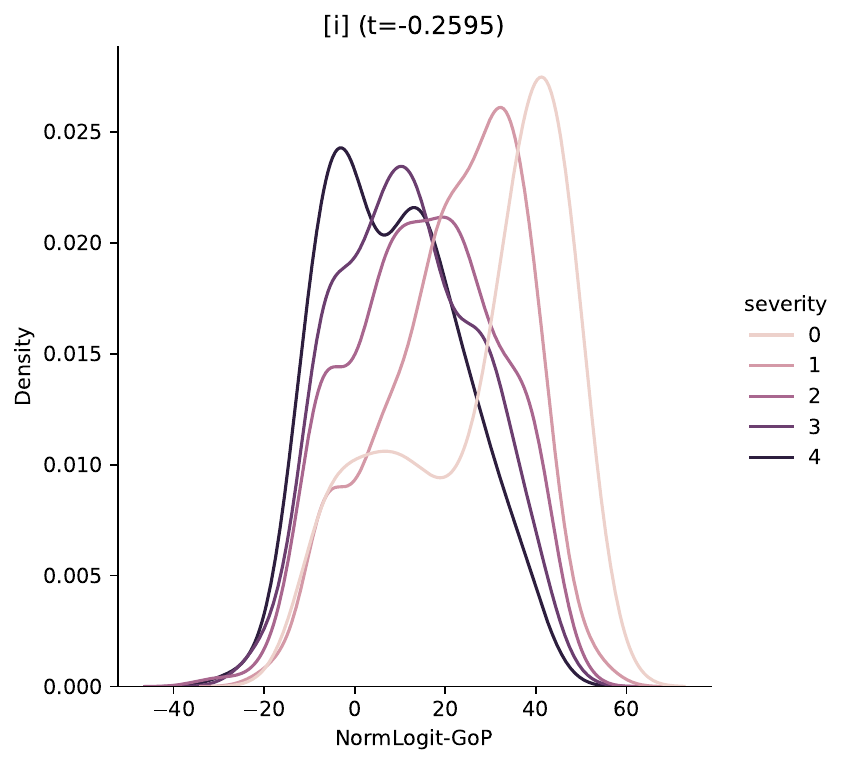}}
\subfloat[Korean phoneme \textipa{/m/}.]{\includegraphics[width=0.43\columnwidth]{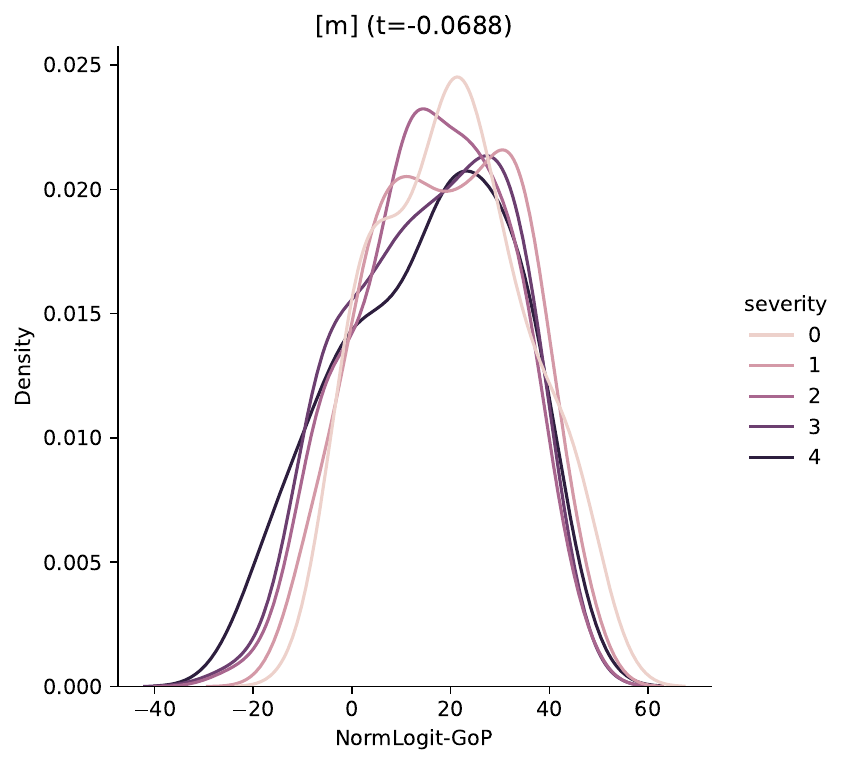}}
\vspace{1em}
\caption{
Kendall's $\tau$ distributions for two phonemes by severity.
0:healthy, 1:mild, 2:mild-to-mod., 3:mod.-to-sev., 4:severe.
}
\vspace{-1em}
\label{figs:phones}
\end{figure}

\Cref{figs:phones} illustrates the GoP distribution between two Korean phonemes \textipa{/i/} and \textipa{/m/}.
The distribution of \textipa{/i/} shows significant variation, whereas the distribution of \textipa{/m/} remains consistent across all severity levels. This suggests that certain phonemes have a greater impact on severity levels based on speech intelligibility, aligning with previous research findings \citep{quintas2022automatic}. Identifying phoneme pronunciations that strongly correlate with speech intelligibility can be valuable for diagnosis and treatment planning.

We performed a quantitative analysis to determine which phoneme scores are highly correlated with speech intelligibility across different languages. Kendall's $\tau$ was calculated for each phoneme, comparing our best-performing Prior+MaxLogit GoP score with the intelligibility scores. The top-5 phonemes based on their correlation are as follows for English:
\textipa{/a\textsci/},\textipa{/\textesh/},\textipa{/a\textupsilon/},\textipa{/z/},\textipa{/\textdyoghlig/}; Korean:\textipa{/i/},\textipa{/s/},\textipa{/n/},\textipa{/a/},\textipa{/\textturnv/}; Tamil:\textipa{/\textrtails/},\textipa{/h/},\textipa{/\textteshlig/},\textipa{/z/},\textipa{/a\textsci/}.
In summary, fricative sounds (\textipa{/s/},\textipa{/\textrtails/},\textipa{/\textesh/},\textipa{/z/}) show a strong correlation with speech intelligibility across languages, consistent with previous findings \citep{hernandez2019acoustic}.
Affricates (\textipa{/\textteshlig/},\textipa{/\textdyoghlig/}) and diphthongs (\textipa{/a\textsci/},\textipa{/a\textupsilon/}) appear in the top-5 phoneme list for both English and Tamil. This can be attributed to the complexity of articulating affricates and diphthongs, which poses challenges for accurate pronunciation in speakers with lower speech intelligibility.
Conversely, Korean exhibits higher correlations for nasal (\textipa{/n/}) and monophthongs (\textipa{/a/}, \textipa{/i/}, \textipa{/\textturnv/}). This may be attributed to the movements of articulators, such as the tongue, velum, and jaw.

\section{Conclusion}
This paper introduces an enhanced GoP method for assessing dysarthria speech intelligibility by incorporating UQ methods. To address the issue of overconfidence in modern neural networks, particularly with disordered speech, the UQ methods tested include (1) normalization of phoneme prediction and (2) modification of the scoring function. Experiments were conducted on dysarthric speech datasets in English, Korean, and Tamil. The results showed that the prior normalized MaxLogit GoP achieved the best performance, surpassing both traditional GoP methods and other proposed GoP variants. Additionally, an analysis was conducted to determine which phoneme pronunciation scores are highly correlated with speech intelligibility, further demonstrating the utility of the proposed method.


\chapter{Clinical Knowledge-Driven Voice Biomarkers for Dysarthric Speech Assessment in a Multilingual Cohort }\label{chapter:analysis}

\section{Background and Research questions}
In \Cref{chap:paralinguistics}, we concluded that paralinguistic features can be effectively employed for multilingual dysarthria severity analysis. While the results gave valuable insights, these features present a challenge for clinicians due to the complexity of interpreting each acoustic characteristic. In \Cref{chapter:pronunciation}, we focused on the phoneme pronunciation aspect, finding that our proposed Goodness of Pronunciation (GoP) score can effectively assess pronunciation in dysarthric speech across multiple languages. Although this method is highly interpretable, it is limited by its exclusive focus on pronunciation. Consequently, this chapter proposes a clinical knowledge-driven voice biomarker feature set that comprehensively captures the full spectrum of dysarthric speech characteristics, including voice quality, pronunciation, and prosody.

Prior research has primarily focused on identifying clinically-driven features for dysarthria severity assessment within monolingual contexts. This chapter takes a novel approach by analyzing clinically-driven voice biomarkers in a multilingual context, encompassing English, Korean, and Tamil. This multilingual analysis aims to differentiate between language-universal and language-specific features of dysarthria. By identifying these features, we not only contribute to the development of accurate and inclusive dysarthria assessment tools for diverse populations, but also broaden our understanding of how dysarthria manifests across languages.

Our approach involves two primary analyses. The first is a multilingual analysis. We perform a two-tiered validation for each language, consisting of both statistical and clinical validations. This step ensures the reliability and clinical usefulness of the feature sets for each language. Then, we perform a multilingual analysis by comparing the selected feature sets. Through this analysis, we identify both language-universal and language-specific features. The language-specific feature set includes features validated for each language, while the language-universal feature set comprises the intersection of validated features from all three languages.

The second approach involves automatic severity classification using the identified language-universal features. This analysis aims to evaluate the effectiveness of these features in accurately classifying dysarthria severity across different languages. We employ three machine learning classifiers—Support Vector Machine (SVM), Multi-Layer Perceptron (MLP), and eXtreme Gradient Boosting (XGBoost)—to assess classification accuracy.

To achieve these objectives, our investigation is guided by the following research questions:

\begin{itemize}
\item What are the significant clinically-driven features for assessing dysarthria severity across English, Korean, and Tamil?
\item How can we distinguish between language-universal and language-specific features in dysarthria assessment in English, Korean, and Tamil?
\item How effective are the language-universal features in automatic dysarthria severity classification?
\end{itemize}

\section{Clinical-knowledge-driven features}\label{sec:4-feature_extraction}
\begin{table*}[t!]
\caption{Description of extracted voice biomarkers.}
\label{tbl:features}
\centering
\resizebox{\textwidth}{!}{
\begin{tabular}{@{}ccC{1cm}p{7.5cm}p{6cm}@{}}
\toprule
\multicolumn{2}{c}{Feature} & Expected change & Biomarker definition & Dysfunction \\
\midrule
\multirow{8}{*}{Voice Quality} 
& jitter & $\uparrow$ & Short-term fundamental frequency (F0) variability. & Vocal instability. Pitch control difficulties. \\
& PPQ & $\uparrow$ & Smoothed fundamental frequency (F0) variability over cycles. & Vocal instability. Pitch control difficulties. \\
& shimmer & $\uparrow$ & Short-term loudness variability. & Vocal instability. Loudness control difficulties. \\
& APQ & $\uparrow$ & Smoothed loudness variability over cycles. & Vocal instability. Loudness control difficulties.\\
& HNR & $\downarrow$ & Harmonic to noise ratio. &  Incomplete vocal fold closure.  \\
& CPP & $\downarrow$ & Cepstral peak prominence. & Dysphonia; poor voice quality. \\
& \# of VB. & $\uparrow$ & Number of voice breaks. & Vocal fold dysfunction. \\
& percentage of VB. & $\uparrow$ & Proportion of phonation with breaks. & Vocal fold dysfunction. \\
\midrule
\multirow{9}{*}{Pronunciation} 
& PCC & $\downarrow$ & Recognition rate of pronounced consonants (fine-tuned XLS-R)& Imprecise pronunciation of consonants. \\
& PCV & $\downarrow$ & Accuracy of pronounced vowels (fine-tuned XLS-R)& Imprecise pronunciation of vowels. \\
& PCP & $\downarrow$ & Accuracy of pronounced phonemes (fine-tuned XLS-R)& Imprecise pronunciation of phonemes. \\
\cmidrule(lr){2-5}
& VSA$\triangle$ & $\downarrow$ &  Acoustic space of corner vowels (\textipa{/A/}, \textipa{/i/}, \textipa{/u/}) & Limited vowel articulation space. Restricted articulator mobility. \\
& VSA$\square$ & $\downarrow$ &  Acoustic space of corner vowels (\textipa{/A/}, \textipa{/i/} \textipa{/ae/}, \textipa{/u/}) & Limited vowel articulation space. Restricted articulator mobility.  \\
& FCR & $\uparrow$ & Degree to which vowel sounds are centralized.  & Centralization of vowels. \\
& VAI & $\downarrow$ & Degree to which vowel sounds are dispersed. & Reduced vowel dispersion. \\
& F2-Ratio & $\downarrow$ & Ratio of F2 values between \textipa{/A/} and \textipa{/i/}. & Limited tongue movement. \\
\midrule
\multirow{15}{*}{Prosody} 
& speaking rate & $\downarrow$ & Number of syllables divided by speech duration including pauses & Difficulty with muscle movements. \\
& articulation rate & $\downarrow$ & Number of syllables divided by speech duration excluding pauses & Slower movement of speech articulators. \\
& \# of pauses & $\uparrow$ & Number of pauses & Difficulty in maintaining speech flow. \\
& average pause duration & $\uparrow$ & Average length of pauses & Long pauses. \\
\cmidrule(lr){2-5}
& mean F0 & $\updownarrow$ & Average fundamental frequency & Pitch control difficulties. \\
& median F0 & $\updownarrow$ & Median fundamental frequency & Pitch control difficulties.  \\
& std. F0 & $\downarrow$ & Standard deviation of fundamental frequency & Monopitch.\\
& min. F0 & $\updownarrow$ & Minimum fundamental frequency & Pitch control difficulties. \\
& max. F0 & $\updownarrow$ & Maximum fundamental frequency & Pitch control difficulties.\\
\cmidrule(lr){2-5}
& mean energy & $\downarrow$ & Average energy & Weak muscles. \\
& median energy & $\downarrow$ & Median energy & Weak muscles.\\
& std. energy & $\downarrow$ & Standard deviation of energy & Loudness control difficulties. \\
& min. energy & $\updownarrow$ & Minimum energy & Loudness control difficulties. \\
& max. energy & $\updownarrow$ & Maximum energy & Loudness control difficulties. \\
\cmidrule(lr){2-5}
& $\%V$ & $\uparrow$ &  Proportion of voiced segments  & Increased effort to maintain voicing.   \\
& VarcoV & $\updownarrow$ & Standard deviation of the durations of voiced segments divided by mean duration & Irregular speech rhythm.  \\
& VarcoC & $\updownarrow$ &  Standard deviation of the durations of consonantal segments divided by mean duration & Irregular speech rhythm.  \\
& nPVIV & $\updownarrow$  & Duration difference between adjacent voiced segments divided by mean duration &  Irregular speech rhythm. \\
& nPVIC & $\updownarrow$& Duration difference between adjacent consonantal segments divided by mean duration & Irregular speech rhythm.  \\
\bottomrule
\label{tab:feat_tbl}
\end{tabular}
}
\end{table*}

To provide a comprehensive characterization of the speech of dysarthric patients, we configured a set of clinical-knowledge-driven features that reflect voice quality-, pronunciation-, and prosody-related features. Pronunciation-related features encompass phoneme accuracy- and vowel distortion-related features, while prosodic features include measures of fluency-, pitch-, energy-, and rhythm-related features. \Cref{tab:feat_tbl} reports the definition, the dysfunction connected to the characteristics, and the expected change of each biomarker. The expected change of a given biomarker is grounded in clinical research documenting the dysfunctions occurring in dysarthric speech. Details on the feature extraction process are described in \Cref{sec:4-feature_extraction}.

\subsection{Voice quality related features}  \label{ssec:vq}
We extracted eight voice quality features commonly employed in clinical assessments of dysarthria \citep{kent2003voice, chandrashekar2019breathiness, koike1973application}: jitter, Pitch Perturbation Quotient (PPQ), shimmer, Amplitude Perturbation Quotient (APQ), Harmonics-to-Noise Ratio (HNR), Cepstral Peak Prominence (CPP), the number of voice breaks (vb.), and the percentage of voice breaks. Clinicians consider elevated values of jitter, PPQ, shimmer, APQ, and the frequency and percentage of voice breaks as indicators of reduced voice quality. Conversely, reductions in HNR and CPP are strongly correlated with increased severity of dysphonia.

Jitter and shimmer quantify short-term fluctuations in fundamental frequency (F0) and amplitude, respectively. To account for natural variations present in healthy speech, we also calculate PPQ and APQ, which provide normalization for jitter and shimmer, respectively \citep{koike1973application}. HNR assesses the balance between periodic (harmonic) and aperiodic (noisy) components within the voice signal, offering insights into potential hoarseness or breathiness. CPP, derived from the cepstrum and calculated from voiced segments following the established method of \citep{murton2023validation}, quantifies the clarity of the voice's fundamental frequency peak. Finally, we examine voice breaks, characterized by inter-pulse intervals exceeding 17.86 ms (based on a 70 Hz pitch floor). The degree of voice breaks is defined as the cumulative duration of voice breaks relative to the total duration. All voice quality features are extracted using Praat \citep{boersma2011praat}, a widely recognized software in speech analysis. Equations \ref{Equation:absJitter} - \ref{Equation:HNR} describe these voice quality measurements. 

\begin{align}
    &\text{absJitter } = \frac{1}{N-1}\sum_{i=2}^{N} \lvert  T_{i}-T_{i-1} \rvert & \label{Equation:absJitter} \\
    \vspace{2mm}
    &\text{absShimmer } = \frac{1}{N-1}\sum_{i=2}^{N} \lvert  A_{i}-A_{i-1} \rvert & \label{Equation:absShimmer} \\
    \vspace{2mm}
    &\text{absPPQ } = \sum_{i=3}^{N-2} \frac{\lvert T_{i} - \frac{T_{i-2} + T_{i-1} + T_{i} + T_{i+1} + T_{I+2}}{5}\rvert} {N-4} & \label{Equation:absppq} \\ 
    \vspace{2mm}
    &\text{absAPQ } = \sum_{i=3}^{N-2} \frac{\lvert A_{i} - \frac{A_{i-2} + A_{i-1} + A_{i} + A_{i+1} + A_{i+2}}{5}\rvert} {N-4} & \label{Equation:absApq} \\
    \vspace{2mm}
    &\text{HNR (db)} = 10\log\left ( \frac{{E}_{p}}{{E}_{n}} \right ) & \label{Equation:HNR}
\end{align}

where $T_i$ is the duration of the $i$th interval, $A_i$ is the amplitude of the $i$th interval, and $N$ is the number of intervals.

\subsection{Pronunciation related features}  \label{ssec:pronunciation}
\subsubsection{Phoneme accuracy}  \label{sssec:sr}
We extracted three phoneme accuracy features: Consonant Recognition Rate (CRR), Vowel Recognition Rate (VRR), and Phoneme Recognition Rate (PRR). These objective metrics quantify the percentage of correctly articulated consonants, vowels, and phonemes, respectively, within each utterance. Dysarthria, characterized by motor impairments that disrupt articulation, often leads to reduced phoneme-level accuracy \citep{kim2010frequency}. Consequently, we hypothesize that these pronunciation scores will exhibit an inverse correlation with dysarthria severity. 

\begin{table}[h]
\captionsetup{justification=centering}
\caption{Alignment for phoneme correctness extraction}
\label{tab:alignment}
\centering
\resizebox{0.65\textwidth}{!}{
\begin{tabular}{c}
\hline\hline
Canonical phoneme sequence \\
\hline
HH  IY  W  IH  L  AH  L  AW  AH  *  R  EH  L  AY \\
\hline
\hline
Decoded phoneme sequence \\
\hline
SH  IY  W  AO  L  AH  L  AW  AE  N  L  IY  *  AY\\
\hline
\hline
\end{tabular}
}
\end{table}

For extraction, we utilized language-specific fine-tuned models \citep{wav2vec-english, wav2vec-korean, wav2vec2-tamil}, with the XLS-R-300m model \citep{babu22_interspeech} serving as a base model. The decoded and canonical phoneme sequences are aligned using a sequence-to-sequence approach. Scores were then calculated by determining the proportion of correctly recognized units (consonants, vowels, or phonemes) relative to the total number present within the target utterance. The decoded and canonical phoneme sequences are aligned, as described in \Cref{tab:alignment}. 
For example, consider a sentence containing five consonants (HH, L, L, R, L) and eight vowels (IY, W, IH, AH, AW, AH, EH, AY). Of these, two consonants (L, L) and five vowels (IY, W, AH, AW, AY) match the canonical phoneme sequence. Consequently, the CRR is calculated as 2/5100 = 40.00\%, VRR as 5/8100 = 62.50\%, and PRR as 7/15*100 = 53.85\%.

\subsubsection{Vowel distortion}  \label{sssec:vd}
We identified five features to examine possible distortions in vowel articulation: Vowel Space Area (VSA$\triangle$, VSA$\square$), Formant Centralization Ratio (FCR), Vowel Articulatory Index (VAI), and F2-ratio. These features are derived from the formant frequencies (F1, F2) of the corner vowels: \textipa{/i/}, \textipa{/u/}, \textipa{/a/}, and \textipa{/AE/}. VSA calculates the area formed by these vowels on the F1/F2 vowel chart, reflecting the speaker's total articulatory range. FCR and VAI are expressed as ratios of the corner vowels' formants to reduce individual speaker variations. The F2-ratio specifically examines the F2 values of \textipa{/i/} and \textipa{/u/}. Dysarthria often impairs articulatory control, resulting in a narrowed vowel space \citep{lansford2014vowel}. This narrowing is reflected in reduced VSA$\triangle$, VSA$\square$, VAI, and F2-ratio scores, while FCR generally increases with the severity of dysarthria.

For feature extraction, we followed this procedure. First, we generated Textgrids to align phonemes within speech samples. The Montreal Forced Aligner \citep{mcauliffe2017montreal} was used for English and Korean, while existing phonetic annotations were converted into Textgrids for Tamil. Next, we used Praat to extract F1 and F2 from the center of each vowel. These formants were then used to compute the vowel space features based on the provided equations (\cref{eq:tri_VSA} - \cref{eq:F2-Ratio}). When necessary, missing corner vowel formants were estimated using average speaker data. The maximum formant frequencies were set to 5000 Hz for male speakers and 5500 Hz for female speakers. Throughout the feature extraction process, we used Parselmouth \citep{parselmouth}.

\begin{align}
    &\begin{aligned}
        \text{VSA}{\bigtriangleup} = \frac{1}{2} | 
 {F1}_{\textipa{i}}\left ( {F2}_{a} - {F2}_{u} \right ) & + {F1}_{a} \left ( {F2}_{u} - {F2}_{\textipa{i}} \right ) \\
        & + {F1}_{u} \left ( {F2}_{\textipa{i}} - {F2}_{a}\right ) |
    \end{aligned} \label{eq:tri_VSA} \\
    &\begin{aligned}
    \text{VSA}{\Box} = \frac{1}{2} | ( {F2}_{i}{F1}_{ae} + &{F2}_{ae}{F1}_{a} \\
    & + {F2}_{a}{F1}_u + {F2}_{u}{F1}_{i} ) \\
     -({F1}_{i}{F2}_{ae} + &{F1}_{ae}{F2}_{a} \\
    & + {F1}_{a}{F2}_{u} + {F1}_{u}{F2}_{i} ) |
    \end{aligned} \label{eq:square_VSA} \\
    &\text{FCR } = \frac{{F2}_{u}+{F2}_{a}+{F1}_{i}+{F1}_{u}}{{F2}_{i}+{F1}_{a}} & \label{eq:FCR} \\
    &\text{VAI } = 
    \frac{{F2}_{i}+{F1}_{a}}{{F2}_{u}+{F2}_{a}+{F1}_{i}+{F1}_{u}} & \label{eq:VAI} \\
    &\text{F2-Ratio } = \frac{{F2}_{i}}{{F2}_{u}} & \label{eq:F2-Ratio}
\end{align}

\subsection{Prosody related features}
\subsubsection{Fluency} \label{sssec:fluency}
To examine potential disruptions in dysarthric speech fluency, we extracted four main quantitative features: speaking rate, articulation rate, pause frequency, and pause duration. Speaking rate, expressed in syllables per second, measures the overall pace of speech, whereas articulation rate focuses on the speed of articulation by excluding pauses. Pause frequency counts the occurrences of silences exceeding a 0.2ms threshold, and pause duration measures the length of these silences. Dysarthria, marked by neuromuscular impairments that affect speech production, often impacts speech fluency \citep{nishio2006comparison, nip2013kinematic}. This typically results in reduced speaking and articulation rates, as well as increased pause frequency and duration. We used Parselmouth for extracting these features.

\subsubsection{Pitch and Intensity} \label{sssec
}
We assessed pitch and intensity by calculating the mean, standard deviation, minimum, and maximum values for both the F0 and energy contours. We excluded intervals with 0 Hz (F0) or 0 dB (energy), as these likely represented silences. Dysarthric speech often shows an elevated and monotonous pitch \citep{schlenck1993aspects}, along with generally lower intensity due to weakened musculature \citep{tomik2010dysarthria, bayestehtashk2015fully}. However, some research suggests a potential increase in minimum energy \citep{yeo2022multilingual}, possibly because dysarthric speakers exert more effort during speech production. We used Parselmouth for feature extraction.

\subsubsection{Rhythm} \label{sssec:rhythm}
To examine potential rhythmic disruptions in dysarthric speech, we extracted five key metrics: \%V, varcoV, varcoC, nPVIV, and nPVIC. \%V measures the proportion of the speech signal occupied by vowels. VarcoV and varcoC assess the variability in the durations of vocalic and consonantal intervals, respectively. nPVIV and nPVIC calculate the average normalized difference between consecutive intervals. nPVI stands for normalized Pairwise Variability Index. The equations \ref{eq:varcoV} - \ref{eq:nPVI} detail these rhythmic measurements. These metrics provide insights into timing irregularities and rhythmic consistency, which can be affected by dysarthria. Research indicates that rhythmic alterations associated with dysarthria can vary depending on the speaker's native language \citep{hernandez2020dysarthria, yeo2022multilingual}. Understanding these language-specific differences is essential for customized dysarthria assessment and treatment. We utilized Textgrids generated for the previous vowel distortion analysis to streamline our analysis.
\begin{align}
   &\text{VarcoV } = \frac{\Delta\text{V}*100}{\text{meanV}} && \label{eq:varcoV} \\
    &\text{VarcoC } = \frac{\Delta\text{C}*100}{\text{meanC}} && \label{eq:varcoC} \\
  &\text{rPVIs } = \frac{1}{m-1}\sum_{k=1}^{m-1} |{d}_{k} - {d}_{k+1}| && \label{eq:rPVI} \\
    &\text{nPVIs } = \frac{100}{m-1} \sum_{k=1}^{m-1}  \left| \frac{d_k - d_{k+1}}{\frac{d_k+d_{k+1}}{2}} \right| && \label{eq:nPVI}
\end{align}

\section{Proposed Analysis Method}\label{sec:5-validation}

\subsection{Statistical and Clinical Validation}
 \begin{figure*}[h]
  \centering
\includegraphics[width=0.8\textwidth]{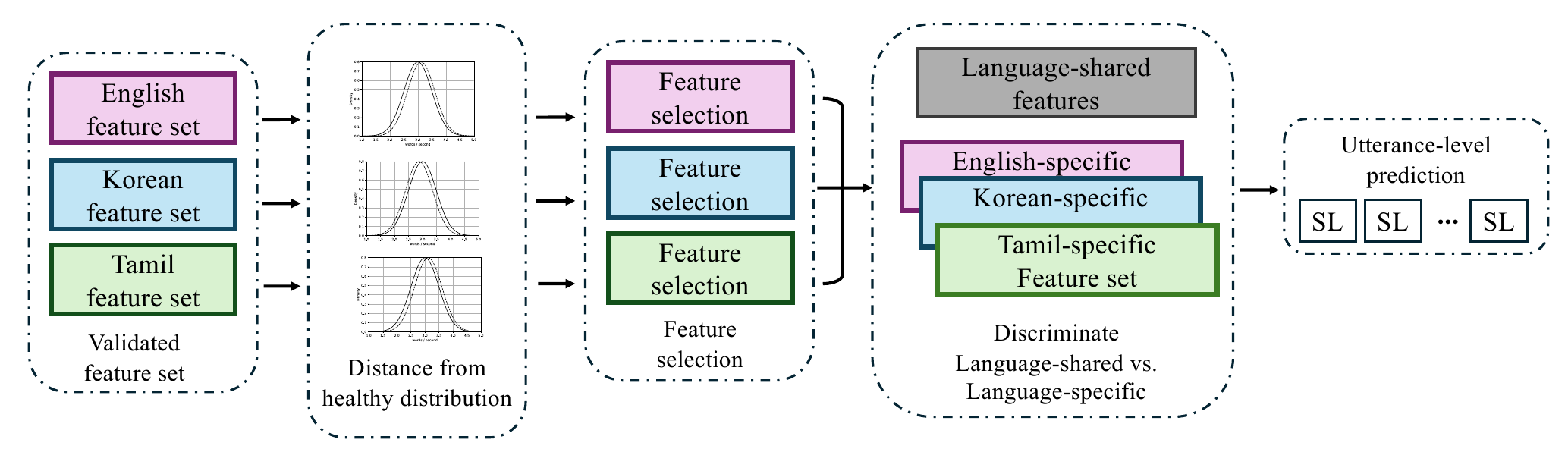}
  \caption{Workflow of Feature validation.}
  \label{fig:overview2} 
\end{figure*}

To ensure the robustness and reliability of our subsequent multilingual analyses, we begin with a rigorous two-tiered feature validation process, comprising both statistical and clinical validation (Figure \ref{fig:overview2}). The statistical validation phase utilizes the Kruskal-Wallis H test and the Kendall rank correlation coefficient (τ). The non-parametric Kruskal-Wallis H test identifies acoustic features that exhibit significant differences across dysarthria severity levels, while the Kendall rank correlation quantifies the strength and direction of the relationship between these features and severity. All statistical analyses are performed using IBM SPSS Statistics 25, with a significance threshold of $p < 0.05$.

Features deemed statistically significant then undergo clinical validation. Here, Kendall's coefficients (τ) are assessed against established clinical knowledge to confirm that the directionality of these features aligns with expected trends in dysarthria severity (Table \ref{tab:feat_tbl}). Only features meeting both statistical and clinical validation criteria are considered validated and retained for further analysis. A comparative analysis of the validated feature sets for each language is then conducted to identify both language-universal (consistent across all languages) and language-specific features.

\subsection{Automatic Dysarthria Severity Classification}\label{ssec:5-multilingual}
The second analysis involves performing automatic severity classification using the language-universal features. We conduct two types of classification experiments: cross-lingual classification and multilingual classification. In cross-lingual classification, models are trained on data from one language and tested on data from another language. In multilingual classification, models are trained using concatenated features from multiple languages.

This method leverages machine learning classifiers to evaluate the effectiveness of these features in accurately predicting dysarthria severity across different languages. By employing classifiers such as Support Vector Machine (SVM), Multi-Layer Perceptron (MLP), and XGBoost, we aim to determine the predictive power of the language-universal features. Each classifier is optimized through a grid search to identify the best hyperparameters, ensuring robust performance. Specifically, for SVM, we evaluate $C$ values (0.1, 1, 10, 100) and $\gamma$ values (1, 0.1, 0.01, 0.001) with a radial basis function (RBF) kernel. For MLP, we consider hidden layer sizes of 50 and 100 neurons, with solvers including Stochastic Gradient Descent (SGD) and Adam, using ReLU activation and a maximum of 5000 iterations. For XGBoost, we examine 300 estimators, maximum depths of 4, 5, and 6, and $\eta$ values of 0.3, 0.4, and 0.5.

To ensure model robustness, each classifier is trained using a Leave-One-Subject-Out (LOSO) cross-validation approach. Classifiers are trained to perform 4-way classification into healthy, mild, moderate, and severe categories. Performance is assessed using average accuracy and F1-score across all speakers.


\section{Results}\label{sec:5-results}
\subsection{Statistical and Clinical Validation}
\Cref{tab:vq_stats} summarizes the feature validation results, presenting the Kruskal-Wallis \textit{H} value and Kendall's $\tau$ coefficient. The \textit{val.} column symbols denote validation levels: $X$ indicates statistically non-significant, $\triangle$ indicates statistically significant but not clinically validated, and $O$ indicates both statistically and clinically validated. The mean value of each feature by severity level is demonstrated in Appendix A.

\begin{table*}[!ht]
\caption{Feature validation results.}
\label{tab:vq_stats}
\centering
\resizebox{\textwidth}{!}{%
\begin{tabular}{cc|c|c|c|c|c|c|c|c|c}
\toprule
\multicolumn{2}{c|}{\multirow{2}{*}{Voice Biomarkers}} & \multicolumn{3}{c|}{English} & \multicolumn{3}{c|}{Korean} & \multicolumn{3}{c}{Tamil} \\
\cmidrule{3-11}
& & \textit{H} & $\tau$ & val. & \textit{H} & $\tau$ & val. & \textit{H} & $\tau$ & val.\\
\midrule
\multirow{8}{*}{Voice Quality} 
& jitter  & 48.39* & 0.03 & X 
            & 41.17* & 0.12* & O
            & 351.40* & -0.13* & $\triangle$ \\
& PPQ & 51.09* & 0.07 & X  
        & 36.16* & 0.02 & X
        & 284.29* & -0.13* & $\triangle$ \\
& shimmer & 182.67* & -0.34* & $\triangle$ 
        & 52.62* & 0.12* & O
        & 484.76* & -0.19* & $\triangle$ \\
& APQ & 174.95* & -0.32* & $\triangle$  
        & 67.11* & 0.15* & O
        & 137.35* & -0.07* & $\triangle$ \\
& HNR & 193.40* & 0.29* & $\triangle$  
        & 11.16 & -0.01 & X
        & 1251.69* & 0.24* & $\triangle$ \\
& CPP & 110.91* & 0.20* & $\triangle$  
        & 120.69* & -0.27* & O
        & 502.78* & -0.18* & O \\
& $\#$ of VB. & 31.11* & 0.18* & O 
        & 115.21* & 0.29* & O
        & 88.82* & -0.07* & $\triangle$ \\
& percentage of VB. & 40.48* & 0.16* & O 
        & 236.39* & 0.42* & O
        & 58.99* & 0.01 & X \\
\midrule
\multirow{9}{*}{Pronunciation} 
& \textbf{CRR} & 340.12* & -0.57* & O 
        & 523.60* & -0.69* & O
        & 5051.99* & -0.66* & O \\
& \textbf{VRR} & 326.19* & -0.56* & O 
        & 515.07* & -0.68* & O
        & 4501.92* & -0.62* & O \\
& \textbf{PRR} & 359.20* & -0.58* & O 
        & 555.87* & -0.71* & O
        & 5497.49* & -0.69* & O \\
\cmidrule{3-11}
& \textbf{VSA$\triangle$} & 125.33* & -0.27* & O
            & 144.58* & -0.33* & O
            & 2322.01* & -0.42* & O \\
& \textbf{VSA$\square$} & 174.47* & -0.37* & O 
            & 91.08* & -0.26* & O
            & 2624.32* & -0.45* & O \\
& \textbf{FCR} & 174.59* & 0.35* & O 
        & 251.74* & 0.44* & O
        & 2895.13* & 0.47* & O \\
& \textbf{VAI} & 174.47* & -0.35* & O  
        & 251.74* & -0.44* & O
        & 2895.13* & -0.47* & O \\
& \textbf{F2-Ratio} & 63.49* & -0.15* & O 
        & 147.83* & -0.33* & O
        & 2225.29* & -0.41* & O \\
\midrule
\multirow{15}{*}{Prosody} 
& \textbf{speaking rate} & 241.63* & -0.45* & O
        & 359.59* & -0.54* & O
        & 759.25* & -0.24* & O \\
& \textbf{articulation rate} & 142.12* & -0.33* & O
        & 42.55 & -0.06 & O
        & 1864.52* & -0.37* & O \\
& $\#$ of pauses & 165.19* & 0.37* & O
        & 207.08* & 0.42* & O
        & 13.73 & 0.01 & X \\
& average pause duration & 155.77* & 0.37* & O 
        & 348.38* & 0.53* & O
        & 41.67* & -0.03 & $\triangle$ \\
\cmidrule{3-11}
& mean F0 & 28.11* & 0.03 & X 
        & 76.92* & 0.23 & O
        & 875.63* & 0.25 & O \\
& \textbf{median F0} & 10.01* & 0.09 & O
        & 34.08* & 0.12 & O
        & 1090.60* & 0.28 & O \\
& std. F0 & 17.12* & 0.00 & X 
        & 153.32* & 0.33 & $\triangle$ 
        & 576.61* & -0.03 & O \\
& min. F0 & 24.33* & -0.03 & X 
        & 84.76* & -0.19 & O
        & 361.69* & 0.06 & O \\
& max. F0 & 24.70* & 0.08 & O 
        & 84.39* & 0.26 & O
        & 518.46* & -0.01 & $\triangle$ \\
\cmidrule{3-11}
& mean energy & 219.01* & 0.48 & $\triangle$  
        & 147.83* & -0.33 & O
        & 1393.16* & -0.19 & O \\
& median energy & 121.74* & 0.33 & $\triangle$  
        & 266.04* & -0.45 & O
        & 439.59* & 0.02 & $\triangle$ \\
& std. energy & 227.75* & 0.48 & $\triangle$  
        & 49.72* & -0.17 & O
        & 1873.16* & -0.25 & O \\
& \textbf{min. energy} & 128.22* & 0.18 & O
        & 81.80* & 0.21 & O
        & 514.47* & 0.05 & O \\
& \textbf{max. energy} & 236.39* & 0.49 & O 
        & 21.13 & -0.08 & O
        & 1681.20* & -0.23 & O \\
\cmidrule{3-11}
& \textbf{$\%V$} & 232.21* & 0.44 & O
        & 144.48* & 0.30 & O
        & 2676.54* & 0.42 & O \\
& VarcoV & 1.95 & 0.27 & X
        & 205.65* & 0.51 & O
        & 537.64* & -0.04 & O \\
& VarcoC & 138.84* & 0.42 & O
        & 0.67 & 0.06 & X
        & 760.69* & 0.11 & O \\
& nPVIV & 204.12* & -0.35 & O
        & 87.10 & 0.01 & X
        & 2798.99* & -0.18 & O \\
& nPVIC & 0.98 & 0.00 & X 
        & 261.14* & -0.45 & O
        & 1313.91* & 0.28 & O \\
\bottomrule
\end{tabular}
}
\end{table*}

\subsubsection{Voice quality-related features}\label{ssec:vq-val}
Voice quality features are expected to reflect the harshness and reduced quality often associated with dysarthria. For English, only two voice break-related features achieved validation. Specifically, the percentage of voice breaks ($H = 40.48, \tau=0.16$) was validated. Jitter ($H = 48.39, \tau=0.03$) and PPQ ($H = 51.09, \tau=0.07$) did not show significant correlations with severity levels in the Kendall rank correlation test. Additionally, shimmer ($H = 182.67, \tau=-0.34$), APQ ($H = 174.95, \tau=-0.32$), HNR ($H = 193.40, \tau=0.29$), and CPP ($H = 110.91, \tau=0.20$) contradicted previous research, failing clinical validation. In the Korean dataset, jitter ($H = 41.17, \tau=0.12$), shimmer ($H = 52.62, \tau=0.12$), APQ ($H = 67.11, \tau=0.15$), and CPP ($H = 120.69, \tau=-0.27$) were validated. However, PPQ ($H = 36.16,\tau=0.02$) and HNR ($H = 11.16, \tau=-0.01$) demonstrated non-significant correlations with dysarthria severity, leading to their exclusion. Tamil's analysis validated only CPP ($H = 502.78,\tau=-0.18$) as a significant voice quality feature, with the percentage of voice breaks ($H = 58.99, \tau=0.01$) showing non-significance in the correlation test. Similar to English, other voice quality features in Tamil, such as shimmer ($H = 484.76, \tau=-0.19$), HNR ($H = 1251.69, \tau=0.24$), demonstrated results that diverge from established findings, prompting their exclusion from classification experiments.

\subsubsection{Pronunciation-related features}\label{ssec:pro-val}
Pronunciation features are designed to capture the misarticulation of consonants and vowels (reduced vowel space) commonly observed in dysarthria. Our analysis revealed consistent validation of all eight pronunciation features (CRR, VRR, PRR, VSA$\triangle$, VSA$\square$, FCR, VAI, F2-Ratio) across all three languages. This validation included both statistical and clinical assessments, demonstrating the robustness of these features in characterizing dysarthria. 

The phoneme accuracy features CRR ($H = 340.12, \tau=-0.57$), VRR ($H = 326.19, \tau=-0.56$), and PRR ($H = 359.20, \tau=-0.58$) were validated across all three languages, indicating their robustness and relevance in assessing dysarthria severity. Korean results for CRR ($H = 523.60, \tau=-0.69$), VRR ($H = 515.07, \tau=-0.68$), and PRR ($H = 555.87, \tau=-0.71$) as well as Tamil results for CRR ($H = 5051.99, \tau=-0.66$), VRR ($H = 4501.92, \tau=-0.62$), and PRR ($H = 5497.47, \tau=-0.69$) confirmed their validity.

The vowel distortion features (VSA$\triangle$, VSA$\square$, FCR, VAI, F2-Ratio) were validated across all three languages. For English, VSA$\triangle$ ($H = 125.33, \tau=-0.27$), VSA$\square$ ($H = 174.47, \tau=-0.37$), FCR ($H = 174.59, \tau=0.35$), VAI ($H = 174.47, \tau=-0.35$), and F2-Ratio ($H = 63.49, \tau=-0.15$) were validated. Korean results for VSA$\triangle$ ($H = 144.58, \tau=-0.33$), VSA$\square$ ($H = 91.08, \tau=-0.26$), FCR ($H = 251.74, \tau=0.14$), VAI ($H = 251.74, \tau=0.44$), and F2-Ratio ($H = 147.83, \tau=-0.33$) confirmed their validity. Tamil validated VSA$\triangle$ ($H = 2322.01, \tau=-0.42$), VSA$\square$ ($H = 2624.32, \tau=-0.45$), FCR ($H = 2895.14, \tau=0.47$), VAI ($H = 2895.14, \tau=-0.47$), and F2-Ratio ($H = 2225.29, \tau=-0.41$).


\subsubsection{Prosody-related features}\label{sssec:prosody-val}
Fluency features are designed to capture the reduced speech rate and the frequency and duration of pauses characteristic of dysarthria. For English, speaking rate ($H = 241.63, \tau=-0.45$), articulation rate ($H = 142.12, \tau=-0.33$), the number of pauses ($H = 165.19, \tau=0.37$), and average pause duration ($H = 155.77, \tau=0.37$) were validated through both statistical and clinical assessments. Korean validated speaking rate ($H = 359.59, \tau=-0.54$), articulation rate ($H = 42.55, \tau=-0.06$), number of pauses ($H = 207.08, \tau=0.42$), and average pause duration ($H = 348.38, \tau=0.54$). However, Tamil only validated articulation rate ($H = 1864.52, \tau=-0.37$), with other fluency features showing non-significance or negative correlations.

Pitch and energy features aim to capture the monotonous pitch, elevated F0, and reduced loudness often associated with dysarthria. For English, median F0 ($H = 10.01, \tau=0.09$), maximum F0 ($H = 24.70, \tau=0.08$), mean energy ($H = 219.01, \tau=0.48$), median energy ($H = 121.74, \tau=0.33$), and standard deviation of energy ($H = 227.75, \tau=0.49$) were validated, while other F0 and energy features were excluded. Korean validated mean F0 ($H = 76.92, \tau=0.03$), median F0 ($H = 34.08, \tau=0.12$), standard deviation of F0 ($H = 153.32, \tau=0.30$), and minimum F0 ($H = 84.76, \tau=0.09$), along with all energy features except for the standard deviation of energy. Tamil validated mean F0 ($H = 875.63, \tau=0.25$), median F0 ($H = 100.98, \tau=0.28$), standard deviation of F0 ($H = 576.61, \tau=0.03$), minimum F0 ($H = 361.69, \tau=0.06$), and maximum F0 ($H = 518.46, \tau=-0.01$), along with all energy features except for median energy.

Rhythm features exhibited language-specific variations in validation. For English, \%V ($H = 232.21, \tau=0.44$), VarcoC ($H = 138.84, \tau=0.42$), and nPVIV ($H = 204.12, \tau=-0.35$) were validated. Korean validated \%V ($H = 144.48, \tau=0.30$), VarcoV ($H = 205.65, \tau=0.51$), and nPVIC ($H = 261.14, \tau=-0.45$). Notably, Tamil demonstrated robust validation across all tested rhythm metrics, including \%V ($H = 2676.54, \tau=0.42$), VarcoV ($H = 537.64, \tau=0.40$), VarcoC ($H = 212.20, \tau=0.67$), nPVIV ($H = 2798.99, \tau=-0.18$), and nPVIC ($H = 1313.91, \tau=0.28$). These findings emphasize the language-specific nature of rhythmic disruptions in dysarthria and highlight the importance of tailored assessment approaches to accurately capture these variations.

\subsubsection{Summary}
Our analysis validated 21, 30, and 24 features for English, Korean, and Tamil, respectively. The multilingual analysis revealed both language-universal and language-specific patterns in feature validation. Notably, no \textbf{voice quality} features exhibited shared patterns across all three languages, challenging the assumption of universal speech motor control mechanisms and warranting further investigation. In contrast, all \textbf{pronunciation} features demonstrated consistent trends across languages, suggesting that limitations in phoneme accuracy and vowel space could serve as potential language-universal dysarthria markers. Regarding prosody-related features, we validated five features across languages: speaking rate, articulation rate, median F0, minimum energy, and maximum energy. \textbf{Fluency} analysis confirmed slower speech rates in severe dysarthria. As for \textbf{pitch}-related features, while median F0 exhibited consistent elevation across languages, \textbf{energy} features showed slight variation. Specifically, maximum energy increased with severity in English but decreased in Korean and Tamil, whereas minimum energy increased with severity across all languages. Rhythm features exhibited considerable variation, with no rhythm features being universally validated across all three languages. 


In summary, our study identified several language-universal characteristics of dysarthria: lower phoneme accuracy (PRR, CRR, VRR), smaller vowel space (lower VSA$\triangle$, VSA$\square$, VAI, F2-Ratio, higher FCR), slower speech rate (speaking rate, articulation rate), elevated F0 (higher median F0), greater intensity when producing speech (greater minimum energy), and longer vowels (\%V). Conversely, lower voice quality, longer and frequent pauses, monotonic speech, softer voice, and irregular rhythm did not exhibit language-universal patterns.

\subsection{Automatic Dysarthria Severity Classification}
Before performing automatic dysarthria severity classification using language-universal features, we first conduct monolingual classification experiments using the validated clinical knowledge-driven features for each language. Subsequently, we perform two types of experiments that train and test across different languages: cross-lingual classification and multilingual classification.

\begin{table}[t]
\centering
\caption{Monolingual classification performances.}
\resizebox{\textwidth}{!}{%
\label{tab:classification_results_mono}
\begin{tabular}{@{}llcccccc@{}}
\toprule
Model & Feature Set & \multicolumn{2}{c}{English} & \multicolumn{2}{c}{Korean} & \multicolumn{2}{c}{Tamil} \\ 
\cmidrule(lr){3-4} \cmidrule(lr){5-6} \cmidrule(lr){7-8}
& & Accuracy & F1-score & Accuracy & F1-score & Accuracy & F1-score \\
\midrule
SVM & eGeMaps &32.59 & 41.63  & 66.25 & 72.30 &  58.87& 67.69 \\
& DisVoice  & 35.81 & 45.47 & 69.32  & 77.14  & 58.26  & 68.94\\
& Proposed & \textbf{36.50} & \textbf{46.20} & \textbf{70.25} & \textbf{78.28}& \textbf{60.38} &\textbf{69.13} \\
\addlinespace
MLP  & eGeMaps & 39.13 & 47.07 & 62.75& 70.36 & 57.04 & 64.88 \\
& DisVoice &\textbf{ 40.70 }& \textbf{51.33}& 64.92 &73.54 & 58.17& 68.97 \\
& Proposed & 37.66 & 47.07 & \textbf{67.88} & \textbf{76.11}  & \textbf{59.45} & \textbf{68.21}\\
\addlinespace
 XGBoost  & eGeMaps &33.01 & 40.20 &67.00 & 74.17 & 54.89 & 63.12 \\
& DisVoice & 36.68 & 45.54& 62.51 & 69.76 & 58.17 &66.87 \\
& Proposed & \textbf{46.04} & \textbf{56.55}& \textbf{73.00} & \textbf{80.36} & \textbf{63.54} & \textbf{71.88}\\
\bottomrule
\end{tabular}
}\label{tab:5-mono}
\end{table}

\subsubsection{Monolingual classification}
To validate the efficacy of our clinically-driven feature set, we compared it against established baseline feature sets, including eGeMAPS and DisVoice. As \Cref{tab:5-mono} illustrates, our proposed feature set outperformed the baseline features across all languages and classifiers, with the exception of the experiment using DisVoice features with the MLP classifier. Among the baseline features, the DisVoice feature set demonstrated superior performance compared to the eGeMAPS feature set. The highest performance was achieved using our proposed feature set with the XGBoost classifier, attaining 46.04\% classification accuracy and 56.55\% F1-score for English, 73.15\% accuracy and 80.36\% F1-score for Korean, and 63.54\% accuracy and 71.88\% F1-score for Tamil. These results validate the effectiveness of our clinical-knowledge-driven features. 

Moreover, it is noteworthy that our proposed feature set has a much smaller dimensionality compared to the baseline sets, with eGeMAPS comprising 88 features and DisVoice 618 features. This further underscores the efficiency and effectiveness of our features, highlighting the importance of employing clinical knowledge for effective and efficient automatic dysarthria severity classification.

\begin{table}[t]
\caption{Cross-lingual classification performances.}
\label{tab:classification_results}
\centering
\resizebox{\textwidth}{!}{%
\begin{tabular}{@{}llcc|cc|cc@{}}
\toprule
Train & Test & \multicolumn{2}{c|}{SVM} & \multicolumn{2}{c|}{MLP} & \multicolumn{2}{c}{XGBoost} \\
      &      & Accuracy & F1-score & Accuracy & F1-score & Accuracy & F1-score \\
\midrule
\textbf{English} & \textbf{English} & \textbf{28.91}& \textbf{39.17}  & \textbf{30.51} & \textbf{41.61} & 31.41 & \textbf{43.19} \\
Korean & English  & 13.16 & 3.60  & 14.04 & 8.29 & 21.93 & 19.74 \\
Tamil & English & 15.44& 4.15 & 28.08& 20.75 & \textbf{44.21} &36.34 \\
Korean+Tamil & English  & 13.16 & 3.66 &23.23 &23.23 & 38.77 & 34.66 \\
\midrule
\textbf{Korean} & \textbf{Korean}  & \textbf{69.25} & \textbf{78.32}   & \textbf{68.75} &\textbf{78.08} & \textbf{73.38 }& \textbf{82.03}\\
English & Korean  & 35.25& 34.88  & 44.25& 47.26&38.12  & 39.18 \\
Tamil & Korean  &61.25 &55.36  &61.88 &59.19 & 63.00 & 61.62\\
English+Tamil  & Korean & 60.00 & 58.73 & 62.38& 61.97  & 61.88 & 61.51 \\
\midrule
\textbf{Tamil} &\textbf{Tamil}  & \textbf{63.22}& \textbf{73.14} &\textbf{63.05} & \textbf{72.97}& \textbf{64.99} &\textbf{74.24} \\
English & Tamil  & 37.35 & 39.39  &36.31 & 39.56 & 39.97  & 41.75 \\
Korean & Tamil  &55.78 & 54.24&53.89 &53.92 & 55.83  & 54.97\\
English+Korean & Tamil & 52.44 & 51.04 &55.71 & 54.91 &56.07 & 55.31 \\
\bottomrule
\end{tabular}%
}\label{tab:5-cross}
\end{table}

\subsubsection{Cross-lingual classification}
\Cref{tab:5-cross} presents the cross-lingual classification performances using three different machine learning classifiers: Support Vector Machine (SVM), Multi-Layer Perceptron (MLP), and XGBoost. The classification tasks involve training models on one language and testing them on another, highlighting the generalization capability of the features across languages.

When models are trained and tested within the same language, the highest performance is observed. For instance, the English-trained XGBoost model achieves an accuracy of 31.41\% and an F1-score of 43.19\% on English test data. Similarly, the Korean-trained XGBoost model achieves an impressive 73.38\% accuracy and 82.03\% F1-score on Korean test data, while the Tamil-trained XGBoost model shows 64.99\% accuracy and 74.24\% F1-score on Tamil test data. These results indicate that models perform best when the training and testing languages are identical.

Cross-lingual performances, however, generally show a decrease in accuracy and F1-score. For example, when trained on Korean and tested on English, the SVM model achieves only 13.16\% accuracy and 3.60\% F1-score. A similar drop is seen in the Korean-to-Tamil and English-to-Korean scenarios. Interestingly, the XGBoost model trained on Tamil and tested on English demonstrates relatively higher performance with 44.21\% accuracy and 36.34\% F1-score, suggesting some transferability of features from Tamil to English. 


\subsubsection{Multilingual classification}
\Cref{tab:5-multilingual} presents the multilingual classification performances for dysarthria severity using three machine learning classifiers. The table compares the performance of monolingual classification against multilingual classification.

For English, the monolingual SVM classifier achieves an accuracy of 28.91\% and an F1-score of 39.17\%, while the multilingual SVM classifier significantly improves these metrics, achieving an accuracy of 61.90\% and an F1-score of 73.33\%. Similarly, the monolingual MLP classifier achieves 30.51\% accuracy and 41.61\% F1-score, with the multilingual MLP classifier showing improved performance at 61.79\% accuracy and 66.67\% F1-score.

\begin{table}[htbp]
\caption{Multilingual classification performances.}
\label{tab:5-multilingual}
\centering
\resizebox{\textwidth}{!}{%
\begin{tabular}{@{}lcccccc@{}}
\toprule
Experiments & \multicolumn{2}{c}{SVM} & \multicolumn{2}{c}{MLP} & \multicolumn{2}{c}{XGBoost} \\
     & Accuracy & F1-score & Accuracy & F1-score & Accuracy & F1-score \\
\midrule
English-mono & 28.91& 39.17  & 30.51 & 41.61 & 31.41 &43.19 \\
\textbf{English-multi}  & \textbf{61.90} & \textbf{73.33}& \textbf{61.79} & \textbf{66.67} & \textbf{72.96}& \textbf{75.66}  \\
\textbf{Korean-mono}   & \textbf{69.25} & \textbf{78.32}   & \textbf{68.75} & \textbf{78.08} & \textbf{73.38}& \textbf{82.03}\\
Korean-multi  & 62.00 & 63.75& 64.88 & 70.00 & 67.00 & 66.42 \\
\textbf{Tamil-mono}  & \textbf{63.22}& \textbf{73.14} &\textbf{63.05} & \textbf{72.97}& \textbf{64.99} &\textbf{74.24} \\
Tamil-multi & 56.77 &69.57 & 55.74 &69.57  & 63.29& 62.77   \\
\bottomrule
\end{tabular}%
}
\end{table}


For Korean, the monolingual classifiers demonstrate strong performance, with the SVM classifier achieving 69.25\% accuracy and 78.32\% F1-score, and the MLP classifier achieving 68.75\% accuracy and 78.08\% F1-score. The multilingual classifiers, however, show a slight decrease in performance, with the SVM classifier achieving 62.00\% accuracy and 63.75\% F1-score, and the MLP classifier achieving 64.88\% accuracy and 70.00\% F1-score.

For Tamil, the monolingual classifiers again demonstrate strong performance, with the SVM classifier achieving 63.22\% accuracy and 73.14\% F1-score, and the MLP classifier achieving 63.05\% accuracy and 72.97\% F1-score. The multilingual classifiers show a decrease in performance, with the SVM classifier achieving 56.77\% accuracy and 69.57\% F1-score, and the MLP classifier achieving 55.74\% accuracy and 69.57\% F1-score.

\subsubsection{Summary}
This study provides a comprehensive evaluation of dysarthria severity classification through monolingual, cross-lingual, and multilingual experiments. The findings indicate that using datasets from other languages does not consistently enhance classification performance, highlighting the inherent challenges in feature transfer, even with language-universal features. Specifically, the performance hierarchy observed was Monolingual $>$ Multilingual $>$ Cross-lingual for Korean and Tamil, while English displayed a different trend: Multilingual $>$ Monolingual $>$ Cross-lingual.

Cross-lingual classification experiments (\Cref{tab:5-cross}) demonstrated significant challenges, with a notable drop in performance when models were trained in cross-lingual settings. This underscores the difficulty in transferring models across languages. Monolingual experiments consistently exhibited the highest performance across all three languages in terms of F1-score. Although classification accuracy was higher in the Tamil-to-English and Korean+Tamil-to-English settings compared to monolingual settings, suggesting partial feature transferability, the high imbalance in severity levels within the English dataset necessitates using the F1-score as a more reliable performance metric. 

In multilingual classification, models trained on combined datasets from multiple languages exhibited mixed results. For Korean and Tamil, multilingual models underperformed compared to their monolingual counterparts, indicating substantial challenges in achieving effective classification performance. The performance degradation of multilingual and cross-lingual classifiers, even with language-universal features, indicates the necessity to consider language-specific features. Conversely, English multilingual models significantly improved classification performance compared to monolingual models, demonstrating the potential effectiveness of multilingual experiments. This discrepancy is likely due to the imbalanced structure of the English dataset, where the moderate severity class has significantly fewer utterances compared to the healthy, mild, and severe classes. With limited data for the moderate class, the English monolingual classifier may not have been fully trained to distinguish it. However, in a multilingual setting, data from Korean and Tamil can augment the dataset, leading to better performance. Therefore, the observed trends in English cannot be conclusively attributed to language differences alone, and further analysis with a more balanced dataset is necessary to draw definitive conclusions. Multilingual classification appears more effective than cross-lingual classification in all three languages, likely due to the increased availability of training data. Including target data in the training dataset likely aids the classifier in optimizing more efficiently.



\section{Conclusion}
In this chapter, we presented a comprehensive analysis of dysarthria severity classification using clinical-knowledge-driven feature sets for English, Korean, and Tamil, validated through statistical and clinical validation. We also performed automatic dysarthria severity classification experiments using language-universal features. According to the experimental results, monolingual models consistently outperformed both cross-lingual and multilingual models for Korean and Tamil, highlighting the importance of language-specific approaches. Conversely, English showed a different trend where multilingual models performed better than monolingual models, suggesting potential benefits from incorporating diverse linguistic data. These findings underscore the need to balance language-universal and language-specific features to develop effective and efficient dysarthria severity classification systems. To conclude, the contributions of this work are as follows:

\begin{itemize}
\item Propose clinically-knowledge-driven feature sets that are statistically and clinically validated for English, Korean, and Tamil dysarthria severity classification.
\item Distinguish between language-universal and language-specific features for English, Korean, and Tamil dysarthric speech assessment.
\item Experimentally compare the efficacy of language-universal features in monolingual, cross-lingual, and multilingual scenarios.
\end{itemize}

\chapter{Incorporating Language Characteristics in Automated Dysarthria Severity Classification}\label{chapter:classification}

\section{Background and Research questions}
Multilingual dysarthria severity classifiers hold immense potential to address critical resource gaps by offering assessment tools for languages lacking dedicated speech analysis infrastructure. Recent research has focused on identifying language-universal features – acoustic characteristics demonstrating consistent behavior across languages \citep{kovac2021multilingual, favaro2023multilingual, kovac2022exploring}. Examples include compressed vowel space \citep{kim2017cross}, pause-based metrics \citep{kovac2021multilingual, kovac2022exploring, favaro2023multilingual}, reduced second formant prominence, mono-pitch \citep{kovac2022exploring, favaro2023multilingual}, and deviations in speech rhythm \citep{favaro2023multilingual}.

However, we posit that integrating language-specific features is crucial for accurate multilingual dysarthria assessment. Speech intelligibility is heavily influenced by a speaker's native language. Previous studies have demonstrated performance degradation in multilingual classification settings compared to controlled monolingual scenarios \citep{hazan2012early, orozco2016automatic, moro2019phonetic}. Cross-lingual experiments further highlight this dependency, revealing that even seemingly universal patterns like rhythmic metrics can fail to transfer effectively between languages \citep{perezautomatic}. Importantly, this notion of "language universality" is limited to the specific languages analyzed in each study, and these features may not be universally applicable across all languages. These findings emphasize the limitations of relying solely on language-universal features and underscore the necessity of incorporating language-specific characteristics for optimal multilingual dysarthria assessment. Our findings in Chapter \ref{chapter:analysis} further support this notion.

This section introduces a methodology that strategically integrates language-shared and language-specific features for classifying dysarthria severity in English, Korean, and Tamil. We use the validated feature set, proposed in our previous chapter, \ref{chapter:analysis}. We then calculate distances between the dysarthric features and feature distributions observed in healthy controls. Feature selection helps determine language-specific feature sets, and cross-lingual comparison allows us to distinguish between language-shared and language-specific characteristics. Finally, we conduct automatic severity classification utilizing both feature sets. Our key research question is:

\begin{enumerate}
\item How can the strategic combination of language-universal and language-specific features maximize the performance of multilingual dysarthria severity classification?
\end{enumerate}

\section{Proposed Method}
\begin{figure}[ht] 
\centering
\includegraphics[width=0.8\columnwidth]{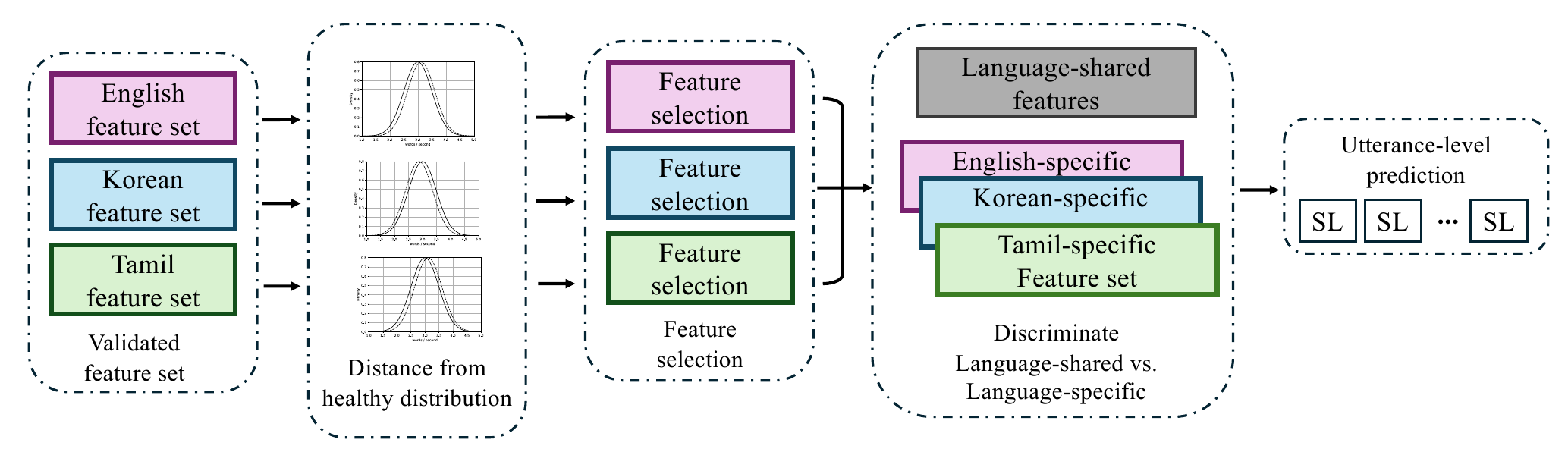}
\caption{
Proposed multilingual dysarthria severity classification.
}
\label{figs:6-method}
\end{figure}

\subsection{Feature Extraction and Validation}
\Cref{figs:6-method} illustrates the proposed multilingual classification method for dysarthria severity assessment. For the feature sets, we employ features that have undergone rigorous statistical and clinical validation as detailed in \Cref{chapter:analysis}. Accordingly, this chapter utilizes 21 validated features for English, 30 for Korean, and 24 for Tamil. 

\subsection{Automatic severity classification}
\subsubsection{Calculation of distance metric}
To measure the degree of acoustic atypicality in dysarthric speech, we calculate a distance metric adapted from \citep{wagner2023careful}. For each acoustic feature, we first determine the mean ($\mu_h$) and standard deviation ($\sigma_{h}$) within a healthy control group. The distance ($f_{i}$) of a data point from this healthy distribution is then calculated using the following formula, where $f_{i}$ represents the feature value of the $i$th utterance:
\begin{equation}
f_i = 
\begin{cases}
\frac{\sigma_{h_i}}{| \mu_{h_i} - f_i |} & \text{if } |f_i - \mu_{h_i}| > \sigma_{h_i} \\
1 & \text{otherwise} 
\end{cases}
\end{equation}

This metric scales deviations that exceed one standard deviation from the healthy control distribution by the inverse of their distance from the healthy mean. Deviations within one standard deviation are assigned a distance of 1, minimizing the effect of minor fluctuations and normalizing the impact of features with different scales. These calculated distances are intended to represent the degree of atypicality in dysarthric speech compared to healthy controls.

\subsubsection{Feature selection}
Even after thorough statistical and clinical validation, the individual contributions of features in an automatic dysarthria severity classification system may need further refinement. To identify the most discriminative acoustic feature sets for each language, we utilize feature selection with the XGBoost algorithm. This process is crucial for achieving optimal classification performance, preventing model overfitting, and highlighting the relative importance of each feature.

\subsubsection{Classification}
\label{sssec:proposed}
By comparing feature selection results across different languages, we strategically identify two key feature categories: language-universal features and language-specific features. This distinction is crucial for developing a robust multilingual classifier that effectively integrates both shared and language-specific characteristics. To optimize model performance, we tailor the feature input during the classification process. As shown in Table \ref{tbl:proposed}, features may be shared across multiple languages or unique to a specific language. During multilingual classification, we discard irrelevant feature values for the target language. This approach maximizes the training samples for shared features, leverages the power of language-specific features, and prevents interference from unrelated features, thereby enhancing classification accuracy. Formally, let a language be denoted by $l$ and an optimal feature set by $U$. The language-specific feature set is represented by $F_l \subset U$, and the language-specific dataset with selected features in $F$ is $D_F^l$. The proposed feature set is then defined as: $\bigcup_{l \in {\texttt{ko}, \texttt{en}, \texttt{ta}}} D_{F_l}^l$.
\begin{table}[h]
\caption{
Tabular for the Proposed experiment
}
\label{tbl:proposed}

\centering
\resizebox{0.5\textwidth}{!}
{

\begin{tabular}{cccc}
\hline

Language & A & B & C \\
\hline
English & O & N/A & N/A \\
Korean & O & O & N/A \\
Tamil & O & O & O \\
\hline


\end{tabular}

}
\end{table}

\section{Classification experiment setting}\label{sec:setting}

\subsection{Feature selection}
\begin{figure}[ht]
    \centering
    \captionsetup{justification=centering}
    \includegraphics[width=0.5\textwidth ]{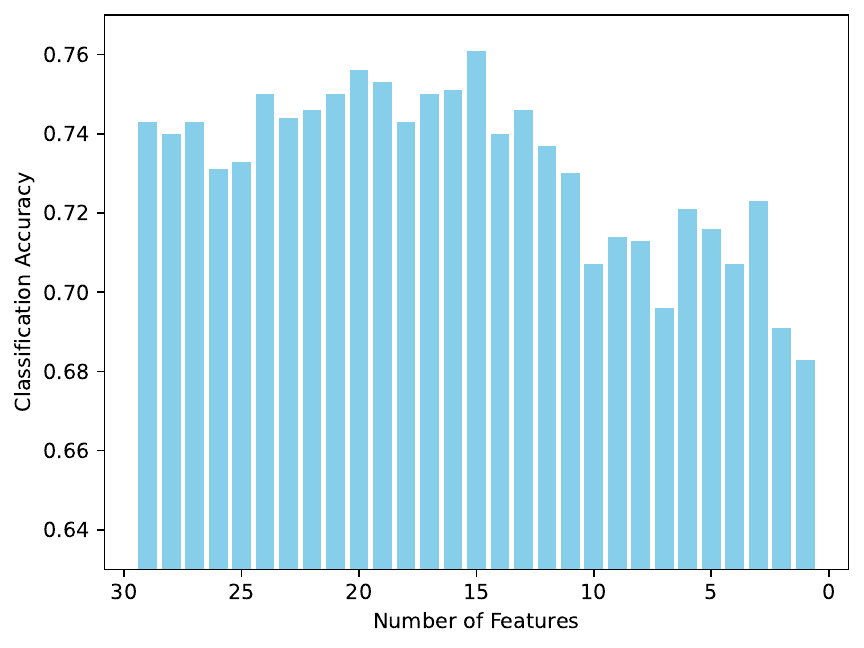}
    \caption{Classification accuracy \\ using different number of features (Korean)}
    \label{fig:iterative}
\end{figure}

\begin{figure*}[t] 
\centering
\subfloat[English]{\includegraphics[width=0.33\textwidth]{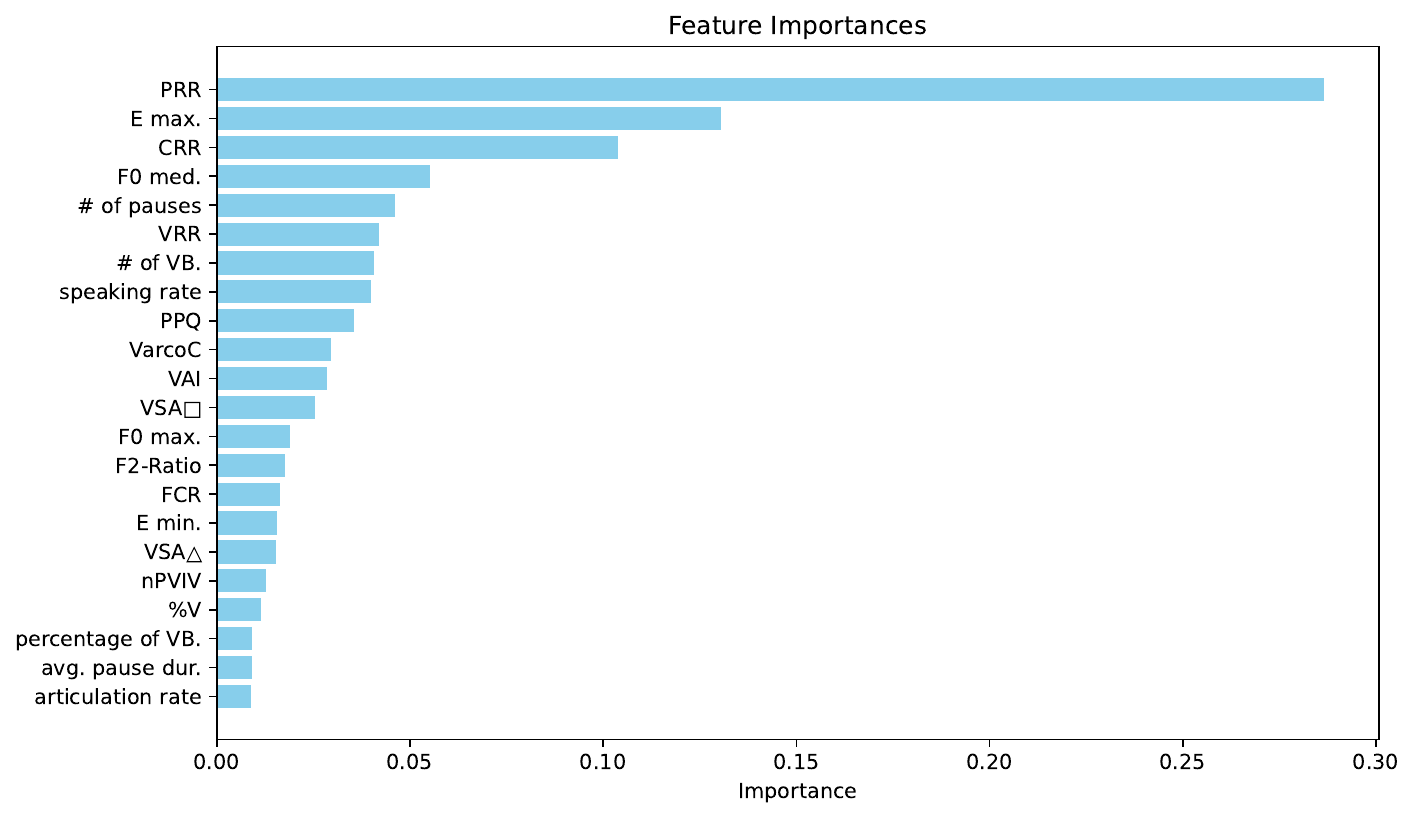}}
\subfloat[Korean]{\includegraphics[width=0.33\textwidth]{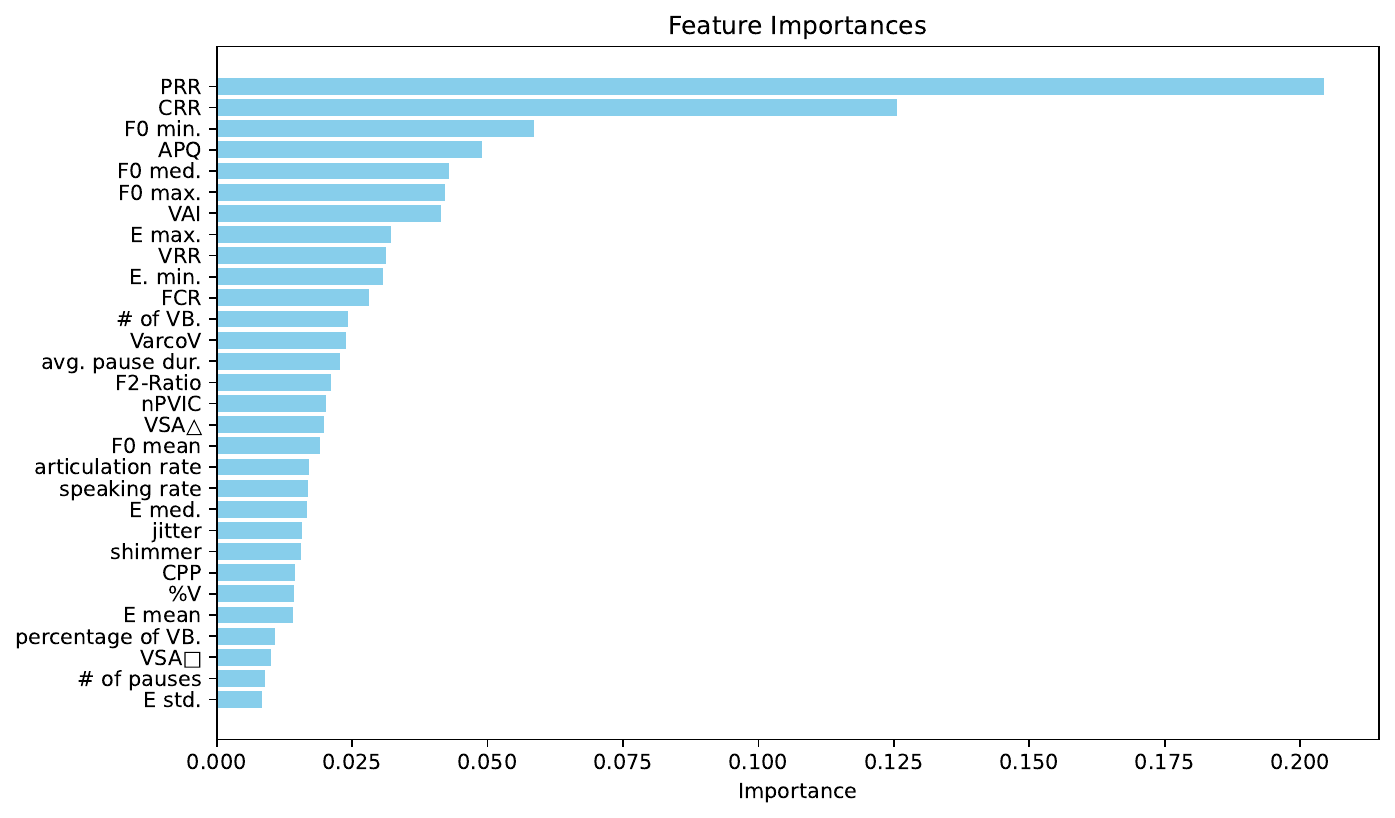}}
\subfloat[Tamil]{\includegraphics[width=0.33\textwidth]{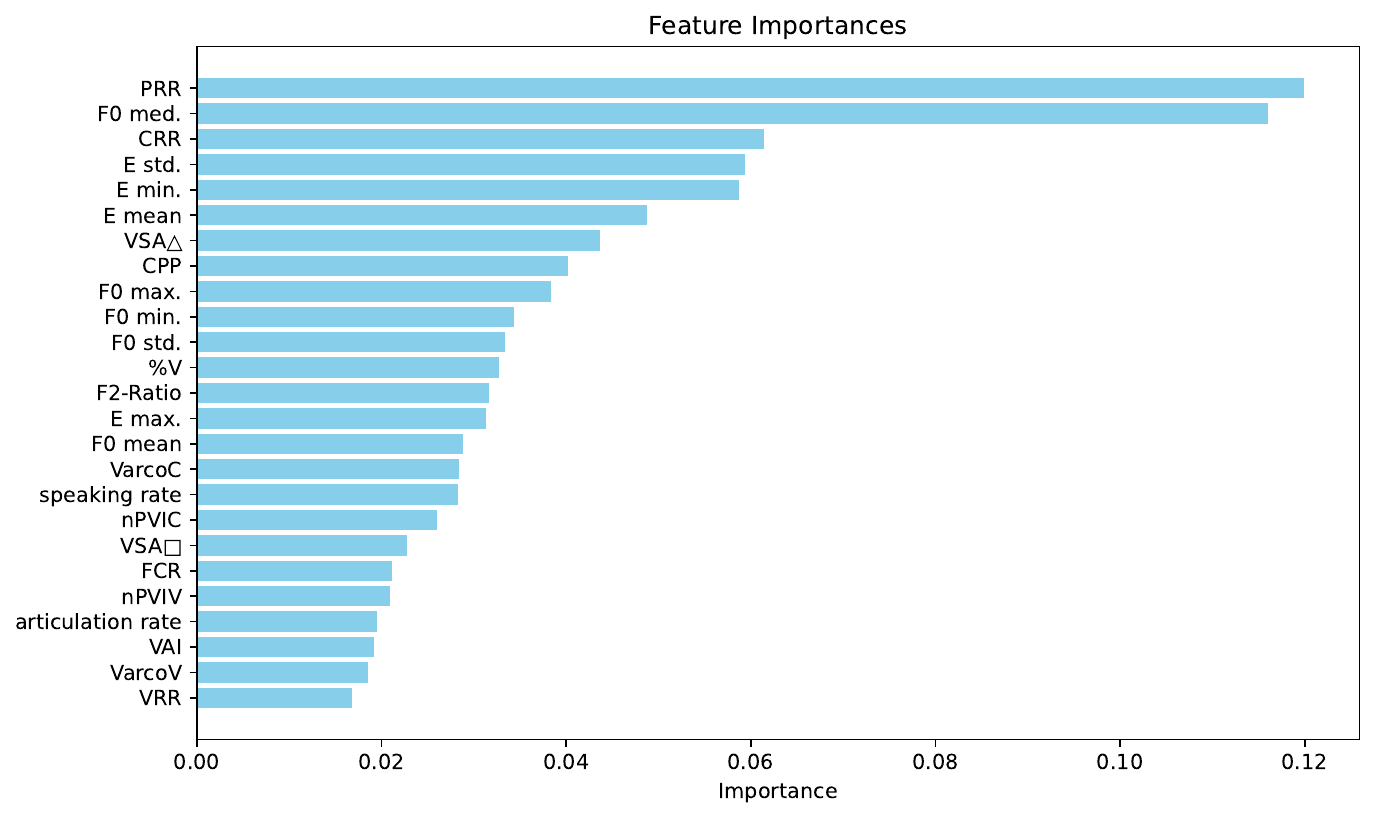}}
\caption{Gain-based importance scores}
\label{fig:gain_scores}
\end{figure*}

We implement a two-step feature selection process using the XGBoost algorithm. First, we determine feature importance scores with XGBoost within a Leave-One-Speaker-Out (LOSO) cross-validation framework. This iterative process involves holding out data from one speaker for testing while training the model on the remaining speakers, then averaging the feature importance scores across all speaker folds. We use a gain-based metric for feature importance, as it reflects a feature's contribution to classification performance by measuring the reduction in node impurity. Figure \ref{fig:gain_scores} shows the averaged importance scores for each feature.
Next, we utilize this feature importance information for iterative feature selection. We systematically remove the least important feature and reassess classification accuracy, enabling us to identify the optimal feature subset that maximizes accuracy for each language. This iterative process begins with the pre-validated feature set described in \Cref{chapter:analysis}. Figure \ref{fig:iterative} illustrates how, for Korean, classification accuracy can be negatively affected by using either too many or too few features.

\subsection{Automatic severity classification}
We chose the XGBoost algorithm for automatic severity classification due to its numerous strengths, including its ability to handle non-linear feature interactions, manage missing values, and reduce bias and underfitting \citep{ogunleye2019xgboost,bao2020multi,athanasiou2020explainable}. Its capability to effectively process datasets with missing values was especially important for our analysis, allowing us to strategically discard irrelevant feature values without compromising model performance. All experiments were conducted on an NVIDIA A100 80GB server.

We optimized each XGBoost classifier through grid search. With the number of estimators set at 300, we searched for the optimal learning rate (eta) within the range of 0.3 to 0.5, while keeping the other hyperparameters at their default settings. To ensure robust results and address potential dataset imbalance, we used a Leave-One-Subject-Out Cross Validation (LOSOCV) approach. This provided a more rigorous evaluation and improved the model's generalizability.

\begin{table*}[t!]
\caption{Optimal feature sets. Bold indicates language-universal features, while underline indicates features shared across two languages only.} 
\label{tab:gain}
\centering
\resizebox{0.9\textwidth}{!}{
\begin{tabular}{@{}c|C{3cm}|C{3cm}|C{3cm}@{}}
\toprule
Category  & English (10) & Korean (16) & Tamil (7) \\
\midrule
Voice quality  
 & PPQ,  \underline{\# of VB.} & APQ, \underline{\# of VB.} & - \\
Phoneme correctness  
 & \textbf{CRR}, \underline{VRR}, \textbf{PRR} & \textbf{CRR}, \underline{VRR}, \textbf{PRR} & \textbf{CRR, PRR} \\
Vowel distortion  
 & - & VAI, FCR, F2-Ratio & VSA $\triangle$\\
Pitch     & \textbf{med. F0} & \textbf{med.}/min./max. \textbf{F0} & \textbf{med.} \textbf{F0} \\
Loudness   & \underline{max}. energy & \underline{min.}/\underline{max.} energy & mean/\underline{min.}/std. energy\\
Fluency    & \# of pauses, speaking rate & avg. pause dur. & - \\
Rhythm  & VarcoC & VarcoV, nPVIC & - \\
\bottomrule
\end{tabular}
}\label{tab:optimal}
\end{table*}

\subsection{Baseline experiments}
\subsubsection{Feature sets}
To validate the extracted voice biomarkers, we performed classification experiments using traditional acoustic features. We used Mel Frequency Cepstral Coefficients (MFCCs) to characterize the spectral shape of the speech signal. The mean and standard deviation of 12-dimensional MFCCs and log energy were extracted using the librosa library \citep{mcfee2015librosa}. Additionally, we assessed performance with eGeMAPS, an extensive set of 88 features designed for general acoustic analysis \citep{eyben2010opensmile}.

\subsubsection{Multilingual feature set design}
To validate the effectiveness of this approach, we conducted baseline experiments using different combinations of language-specific feature sets. In the \textit{Intersection} experiment, we used only the features that were selected across all three languages (language-universal features). Conversely, the \textit{Union} experiment included all features selected for any of the languages. Table \ref{tbl:other_exp}illustrates the dataframe structures for these experiments.
Using the same notation from \Cref{sssec:proposed}, the feature sets are defined as follows:

\begin{enumerate}
\item  
(intersection) $\bigcup_{l \in \{\texttt{en}, \texttt{ko}, \texttt{ta}\}} D^l_{F_{\texttt{en}} \cap F_{\texttt{ko}} \cap F_{\texttt{ta}}} $ 
\item  
(union) $\bigcup_{l \in \{\texttt{en}, \texttt{ko}, \texttt{ta}\}} D^l_{F_{\texttt{en}} \cup F_{\texttt{ko}} \cup F_{\texttt{ta}}}$ 
\end{enumerate}
\newcommand{\mytab}{
    \begin{tabular}{cccc}
    \hline
    Language & A & B & C \\
    \hline
    English & O & N/A & N/A \\
    Korean & O & N/A & N/A \\
    Tamil & O & N/A & N/A \\
    \hline
    \end{tabular}
}

\newcommand{\mytabs}{
    \begin{tabular}{cccc}
    \hline
    Language & A & B & C \\
    \hline
    English & O & O & O \\
    Korean & O & O & O \\
    Tamil & O & O & O \\
    \hline
    \end{tabular}
}

\begin{table}[h]
    \caption{Tabular for Intersection and Union experiments}
    \subfloat[][Intersection]{\mytab}%
    \qquad
    \subfloat[][Union]{\mytabs}%
    \label{tbl:other_exp}
\end{table}

\subsubsection{Baseline classifiers}
We performed baseline experiments using Support Vector Machine (SVM) and Multi-Layer Perceptron (MLP) classifiers. Since these classifiers cannot handle missing values, we restricted the experiments to intersection and union feature sets. Each classifier was optimized through a grid search process. For SVM, we used a radial basis function (rbf) kernel and searched for the optimal hyperparameters $C$ (ranging from $10^{-1}$ to $10^3$) and $\gamma$ (from $10^{-3}$ to $1$). For MLP, we explored hidden layer sizes of ($50$,) and ($100$,) along with Adam and stochastic gradient descent (SGD) optimizers. All MLP experiments used the ReLU activation function with a maximum iteration setting of $5000$.

\subsection{Evaluation}
We evaluated classification performance using the weighted F1-score averaged across speakers. Initially, the model was trained to predict the severity level of individual utterances. We then calculated the weighted F1-score for each speaker and averaged these speaker-level performances to obtain the final performance score. We selected the weighted F1-score as the evaluation metric due to the significant imbalance in the dataset.

\section{Experimental Results}
\subsection{Feature selection}
Table \ref{tab:gain} outlines the optimal feature sets for each language, chosen for their ability to achieve the highest classification accuracy. For English, the selected features are PPQ, number of voice breaks, CRR, VRR, PRR, median F0, maximum energy, number of pauses, speaking rate, and VarcoC. The optimal features for Korean include APQ, number of voice breaks, CRR, VRR, PRR, VAI, FCR, F2-Ratio, median/minimum/maximum F0, minimum/maximum energy, average pause duration, VarcoV, and nPVIC. For Tamil, the optimal set comprises CRR, PRR, VSA$\triangle$, median F0, and mean/minimum/standard deviation energy.
We identified a language-universal feature set that includes CRR, PRR, and median F0, which was used in our baseline intersection experiment. Additionally, we noted shared features between language pairs: number of voice breaks and maximum energy between English and Korean, and minimum energy between Korean and Tamil. Other features are considered language-specific.

\subsection{Automatic severity assessment}
\subsubsection{Efficacy of the hand-crafted feature sets}
Before proceeding with multilingual classification, we conducted monolingual classification experiments to validate the effectiveness of our handcrafted features in capturing the characteristics of dysarthric speech. We compared the statistically and clinically validated feature sets (Handcrafted-a) and the optimal feature sets obtained through feature selection (Handcrafted-b) against baseline features, MFCCs, and eGeMAPS.

As shown in \Cref{tab:monolingual}, both Handcrafted-a and Handcrafted-b feature sets significantly outperformed the baselines, achieving higher F1-scores across all three languages (English, Korean, and Tamil) and all three classifiers (SVM, MLP, XGBoost). Notably, the feature selection process further improved performance, with the Handcrafted-b feature set consistently yielding superior F1-scores compared to Handcrafted-a. This underscores the effectiveness of feature selection in refining feature sets for optimal dysarthria classification. The highest overall performance was achieved using the XGBoost classifier with the Handcrafted-b feature set, resulting in average F1-scores of 63.52\%, 80.80\%, and 69.96\% for English, Korean, and Tamil respectively, with an overall average of 71.22\%.

\begin{table}[t]
\caption{Monolingual classification results. Best performances are indicated in bold.}
\label{tab:comparative_analysis}
\centering
\resizebox{0.7\textwidth}{!}{%
\begin{tabular}{@{}clcccc@{}}
\toprule
Classifier & Feature Set & English & Korean & Tamil & Average \\ \midrule
\multirow{4}{*}{SVM}
 & MFCCs & 51.62 & 58.68 & 51.64 & 53.98 \\
 & eGeMaps & 43.59 & 70.58 & 58.14 & 57.44 \\
 & Hand-crafted-a & 49.05 & 80.09 & 57.00 & 62.05 \\
 & Hand-crafted-b & 54.00 & 81.23 & 67.11 & 67.45 \\ \midrule
\multirow{4}{*}{MLP}
 & MFCCs & 53.18 & 61.46 & 53.11 & 55.92 \\
 & eGeMaps & 44.58 & 68.91 & 54.92 & 56.14 \\
 & Hand-crafted-a & 51.00 & 78.50 & 45.00 & 58.17 \\
 & Hand-crafted-b & 54.28 & 80.32 & 68.04 & 67.55 \\ \midrule
\multirow{4}{*}{XGBoost}
 & MFCCs & 51.53 & 64.11 & 54.59 & 56.74 \\
 & eGeMaps & 40.14 & 67.71 & 54.87 & 54.24 \\
 & Hand-crafted-a & 58.95 & 80.29 & 62.07 & 67.10 \\
 &\textbf{Hand-crafted-b} & \textbf{63.52} & \textbf{81.70 }& \textbf{69.95} & \textbf{71.72} \\ \bottomrule
\end{tabular}%
}\label{tab:monolingual}
\end{table}

\begin{table}[t!]
\caption{Multilingual classification results. Best performances are indicated in bold.}
\label{tab:classification_results_cross}
\centering
\resizebox{0.7\textwidth}{!}{%
\begin{tabular}{@{}clccccc@{}}
\toprule
Classifier& Feature Set & English & Korean & Tamil & Average \\ \midrule
\multirow{4}{*}{SVM} 
 & MFCCs & 49.13 & 57.65 & 53.02 & 53.27 \\
  & eGeMaps & 61.91 & 66.35  & 58.41 &  62.56\\
 & Intersection & 54.17 & 78.82 & 60.28 & 64.42 \\
 & Union & 63.10 & 77.12 & 63.99 & 68.07 \\ \midrule
\multirow{4}{*}{MLP} 
 & MFCCs & 54.77 & 62.58 & 51.72 & 56.36 \\
   & eGeMaps & 64.76 & 67.22 & 57.75& 63.24 \\
 & Intersection & 54.53 & 78.00 & 62.50 & 65.01 \\
 & Union & 58.57 & 75.03 & 66.54 & 66.71 \\ \midrule
\multirow{5}{*}{XGBoost} 
 & MFCCs & 55.98 & 59.03 & 52.35 & 55.79 \\
   & eGeMaps & 62.11 & 68.16 & 58.10 & 62.79 \\
 & Intersection & 56.36 & 75.31 & 67.73 & 66.47 \\
 & Union & 60.82 & \textbf{80.51} & 70.70 & 70.68 \\
 & \textbf{Proposed} & \textbf{63.68} & 80.20 & \textbf{71.34} & \textbf{71.74} \\ \bottomrule
\end{tabular}%
}\label{tab:multilingual}
\end{table}

\subsubsection{Comparison between multilingual classifications}
Table \ref{tab:multilingual} provides a comprehensive comparison of classification performance for multilingual dysarthria severity assessment, including baseline features (MFCCs, eGeMAPS), intersection experiments (using only language-universal features), union experiments (combining all language-specific features), and our proposed method. The results consistently highlight the superiority of our proposed approach, which strategically integrates both language-universal and language-specific features.

Intersection experiments, which rely solely on language-universal features, represent the most conventional approach to multilingual dysarthric speech assessment. Although union experiments, which combine all language-specific features, are another conventional method, they have not been widely adopted in previous research. Our findings indicate that both these conventional methods are less effective than our proposed approach.

The performance ranking across different classifiers remained consistent, with the union experiment outperforming the intersection experiment, followed by eGeMAPS and MFCCs. This confirms our hypothesis that relying only on language-universal features is not sufficient for optimal classification. Notably, the proposed method achieved the highest accuracy for English (63.68\%) and Tamil (71.34\%), while remaining competitive for Korean (80.20\%) compared to the union experiment (80.51\%). With an average accuracy of 71.74\% across all languages, our proposed method showed relative improvements of 7.33\% and 0.93\% over the intersection and union experiments, respectively.

These results underscore the importance of incorporating language-specific features for enhanced multilingual dysarthria severity classification. Our proposed method outperforms the two conventional approaches by effectively leveraging both language-universal and language-specific features, leading to superior performance in multilingual dysarthria severity classification.

\subsubsection{Comparison between monolingual and multilingual classifications}
To further confirm the effectiveness of our proposed multilingual approach, we compared its performance with the best monolingual experiments (using the Handcrafted-b feature set) from Table \ref{tab:monolingual}. The multilingual approach yielded comparable results overall, with slight accuracy improvements for English and Tamil, and a minor decrease for Korean. The average performance across all languages was almost identical between the monolingual and multilingual experiments, at 71.72\% and 71.74\%, respectively. Notably, our proposed method is the only approach in this study that matches the performance of the best monolingual classifications. This underscores the effectiveness and potential of our method for assessing dysarthria severity in a multilingual context.

\section{Discussion}
\subsection{Effect of Tamil dataset sizes}
As shown in Table \ref{tab:dataset}, the Tamil dataset is considerably larger than the English and Korean datasets. To prevent our classifiers from overfitting to the Tamil dataset, we conducted multilingual experiments using varying proportions of utterances, ranging from 20\% to 100\%. Utterances were selected incrementally, with larger ratios encompassing all utterances from smaller ratios. 

\begin{figure}[t]
    \centering
    \captionsetup{justification=centering}
    \includegraphics[width=0.45\textwidth ]{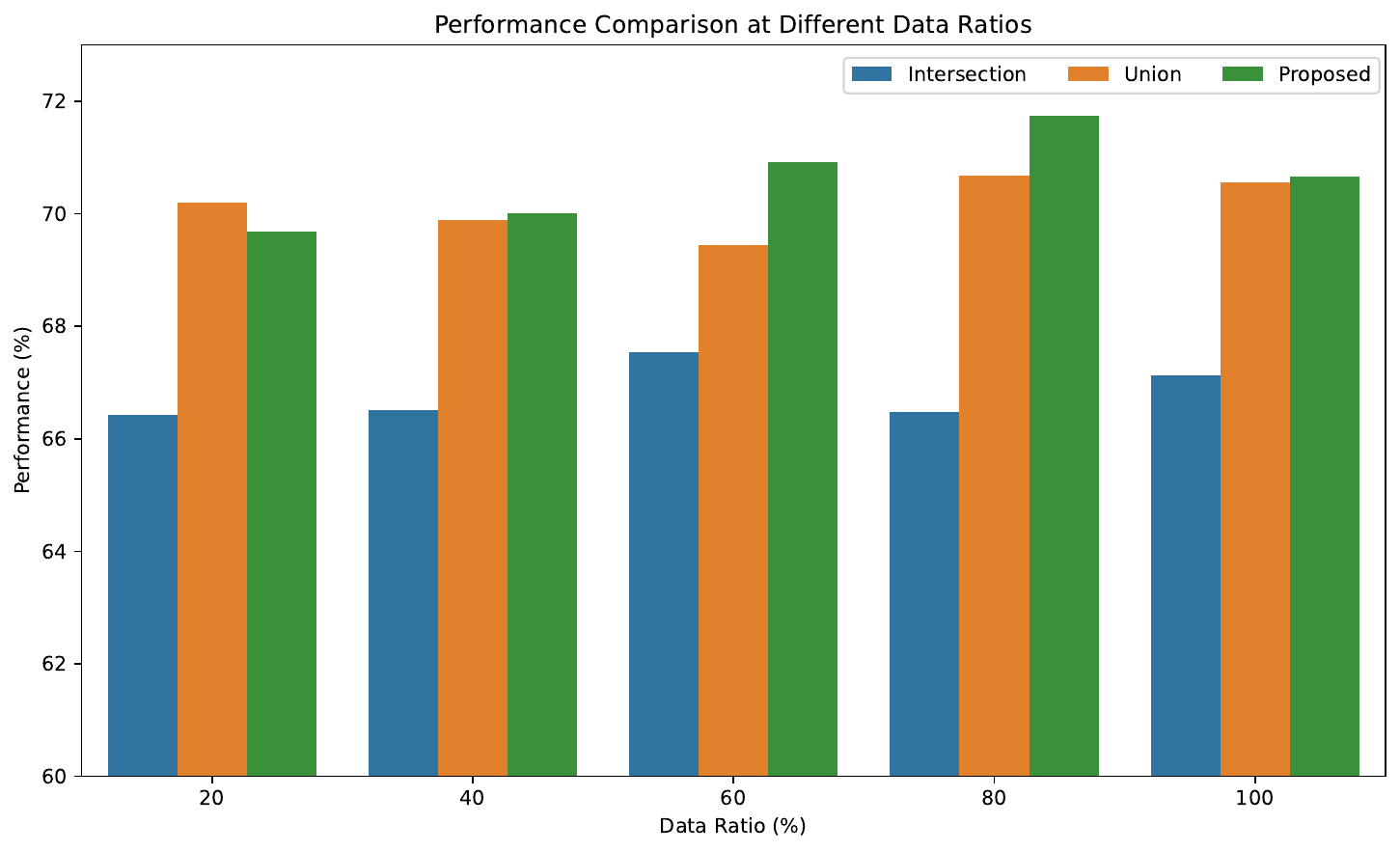}
    \caption{Classification performances \\ using different ratio of Tamil datasets.}
    \label{fig:importance}
\end{figure}

As presented in \Cref{fig:importance}, the intersection experiment, which relies solely on language-universal features, consistently performed worse than both the union and proposed experiments. This emphasizes the necessity of incorporating language-specific features to achieve optimal classification accuracy. Our proposed method, which strategically integrates both language-universal and language-specific features, demonstrated superior performance across most utterance ratios. The exception was the 20\% ratio, where the union experiment outperformed it. We hypothesize that this exception is due to the need for a sufficiently large dataset for our proposed method to effectively learn how to handle missing values, which are strategically excluded to mitigate the negative impact of irrelevant features. The highest classification performance was achieved using 80\% of the utterances, forming the basis for the results presented in this paper.

\subsection{Comparison to the previous study}
To more clearly articulate the contribution of our work, we compare our methods and results with those from the previous study by \citep{yeo2022cross}, the only study that performed multilingual classification using the same datasets: English TORGO, Korean QoLT, and Tamil SSNCE. The three main differences are in validation, feature selection, and classification.

Firstly, we implemented a two-step validation process that includes both statistical and clinical validation to ensure our classification systems use feature sets that are clinically useful and interpretable, which is essential for clinical diagnosis. Secondly, to prevent the data leakage present in the previous study, which resulted in overly optimistic outcomes, we used a Leave-One-Speaker-Out (LOSO) cross-validation scheme for feature importance scoring and selection. This approach enhances the reliability and clinical applicability of our model. Lastly, we introduced a distance-based metric to refine our feature representation by capturing the degree of feature atypicality, in contrast to the raw values used in \citep{yeo2022cross}.

According to the classification results, \citep{yeo2022cross} achieved F1-scores of 59.03\% for English, 76.40\% for Korean, and 46.91\% for Tamil, with an overall average of 60.78\% in a monolingual setting. For multilingual classification, their experiment using both language-universal and language-specific features yielded the highest performances, with scores of 69.46\% for English, 78.13\% for Korean, and 53.84\% for Tamil, resulting in an overall average of 67.14\%. In comparison, our study achieved significant improvements in monolingual classifications for all three languages, with F1-scores of 63.52\% for English, 81.70\% for Korean, and 69.95\% for Tamil, leading to an overall average of 71.72\%. Additionally, our multilingual experiments generally outperformed the previous study, with scores of 63.68\% for English, 80.20\% for Korean, and 71.34\% for Tamil, culminating in an average score of 71.74\%. While Korean and Tamil showed better performance, English performed slightly worse. We attribute this to the elimination of data leakage, which had a more significant impact on English due to its smaller dataset size.

To summarize, our study generally achieves better scores compared to the previous study. Interestingly, despite the elimination of data leakage, we achieved higher performances than \citep{yeo2022cross}. We believe this improvement is due to our distance-based feature, which directly introduces a classifier with a degree of atypicality. However, based on the experimental results, distance-based metrics appear to be more effective in a monolingual setting than in a multilingual setting, with the former showing a greater relative increase in performance (18\% vs. 6.85\%). Additionally, the relative increase from monolingual to multilingual experiments was much lower than in the previous study, likely due to greater benefits in monolingual settings. We hypothesize that this discrepancy is due to the inherent linguistic diversity across languages, resulting in distinct manifestations of dysarthric speech patterns that may not be fully captured by a universal distance metric. Further investigation is needed to explore the optimal use of distance-based features in multilingual dysarthria assessment.

This study significantly contributes to the field by validating the importance of incorporating both shared and language-specific features in multilingual dysarthria classification. Our findings highlight that solely relying on language-universal features, a common practice in multilingual dysarthria assessment, may not fully capture the nuances of dysarthric speech across different linguistic contexts. By integrating language-specific features, we demonstrate an improved ability to model the unique manifestations of dysarthria within each language, thereby enhancing overall classification performance.

\section{Conclusion}
This study highlights the significance of incorporating both commonly shared and language-specific features for multilingual dysarthria severity classification. We propose a method that leverages both types of features, treating language-sensitive features irrelevant to a specific language as missing values. Our experimental results demonstrate that this approach outperforms baseline models that use only shared or all combined features across English, Korean, and Tamil. Our method shows the benefits of combining language-universal features for broad learning with language-specific features for nuanced adaptation, while avoiding interference from irrelevant features of other languages.

To conclude, the contributions of this work are as follows:

\begin{itemize}
\item Propose a novel method integrating both language-universal and language-specific features to improve multilingual dysarthria severity classification.
\item Highlight the importance of language-specific characteristics in achieving optimal performance in multilingual dysarthria classification.
\end{itemize}

A limitation of this study is the potential issue of XGBoost's default handling of missing values. While our method produces the best results given the choices made, it doesn't guarantee that these choices, which may include default directions, are optimal \citep{xgboost_missing}. More missing values could increase the likelihood of incorrect model choices. Thus, since our method uses a varying number of features per language, English, with the most dropped features, may have been particularly affected. Future research aims to address this limitation. Additionally, future studies will explore the generalizability of our model with different datasets and languages. We will also investigate the use of deep neural networks designed to exploit both shared and language-specific features for multilingual classification. Finally, while this study was limited by data availability, considering dysarthric subtypes is essential for future research. 
\chapter{Conclusion and Future Directions}
\label{chp:conclusion}

\section{Thesis Results}
In this thesis, we delved into the complexities of dysarthria assessment in a multilingual context. Through a comprehensive analysis of dysarthric speech in English, Korean, and Tamil, we confirmed the presence of both language-universal and language-specific features in dysarthric speech characteristics. 

Our proposed multilingual dysarthria severity classification model, leveraging both language-universal and language-specific features within an XGBoost framework, demonstrated a notable improvement over traditional multilingual approaches that rely solely on language-universal features. This finding underscores the importance of incorporating language-specific information to achieve optimal performance in automatic dysarthria assessment. This result highlights the potential of our approach to enhance the accuracy and efficiency of dysarthria severity assessment across diverse linguistic populations.

\section{Thesis Contributions}
Research into the influence of language on dysarthria characteristics is a relatively recent development compared to similar studies in aphasia \citep{moya2023raising, kim2024does}. Aphasia research from the 1980s highlighted the significant role of language differences in shaping aphasia manifestations, often more so than the type of aphasia itself \citep{menn1990agrammatic, vaid1991sentence, wulfeck1989pragmatics}. In contrast, multilingual studies in dysarthria have been limited, partly due to the assumption that speech motor control and its associated disorders are universal across languages \citep{miller2014introduction}.
This thesis makes several significant contributions to the field of dysarthria assessment:

\begin{itemize}
\item \textbf{Multilingual Analysis of Dysarthric Speech}: We conducted a comprehensive multilingual analysis of dysarthric speech across English, Korean, and Tamil, three distinct languages in terms of phonemic and prosodic structures. This analysis aimed to identify both language-universal and language-specific features of dysarthria. These features provide valuable insights into how dysarthria affects speech across languages and inform the development of more effective multilingual assessment tools.
\item \textbf{Clinical-knowledge-driven features}: To assess dysarthria severity across languages, we explored and validated a set of features informed by clinical knowledge of dysarthria. These features bridge the gap between the acoustic characteristics of speech and their clinical interpretations for multilingual dysarthria assessment. We focused on three speech dimensions: voice quality, pronunciation, and prosody. Through our multilingual analysis, we identified which features are likely to be universal across languages (e.g., misarticulation of phonemes) and which may be influenced by specific language characteristics (e.g., phonation and prosodic elements).
\item \textbf{Multilingual Dysarthria Severity Classification Method Considering Language Characteristics}: We proposed a novel multilingual dysarthria severity classification method. This method leveraged a comprehensive set of clinically-driven features, incorporating both language-universal and language-specific characteristics. By integrating these features within an XGBoost framework, known for its effectiveness in handling complex feature interactions, we addressed the critical role of language-specific information for accurate assessment. This method has the potential to outperform traditional multilingual classification approaches that solely relied on language-universal features, emphasizing the necessity of considering language characteristics for accurate dysarthria severity classification.
\end{itemize}

\section{Recommendations for Future Work}
This thesis establishes a novel paradigm for multilingual dysarthria assessment, paving the way for significant advancements in the speech pathology field. Here, we outline two key areas for future exploration that hold immense potential to further strengthen the multilingual approach of this work: analysis-focused approach and model-focused approach.

First, future research should prioritize acquiring and analyzing gender-balanced and severity-balanced datasets across multiple languages. Balanced datasets are crucial for ensuring the clinical utility and generalizability of dysarthria assessment tools, allowing for a nuanced understanding of dysarthria across different demographics and severity levels. Existing datasets often suffer from gender and severity imbalances, which hinder robust statistical analysis and may introduce bias. Balanced datasets improve the accuracy of multilingual analysis by providing an equitable representation of different demographics and severity levels. This enhancement leads to more reliable and unbiased statistical analyses, ultimately improving the robustness and applicability of these methods across diverse populations. Collecting new, balanced datasets will lead to better multilingual analysis, and ultimately enhance the robustness and applicability of these methods for diverse populations.

Additionally, analyzing parallel datasets, which include recordings from individuals with and without dysarthria speaking the same sentences, is recommended. This approach, achievable through careful subsampling, can improve the understanding of dysarthric speech characteristics and facilitate multilingual context analysis by controlling for inherent phonetic and prosodic differences between languages.

As for the model-focused approach, future work should integrate advanced machine learning techniques tailored for multilingual dysarthria severity classification. Exploring alternative deep learning architectures, such as transformer-based models or convolutional neural networks, can lead to significant performance improvements. Developing multi-task learning frameworks to address severity classification and phoneme-level pronunciation assessment concurrently in multiple languages will further enhance model utility.

Incorporating self-supervised learning approaches can leverage large amounts of unlabeled multilingual data, improving model training. Transfer learning and domain adaptation techniques will allow models trained in one language to be effectively adapted to others, supporting the development of novel multilingual assessment tools. These advancements will ensure equitable access to accurate diagnosis across diverse linguistic populations.

By focusing on these analysis-focused and model-focused approaches, future research can build on this study's findings to create more inclusive, efficient, and precise dysarthria assessment tools, ultimately improving clinical outcomes and quality of life for individuals with dysarthria worldwide.

\appendix
\chapter{Classification accuracies with paralinguistic features.}
\begin{center}
\begin{longtable}{llrrr}
\caption{Performance comparison of kNN, SVM, MLP, RF, and xgboost classifiers on various paralinguistic feature sets for English dysarthric speech data.} \\
\toprule
Classifier & Feature & Feats & Accuracy & F1 \\
\midrule
\endfirsthead

\multicolumn{5}{c}{{\tablename\ \thetable{} -- continued from previous page}} \\
\toprule
Classifier & Feature & Feats & Accuracy & F1 \\
\midrule
\endhead

\bottomrule
\endfoot

\bottomrule
\endlastfoot

kNN 
 & lasso & 9 & 46.23 & 58.70 \\
 & lasso+filter & 7 & 48.60 & 61.31 \\
 & lasso+RFE & 6 & 46.00 & 57.75 \\
 & lasso+ETC & 3 & 42.09 & 54.52 \\
 & elastic & 10 & 50.37 & 62.42 \\
 & elastic+filter & 8 & 46.55 & 59.20 \\
 & elastic+RFE & 5 & 41.93 & 55.02 \\
 & elastic+ETC & 3 & 42.09 & 54.52 \\
 & clustering & 135 & 46.07 & 58.86 \\
 & clustering+filter & 18 & \textbf{51.30} & \textbf{63.02} \\
 & clustering+RFE & 8 & 47.04 & 57.46 \\
 & clustering+ETC & 17 & 51.17 & 62.30 \\
\midrule
SVM 
 & lasso & 9 & 46.17 & 58.90 \\
 & lasso+filter & 7 & 47.01 & 58.36 \\
 & lasso+RFE & 6 & 47.25 & 58.43 \\
 & lasso+ETC & 3 & 46.12 & 58.81 \\
 & elastic & 10 & 50.88 & 63.26 \\
 & elastic+filter & 8 & 47.90 & 58.64 \\
 & elastic+RFE & 5 & 45.79 & 57.50 \\
 & elastic+ETC & 3 & 46.12 & 58.81 \\
 & clustering & 135 & 34.31 & 44.30 \\
 & clustering+filter & 18 & 44.36 & 57.46 \\
 & clustering+RFE & 8 & 43.68 & 52.17 \\
 & clustering+ETC & 17 & 42.14 & 50.80 \\
\midrule
MLP 
 & lasso & 9 & 47.13 & 60.97 \\
 & lasso+filter & 7 & 49.35 & 61.38 \\
 & lasso+RFE & 6 & 43.79 & 57.07 \\
 & lasso+ETC & 3 & 45.23 & 58.01 \\
 & elastic & 10 & 47.64 & 61.58 \\
 & elastic+filter & 8 & 48.73 & 62.19 \\
 & elastic+RFE & 5 & 45.22 & 58.59 \\
 & elastic+ETC & 3 & 43.19 & 56.28 \\
 & clustering & 135 & 41.77 & 53.95 \\
 & clustering+filter & 18 & 47.11 & 59.86 \\
 & clustering+RFE & 8 & 40.54 & 50.51 \\
 & clustering+ETC & 17 & 47.69 & 57.26 \\
\midrule
RF 
 & lasso & 9 & 43.35 & 55.37 \\
 & lasso+filter & 7 & 44.69 & 56.07 \\
 & lasso+RFE & 6 & 45.91 & 57.74 \\
 & lasso+ETC & 3 & 45.16 & 57.19 \\
 & elastic & 10 & 46.37 & 57.85 \\
 & elastic+filter & 8 & 45.27 & 56.38 \\
 & elastic+RFE & 5 & 45.64 & 57.88 \\
 & elastic+ETC & 3 & 44.64 & 56.49 \\
 & clustering & 135 & 28.07 & 33.69 \\
 & clustering+filter & 18 & 39.03 & 50.42 \\
 & clustering+RFE & 8 & 34.53 & 40.92 \\
 & clustering+ETC & 17 & 36.55 & 43.38 \\
\midrule
xgboost 
 & lasso & 9 & 44.58 & 58.49 \\
 & lasso+filter & 7 & 44.59 & 58.09 \\
 & lasso+RFE & 6 & 41.17 & 53.46 \\
 & lasso+ETC & 3 & 40.85 & 55.12 \\
 & elastic & 10 & 44.19 & 58.76 \\
 & elastic+filter & 8 & 45.62 & 59.33 \\
 & elastic+RFE & 5 & 40.93 & 53.60 \\
 & elastic+ETC & 3 & 40.85 & 55.12 \\
 & clustering & 135 & 38.22 & 46.88 \\
 & clustering+filter & 18 & 47.56 & 61.42 \\
 & clustering+RFE & 8 & 40.91 & 50.99 \\
 & clustering+ETC & 17 & 46.38 & 56.60 \\
\end{longtable}
\end{center}
\begin{center}
\begin{longtable}{llrrrr}
\caption{Performance comparison of kNN, SVM, MLP, RF, and xgboost classifiers on various feature sets for Korean dysarthric speech data.} \\
\toprule
Classifier & Feature & Feats & Accuracy & F1 \\
\midrule
\endfirsthead

\multicolumn{5}{c}{{\tablename\ \thetable{} -- continued from previous page}} \\
\toprule
Classifier & Feature & Feats & Accuracy & F1 \\
\midrule
\endhead

\bottomrule
\endfoot

\bottomrule
\endlastfoot

kNN 
 & lasso & 9 & 46.23 & 58.70 \\
 & lasso+filter & 7 & 48.60 & 61.31 \\
 & lasso+RFE & 6 & 46.00 & 57.75 \\
 & lasso+ETC & 3 & 42.09 & 54.52 \\
 & elastic & 10 & 50.37 & 62.42 \\
 & elastic+filter & 8 & 46.55 & 59.20 \\
 & elastic+RFE & 5 & 41.93 & 55.02 \\
 & elastic+ETC & 3 & 42.09 & 54.52 \\
 & clustering & 135 & 46.07 & 58.86 \\
 & clustering+filter & 18 & 51.30 & 63.02 \\
 & clustering+RFE & 8 & 47.04 & 57.46 \\
 & clustering+ETC & 17 & 51.17 & 62.30 \\
\midrule
SVM 
 & lasso & 9 & 46.17 & 58.90 \\
 & lasso+filter & 7 & 47.01 & 58.36 \\
 & lasso+RFE & 6 & 47.25 & 58.43 \\
 & lasso+ETC & 3 & 46.12 & 58.81 \\
 & elastic & 10 & 50.88 & 63.26 \\
 & elastic+filter & 8 & 47.90 & 58.64 \\
 & elastic+RFE & 5 & 45.79 & 57.50 \\
 & elastic+ETC & 3 & 46.12 & 58.81 \\
 & clustering & 135 & 34.31 & 44.30 \\
 & clustering+filter & 18 & 44.36 & 57.46 \\
 & clustering+RFE & 8 & 43.68 & 52.17 \\
 & clustering+ETC & 17 & 42.14 & 50.80 \\
\midrule
MLP 
 & lasso & 9 & 47.13 & 60.97 \\
 & lasso+filter & 7 & 49.35 & 61.38 \\
 & lasso+RFE & 6 & 43.79 & 57.07 \\
 & lasso+ETC & 3 & 45.23 & 58.01 \\
 & elastic & 10 & 47.64 & 61.58 \\
 & elastic+filter & 8 & 48.73 & 62.19 \\
 & elastic+RFE & 5 & 45.22 & 58.59 \\
 & elastic+ETC & 3 & 43.19 & 56.28 \\
 & clustering & 135 & 41.77 & 53.95 \\
 & clustering+filter & 18 & 47.11 & 59.86 \\
 & clustering+RFE & 8 & 40.54 & 50.51 \\
 & clustering+ETC & 17 & 47.69 & 57.26 \\
\midrule
RF 
 & lasso & 9 & 43.35 & 55.37 \\
 & lasso+filter & 7 & 44.69 & 56.07 \\
 & lasso+RFE & 6 & 45.91 & 57.74 \\
 & lasso+ETC & 3 & 45.16 & 57.19 \\
 & elastic & 10 & 46.37 & 57.85 \\
 & elastic+filter & 8 & 45.27 & 56.38 \\
 & elastic+RFE & 5 & 45.64 & 57.88 \\
 & elastic+ETC & 3 & 44.64 & 56.49 \\
 & clustering & 135 & 28.07 & 33.69 \\
 & clustering+filter & 18 & 39.03 & 50.42 \\
 & clustering+RFE & 8 & 34.53 & 40.92 \\
 & clustering+ETC & 17 & 36.55 & 43.38 \\
\midrule
xgboost
 & lasso & 9 & 44.58 & 58.49 \\
 & lasso+filter & 7 & 44.59 & 58.09 \\
 & lasso+RFE & 6 & 41.17 & 53.46 \\
 & lasso+ETC & 3 & 40.85 & 55.12 \\
 & elastic & 10 & 44.19 & 58.76 \\
 & elastic+filter & 8 & 45.62 & 59.33 \\
 & elastic+RFE & 5 & 40.93 & 53.60 \\
 & elastic+ETC & 3 & 40.85 & 55.12 \\
 & clustering & 135 & 38.22 & 46.88 \\
 & clustering+filter & 18 & 47.56 & 61.42 \\
 & clustering+RFE & 8 & 40.91 & 50.99 \\
 & clustering+ETC & 17 & 46.38 & 56.60 \\
\midrule
kNN 
 & lasso & 19 & 59.35 & 69.92 \\
 & lasso+filter & 10 & 59.10 & 68.50 \\
 & lasso+RFE & 10 & 60.56 & 69.97 \\
 & lasso+ETC & 3 & 57.36 & 68.09 \\
 & elastic & 26 & 60.99 & 71.36 \\
 & elastic+filter & 11 & 57.97 & 68.80 \\
 & elastic+RFE & 12 & 58.48 & 68.19 \\
 & elastic+ETC & 5 & 58.02 & 68.74 \\
 & clustering & 120 & 55.65 & 66.35 \\
 & clustering+filter & 18 & 57.58 & 68.48 \\
 & clustering+RFE & 14 & 60.27 & 70.83 \\
 & clustering+ETC & 25 & 61.79 & 71.83 \\
\midrule
SVM 
 & lasso & 19 & 64.70 & 74.31 \\
 & lasso+filter & 10 & 61.07 & 70.10 \\
 & lasso+RFE & 10 & 62.72 & 70.57 \\
 & lasso+ETC & 3 & 55.21 & 64.07 \\
 & elastic & 26 & 65.54 & 75.67 \\
 & elastic+filter & 11 & 64.45 & 73.06 \\
 & elastic+RFE & 12 & 63.74 & 72.27 \\
 & elastic+ETC & 5 & 60.74 & 68.91 \\
 & clustering & 120 & 66.97 & 75.60 \\
 & clustering+filter & 18 & 60.54 & 70.50 \\
 & clustering+RFE & 14 & 64.75 & 73.02 \\
 & clustering+ETC & 25 & 68.25 & 76.37 \\
\midrule
MLP 
 & lasso & 19 & 62.92 & 72.76 \\
 & lasso+filter & 10 & 64.28 & 73.28 \\
 & lasso+RFE & 10 & 64.83 & 73.22 \\
 & lasso+ETC & 3 & 58.82 & 68.03 \\
 & elastic & 26 & 63.60 & 74.48 \\
 & elastic+filter & 11 & 63.23 & 72.25 \\
 & elastic+RFE & 12 & 64.63 & 73.96 \\
 & elastic+ETC & 5 & 62.64 & 70.63 \\
 & clustering & 120 & 64.15 & 74.11 \\
 & clustering+filter & 18 & 58.32 & 69.33 \\
 & clustering+RFE & 14 & 62.75 & 72.99 \\
 & clustering+ETC & 25 & 63.41 & 72.50 \\
\midrule
RF 
 & lasso & 19 & 62.17 & 69.97 \\
 & lasso+filter & 10 & 61.25 & 69.64 \\
 & lasso+RFE & 10 & 62.26 & 70.43 \\
 & lasso+ETC & 3 & 58.17 & 68.51 \\
 & elastic & 26 & 61.64 & 69.50 \\
 & elastic+filter & 11 & 62.04 & 70.30 \\
 & elastic+RFE & 12 & 64.23 & 72.55 \\
 & elastic+ETC & 5 & 60.41 & 69.61 \\
 & clustering & 120 & 58.56 & 65.68 \\
 & clustering+filter & 18 & 62.56 & 71.30 \\
 & clustering+RFE & 14 & 66.09 & 73.75 \\
 & clustering+ETC & 25 & 65.50 & 72.73 \\
\midrule
xgboost
 & lasso & 19 & 59.96 & 69.07 \\
 & lasso+filter & 10 & 57.47 & 66.83 \\
 & lasso+RFE & 10 & 58.81 & 67.81 \\
 & lasso+ETC & 3 & 52.40 & 64.45 \\
 & elastic & 26 & 59.08 & 68.57 \\
 & elastic+filter & 11 & 57.52 & 67.35 \\
 & elastic+RFE & 12 & 57.03 & 67.36 \\
 & elastic+ETC & 5 & 55.12 & 66.29 \\
 & clustering & 120 & 59.19 & 68.54 \\
 & clustering+filter & 18 & 56.98 & 67.43 \\
 & clustering+RFE & 14 & 60.74 & 69.87 \\
 & clustering+ETC & 25 & 61.64 & 70.78 \\
\end{longtable}
\end{center}
\begin{center}
\begin{longtable}{llrrrr}
\caption{Performance comparison of kNN, SVM, MLP, RF, and xgboost classifiers on various feature sets for Tamil dysarthric speech data.} \\
\toprule
Classifier & Feature & Feats & Accuracy & F1 \\
\midrule
\endfirsthead

\multicolumn{5}{c}{{\tablename\ \thetable{} -- continued from previous page}} \\
\toprule
Classifier & Feature & Feats & Accuracy & F1 \\
\midrule
\endhead

\bottomrule
\endfoot

\bottomrule
\endlastfoot

kNN 
 & lasso & 9 & 46.23 & 58.70 \\
 & lasso\_filter & 7 & 48.60 & 61.31 \\
 & lasso\_RFE & 6 & 46.00 & 57.75 \\
 & lasso\_ETC & 3 & 42.09 & 54.52 \\
 & elastic & 10 & 50.37 & 62.42 \\
 & elastic\_filter & 8 & 46.55 & 59.20 \\
 & elastic\_RFE & 5 & 41.93 & 55.02 \\
 & elastic\_ETC & 3 & 42.09 & 54.52 \\
 & clustering & 135 & 46.07 & 58.86 \\
 & clustering\_filter & 18 & 51.30 & 63.02 \\
 & clustering\_RFE & 8 & 47.04 & 57.46 \\
 & clustering\_ETC & 17 & 51.17 & 62.30 \\
\midrule
SVM 
 & lasso & 9 & 46.17 & 58.90 \\
 & lasso\_filter & 7 & 47.01 & 58.36 \\
 & lasso\_RFE & 6 & 47.25 & 58.43 \\
 & lasso\_ETC & 3 & 46.12 & 58.81 \\
 & elastic & 10 & 50.88 & 63.26 \\
 & elastic\_filter & 8 & 47.90 & 58.64 \\
 & elastic\_RFE & 5 & 45.79 & 57.50 \\
 & elastic\_ETC & 3 & 46.12 & 58.81 \\
 & clustering & 135 & 34.31 & 44.30 \\
 & clustering\_filter & 18 & 44.36 & 57.46 \\
 & clustering\_RFE & 8 & 43.68 & 52.17 \\
 & clustering\_ETC & 17 & 42.14 & 50.80 \\
\midrule
MLP & total & 619 & 52.67 & 65.19 \\
 & lasso & 9 & 47.13 & 60.97 \\
 & lasso\_filter & 7 & 49.35 & 61.38 \\
 & lasso\_RFE & 6 & 43.79 & 57.07 \\
 & lasso\_ETC & 3 & 45.23 & 58.01 \\
 & elastic & 10 & 47.64 & 61.58 \\
 & elastic\_filter & 8 & 48.73 & 62.19 \\
 & elastic\_RFE & 5 & 45.22 & 58.59 \\
 & elastic\_ETC & 3 & 43.19 & 56.28 \\
 & clustering & 135 & 41.77 & 53.95 \\
 & clustering\_filter & 18 & 47.11 & 59.86 \\
 & clustering\_RFE & 8 & 40.54 & 50.51 \\
 & clustering\_ETC & 17 & 47.69 & 57.26 \\
\midrule
RF 
 & lasso & 9 & 43.35 & 55.37 \\
 & lasso\_filter & 7 & 44.69 & 56.07 \\
 & lasso\_RFE & 6 & 45.91 & 57.74 \\
 & lasso\_ETC & 3 & 45.16 & 57.19 \\
 & elastic & 10 & 46.37 & 57.85 \\
 & elastic\_filter & 8 & 45.27 & 56.38 \\
 & elastic\_RFE & 5 & 45.64 & 57.88 \\
 & elastic\_ETC & 3 & 44.64 & 56.49 \\
 & clustering & 135 & 28.07 & 33.69 \\
 & clustering\_filter & 18 & 39.03 & 50.42 \\
 & clustering\_RFE & 8 & 34.53 & 40.92 \\
 & clustering\_ETC & 17 & 36.55 & 43.38 \\
\midrule
xgboost 
 & lasso & 9 & 44.58 & 58.49 \\
 & lasso\_filter & 7 & 44.59 & 58.09 \\
 & lasso\_RFE & 6 & 41.17 & 53.46 \\
 & lasso\_ETC & 3 & 40.85 & 55.12 \\
 & elastic & 10 & 44.19 & 58.76 \\
 & elastic\_filter & 8 & 45.62 & 59.33 \\
 & elastic\_RFE & 5 & 40.93 & 53.60 \\
 & elastic\_ETC & 3 & 40.85 & 55.12 \\
 & clustering & 135 & 38.22 & 46.88 \\
 & clustering\_filter & 18 & 47.56 & 61.42 \\
 & clustering\_RFE & 8 & 40.91 & 50.99 \\
 & clustering\_ETC & 17 & 46.38 & 56.60 \\
\midrule
kNN 
 & lasso & 19 & 59.35 & 69.92 \\
 & lasso\_filter & 10 & 59.10 & 68.50 \\
 & lasso\_RFE & 10 & 60.56 & 69.97 \\
 & lasso\_ETC & 3 & 57.36 & 68.09 \\
 & elastic & 26 & 60.99 & 71.36 \\
 & elastic\_filter & 11 & 57.97 & 68.80 \\
 & elastic\_RFE & 12 & 58.48 & 68.19 \\
 & elastic\_ETC & 5 & 58.02 & 68.74 \\
 & clustering & 120 & 55.65 & 66.35 \\
 & clustering\_filter & 18 & 57.58 & 68.48 \\
 & clustering\_RFE & 14 & 60.27 & 70.83 \\
 & clustering\_ETC & 25 & 61.79 & 71.83 \\
\midrule
SVM 
 & lasso & 19 & 64.70 & 74.31 \\
 & lasso\_filter & 10 & 61.07 & 70.10 \\
 & lasso\_RFE & 10 & 62.72 & 70.57 \\
 & lasso\_ETC & 3 & 55.21 & 64.07 \\
 & elastic & 26 & 65.54 & 75.67 \\
 & elastic\_filter & 11 & 64.45 & 73.06 \\
 & elastic\_RFE & 12 & 63.74 & 72.27 \\
 & elastic\_ETC & 5 & 60.74 & 68.91 \\
 & clustering & 120 & 66.97 & 75.60 \\
 & clustering\_filter & 18 & 60.54 & 70.50 \\
 & clustering\_RFE & 14 & 64.75 & 73.02 \\
 & clustering\_ETC & 25 & 68.25 & 76.37 \\
\midrule
MLP 
 & lasso & 19 & 62.92 & 72.76 \\
 & lasso\_filter & 10 & 64.28 & 73.28 \\
 & lasso\_RFE & 10 & 64.83 & 73.22 \\
 & lasso\_ETC & 3 & 58.82 & 68.03 \\
 & elastic & 26 & 63.60 & 74.48 \\
 & elastic\_filter & 11 & 63.23 & 72.25 \\
 & elastic\_RFE & 12 & 64.63 & 73.96 \\
 & elastic\_ETC & 5 & 62.64 & 70.63 \\
 & clustering & 120 & 64.15 & 74.11 \\
 & clustering\_filter & 18 & 58.32 & 69.33 \\
 & clustering\_RFE & 14 & 62.75 & 72.99 \\
 & clustering\_ETC & 25 & 63.41 & 72.50 \\
\midrule
RF 
 & lasso & 19 & 62.17 & 69.97 \\
 & lasso\_filter & 10 & 61.25 & 69.64 \\
 & lasso\_RFE & 10 & 62.26 & 70.43 \\
 & lasso\_ETC & 3 & 58.17 & 68.51 \\
 & elastic & 26 & 61.64 & 69.50 \\
 & elastic\_filter & 11 & 62.04 & 70.30 \\
 & elastic\_RFE & 12 & 64.23 & 72.55 \\
 & elastic\_ETC & 5 & 60.41 & 69.61 \\
 & clustering & 120 & 58.56 & 65.68 \\
 & clustering\_filter & 18 & 62.56 & 71.30 \\
 & clustering\_RFE & 14 & 66.09 & 73.75 \\
 & clustering\_ETC & 25 & 65.50 & 72.73 \\
\midrule
xgboost 
 & lasso & 19 & 59.96 & 69.07 \\
 & lasso\_filter & 10 & 57.47 & 66.83 \\
 & lasso\_RFE & 10 & 58.81 & 67.81 \\
 & lasso\_ETC & 3 & 52.40 & 64.45 \\
 & elastic & 26 & 59.08 & 68.57 \\
 & elastic\_filter & 11 & 57.52 & 67.35 \\
 & elastic\_RFE & 12 & 57.03 & 67.36 \\
 & elastic\_ETC & 5 & 55.12 & 66.29 \\
 & clustering & 120 & 59.19 & 68.54 \\
 & clustering\_filter & 18 & 56.98 & 67.43 \\
 & clustering\_RFE & 14 & 60.74 & 69.87 \\
 & clustering\_ETC & 25 & 61.64 & 70.78 \\
\midrule
kNN
 & lasso & 66 & 46.59 & 57.57 \\
 & lasso\_filter & 14 & 48.61 & 60.88 \\
 & lasso\_RFE & 23 & 49.54 & 60.97 \\
 & lasso\_ETC & 19 & 51.50 & 63.40 \\
 & elastic & 66 & 46.59 & 57.57 \\
 & elastic\_filter & 14 & 48.61 & 60.88 \\
 & elastic\_RFE & 23 & 49.54 & 60.97 \\
 & elastic\_ETC & 19 & 51.50 & 63.40 \\
 & clustering & 109 & 45.48 & 57.84 \\
 & clustering\_filter & 10 & 44.13 & 55.92 \\
 & clustering\_RFE & 28 & 49.51 & 60.84 \\
 & clustering\_ETC & 25 & 49.94 & 61.47 \\
\midrule
SVM 
 & lasso & 66 & 51.23 & 63.37 \\
 & lasso\_filter & 14 & 49.44 & 60.78 \\
 & lasso\_RFE & 23 & 51.67 & 63.05 \\
 & lasso\_ETC & 19 & 52.35 & 63.85 \\
 & elastic & 66 & 51.23 & 63.37 \\
 & elastic\_filter & 14 & 49.44 & 60.78 \\
 & elastic\_RFE & 23 & 51.67 & 63.05 \\
 & elastic\_ETC & 19 & 52.35 & 63.85 \\
 & clustering & 109 & 51.72 & 63.10 \\
 & clustering\_filter & 10 & 44.64 & 56.22 \\
 & clustering\_RFE & 28 & 50.08 & 61.35 \\
 & clustering\_ETC & 25 & 50.57 & 61.62 \\
\midrule
MLP 
 & lasso & 66 & 49.75 & 62.43 \\
 & lasso\_filter & 14 & 47.16 & 59.26 \\
 & lasso\_RFE & 23 & 51.50 & 63.40 \\
 & lasso\_ETC & 19 & 52.04 & 64.45 \\
 & elastic & 66 & 50.35 & 62.74 \\
 & elastic\_filter & 14 & 48.43 & 60.57 \\
 & elastic\_RFE & 23 & 51.67 & 63.86 \\
 & elastic\_ETC & 19 & 52.37 & 64.70 \\
 & clustering & 109 & 51.53 & 63.45 \\
 & clustering\_filter & 10 & 45.69 & 57.52 \\
 & clustering\_RFE & 28 & 51.43 & 62.95 \\
 & clustering\_ETC & 25 & 50.97 & 62.62 \\
\midrule
RF 
 & lasso & 66 & 48.41 & 58.22 \\
 & lasso\_filter & 14 & 50.00 & 60.52 \\
 & lasso\_RFE & 23 & 49.98 & 59.69 \\
 & lasso\_ETC & 19 & 50.02 & 60.01 \\
 & elastic & 66 & 48.65 & 58.31 \\
 & elastic\_filter & 14 & 49.66 & 59.87 \\
 & elastic\_RFE & 23 & 49.59 & 59.42 \\
 & elastic\_ETC & 19 & 50.05 & 60.09 \\
 & clustering & 109 & 50.39 & 59.20 \\
 & clustering\_filter & 10 & 46.36 & 58.55 \\
 & clustering\_RFE & 28 & 53.86 & 62.48 \\
 & clustering\_ETC & 25 & 54.10 & 62.48 \\
\midrule
xgboost
 & lasso & 66 & 52.10 & 63.42 \\
 & lasso\_filter & 14 & 48.65 & 60.07 \\
 & lasso\_RFE & 23 & 53.34 & 64.50 \\
 & lasso\_ETC & 19 & 53.67 & 64.93 \\
 & elastic & 66 & 52.10 & 63.42 \\
 & elastic\_filter & 14 & 48.65 & 60.07 \\
 & elastic\_RFE & 23 & 53.34 & 64.50 \\
 & elastic\_ETC & 19 & 53.67 & 64.93 \\
 & clustering & 109 & 53.48 & 63.41 \\
 & clustering\_filter & 10 & 45.60 & 57.94 \\
 & clustering\_RFE & 28 & 54.40 & 64.72 \\
 & clustering\_ETC & 25 & 54.05 & 64.56 \\
\end{longtable}
\end{center}

\chapter{Descriptive analysis of voice biomarkers.}
\begin{table}[ht]
\centering
\caption{Mean and Standard Deviation by Severity Level (voice quality)}
\label{tab:features_by_severity}
\resizebox{\textwidth}{!}{%
\begin{tabular}{@{}clcccccccccccccc@{}}
\toprule
\multirow{2}{*}{Lang} & \multirow{2}{*}{Feature} & \multicolumn{2}{c}{Healthy} & \multicolumn{2}{c}{Mild} & \multicolumn{2}{c}{Moderate} & \multicolumn{2}{c}{Severe} \\
\cmidrule(lr){3-4} \cmidrule(lr){5-6} \cmidrule(lr){7-8} \cmidrule(lr){9-10}
& & Mean & Std & Mean & Std & Mean & Std & Mean & Std \\
\midrule
\multirow{8}{*}{EN} 
& HNR & 9.99 & 3.23 & 11.09 & 3.00 & 18.69 & 3.32 & 12.60 & 4.12 \\
& jitter & 1.78 & 0.69 & 1.99 & 0.80 & 1.45 & 0.73 & 2.51 & 1.48 \\
& PPQ & 0.97 & 0.41 & 1.08 & 0.36 & 0.76 & 0.45 & 1.45 & 1.06 \\
& shimmer & 11.55 & 2.72 & 9.71 & 2.47 & 6.02 & 1.47 & 9.36 & 3.64 \\
& APQ & 6.42 & 2.05 & 5.34 & 1.83 & 2.85 & 0.82 & 5.14 & 2.47 \\
& num. of VB. & 6.50 & 3.56 & 7.45 & 4.28 & 9.06 & 5.42 & 11.75 & 9.21 \\
& per. of VB. & 16.95 & 9.68 & 20.09 & 12.59 & 25.84 & 9.91 & 23.06 & 13.86 \\
& CPP & 9.31 & 1.98 & 9.63 & 1.38 & 12.12 & 1.69 & 9.68 & 3.12 \\
\addlinespace
\midrule
\multirow{8}{*}{KO}
& HNR & 16.50 & 2.62 & 17.13 & 2.69 & 16.21 & 3.64 & 17.24 & 2.58 \\
& jitter & 1.62 & 0.55 & 1.40 & 0.57 & 1.74 & 0.74 & 1.70 & 0.82 \\
& PPQ & 0.98 & 0.14 & 0.69 & 0.49 & 0.82 & 0.50 & 0.93 & 0.46 \\
& shimmer & 7.55 & 2.15 & 6.92 & 1.71 & 8.27 & 2.51 & 7.40 & 2.95 \\
& APQ & 3.23 & 1.09 & 3.09 & 0.94 & 3.95 & 1.60 & 3.21 & 1.35 \\
& num. of VB. & 6.13 & 2.90 & 8.79 & 4.63 & 11.02 & 5.37 & 13.54 & 5.35 \\
& per. of VB. & 13.88 & 6.54 & 21.24 & 12.10 & 31.80 & 15.57 & 49.83 & 16.92 \\
& CPP & 13.79 & 1.62 & 13.93 & 1.97 & 12.08 & 2.34 & 12.28 & 1.49 \\
\addlinespace
\midrule
\multirow{8}{*}{TA}
& HNR & 13.14 & 2.57 & 15.89 & 2.57 & 15.75 & 3.22 & 15.63 & 4.58 \\
& jitter & 1.38 & 0.52 & 1.18 & 0.46 & 1.16 & 0.50 & 1.39 & 0.86 \\
& PPQ & 0.67 & 0.48 & 0.52 & 0.54 & 0.44 & 0.51 & 0.65 & 0.72 \\
& shimmer & 8.09 & 1.90 & 7.14 & 1.56 & 6.99 & 2.02 & 7.55 & 3.47 \\
& APQ & 3.60 & 1.05 & 3.36 & 1.16 & 3.33 & 1.20 & 4.15 & 2.53 \\
& num. of VB. & 3.97 & 1.98 & 3.52 & 2.00 & 3.62 & 2.04 & 3.40 & 1.73 \\
& per. of VB. & 19.68 & 8.93 & 17.78 & 9.24 & 20.40 & 10.67 & 19.18 & 7.95 \\
& CPP & 14.38 & 1.24 & 13.38 & 2.47 & 13.00 & 2.59 & 12.98 & 2.66 \\
\bottomrule
\end{tabular}
}
\end{table}
\begin{table}[ht]
\centering
\caption{Mean and Standard Deviation by Severity Level (pronunciation accuracy)}
\label{tab:features_by_severity_pr}
\resizebox{0.85\textwidth}{!}{%
\begin{tabular}{@{}clcccccccccc@{}}
\toprule
\multirow{2}{*}{Lang} & \multirow{2}{*}{Feature} & \multicolumn{2}{c}{Healthy} & \multicolumn{2}{c}{Mild} & \multicolumn{2}{c}{Moderate} & \multicolumn{2}{c}{Severe} \\
\cmidrule(lr){3-4} \cmidrule(lr){5-6} \cmidrule(lr){7-8} \cmidrule(lr){9-10}
& & Mean & Std & Mean & Std & Mean & Std & Mean & Std \\
\midrule
\multirow{3}{*}{EN} 
& CRR & 0.92 & 0.09 & 0.85 & 0.17 & 0.40 & 0.17 & 0.38 & 0.19 \\
& VRR & 0.88 & 0.13 & 0.83 & 0.19 & 0.45 & 0.19 & 0.36 & 0.16 \\
& PRR & 0.91 & 0.09 & 0.84 & 0.16 & 0.43 & 0.15 & 0.39 & 0.14 \\
\addlinespace
\midrule
\multirow{3}{*}{KO}
& CRR & 0.91 & 0.07 & 0.71 & 0.19 & 0.32 & 0.23 & 0.03 & 0.06 \\
& VRR & 0.96 & 0.05 & 0.75 & 0.19 & 0.40 & 0.23 & 0.09 & 0.05 \\
& PRR & 0.94 & 0.05 & 0.73 & 0.18 & 0.35 & 0.22 & 0.05 & 0.04 \\
\addlinespace
\midrule
\multirow{3}{*}{TA}
& CRR & 0.91 & 0.10 & 0.65 & 0.20 & 0.46 & 0.22 & 0.19 & 0.17 \\
& VRR & 0.95 & 0.09 & 0.71 & 0.19 & 0.57 & 0.22 & 0.41 & 0.18 \\
& PRR & 0.93 & 0.07 & 0.68 & 0.16 & 0.52 & 0.18 & 0.30 & 0.14 \\
\bottomrule
\end{tabular}
}
\end{table}
\begin{table}[ht]
\centering
\caption{Mean and Standard Deviation by Severity Level (vowel distortion)}
\label{tab:features_by_severity_vd}
\resizebox{0.85\textwidth}{!}{%
\begin{tabular}{@{}clcccccccccc@{}}
\toprule
\multirow{2}{*}{Lang} & \multirow{2}{*}{Feature} & \multicolumn{2}{c}{Healthy} & \multicolumn{2}{c}{Mild} & \multicolumn{2}{c}{Moderate} & \multicolumn{2}{c}{Severe} \\
\cmidrule(lr){3-4} \cmidrule(lr){5-6} \cmidrule(lr){7-8} \cmidrule(lr){9-10}
& & Mean & Std & Mean & Std & Mean & Std & Mean & Std \\
\midrule
\multirow{3}{*}{EN} 
& VAI & 0.81 & 0.08 & 0.81 & 0.08 & 0.73 & 0.08 & 0.69 & 0.07 \\
& FCR & 1.25 & 0.13 & 1.24 & 0.13 & 1.38 & 0.14 & 1.47 & 0.15 \\
& F2-Ratio & 1.37 & 0.21 & 1.43 & 0.23 & 1.31 & 0.30 & 1.28 & 0.29 \\
\addlinespace
\midrule
\multirow{3}{*}{KO}
& VAI & 0.84 & 0.11 & 0.78 & 0.12 & 0.67 & 0.12 & 0.60 & 0.09 \\
& FCR & 1.21 & 0.19 & 1.32 & 0.25 & 1.55 & 0.31 & 1.71 & 0.30 \\
& F2-Ratio & 1.76 & 0.42 & 1.66 & 0.42 & 1.36 & 0.36 & 1.26 & 0.31 \\
\addlinespace
\midrule
\multirow{3}{*}{TA}
& VAI & 0.83 & 0.12 & 0.71 & 0.10 & 0.67 & 0.11 & 0.61 & 0.09 \\
& FCR & 1.23 & 0.19 & 1.44 & 0.22 & 1.53 & 0.26 & 1.68 & 0.25 \\
& F2-Ratio & 1.95 & 0.54 & 1.57 & 0.38 & 1.44 & 0.38 & 1.28 & 0.33 \\
\bottomrule
\end{tabular}
}
\end{table}

\begin{table}[ht]
\centering
\caption{Mean and Standard Deviation by Severity Level (Pitch and Intensity)}
\label{tab:features_by_severity_pint}
\resizebox{\textwidth}{!}{%
\begin{tabular}{@{}clcccccccccc@{}}
\toprule
\multirow{2}{*}{Lang} & \multirow{2}{*}{Feature} & \multicolumn{2}{c}{Healthy} & \multicolumn{2}{c}{Mild} & \multicolumn{2}{c}{Moderate} & \multicolumn{2}{c}{Severe} \\
\cmidrule(lr){3-4} \cmidrule(lr){5-6} \cmidrule(lr){7-8} \cmidrule(lr){9-10}
& & Mean & Std. & Mean & Std. & Mean & Std. & Mean & Std. \\
\midrule
\multirow{10}{*}{EN} 
& F0 Mean & 149.29 & 24.47 & 149.03 & 27.60 & 144.62 & 48.88 & 164.26 & 36.60 \\
& F0 Med. & 131.98 & 36.92 & 137.46 & 45.12 & 154.05 & 66.91 & 139.42 & 33.41 \\
& F0 Std. & 81.21 & 26.55 & 79.30 & 20.42 & 71.05 & 22.40 & 88.38 & 28.83 \\
& F0 Min. & 51.29 & 1.65 & 52.08 & 2.51 & 51.13 & 1.98 & 52.22 & 3.20 \\
& F0 Max. & 401.95 & 76.24 & 395.00 & 77.90 & 391.96 & 65.75 & 433.02 & 68.35 \\
\addlinespace
& E Mean & 27.70 & 18.87 & 53.26 & 49.30 & 74.34 & 75.22 & 103.55 & 57.11 \\
& E Med. & 13.46 & 13.39 & 20.19 & 27.51 & 41.93 & 43.21 & 55.55 & 52.36 \\
& E Std. & 30.93 & 20.17 & 62.13 & 55.43 & 76.19 & 74.88 & 109.52 & 54.12 \\
& E Min. & 1.80 & 1.12 & 2.81 & 2.79 & 3.38 & 9.24 & 4.44 & 4.14 \\
& E Max. & 119.02 & 72.67 & 226.60 & 174.22 & 274.83 & 231.29 & 380.41 & 148.09 \\
\addlinespace
\midrule
\multirow{5}{*}{KO}
& F0 Mean & 209.00 & 55.46 & 215.93 & 35.61 & 230.78 & 31.94 & 258.92 & 45.87 \\
& F0 Med. & 192.13 & 56.73 & 194.43 & 42.35 & 198.38 & 45.65 & 239.16 & 67.97 \\
& F0 Std. & 99.13 & 26.10 & 118.88 & 21.53 & 129.84 & 18.52 & 139.01 & 21.27 \\
& F0 Min. & 51.75 & 5.87 & 50.31 & 2.52 & 50.18 & 1.38 & 50.02 & 0.18 \\
& F0 Max. & 470.64 & 58.09 & 492.12 & 25.38 & 498.73 & 8.09 & 498.36 & 9.21 \\
\addlinespace
& E Mean & 35.92 & 12.73 & 26.67 & 12.21 & 21.03 & 11.90 & 16.24 & 8.12 \\
& E Med. & 31.00 & 12.81 & 19.08 & 11.70 & 11.03 & 9.65 & 6.04 & 5.68 \\
& E Std. & 33.17 & 12.07 & 27.27 & 12.76 & 24.36 & 13.17 & 22.72 & 9.81 \\
& E Min. & 0.15 & 0.07 & 0.23 & 0.14 & 0.27 & 0.14 & 0.25 & 0.12 \\
& E Max. & 121.95 & 46.55 & 104.91 & 47.48 & 100.29 & 53.06 & 107.68 & 44.98 \\
\addlinespace
\midrule
\multirow{5}{*}{TA}
& F0 Mean & 169.61 & 41.59 & 175.35 & 31.03 & 200.26 & 47.53 & 204.44 & 32.53 \\
& F0 Med. & 173.38 & 45.83 & 185.49 & 34.61 & 214.37 & 58.42 & 225.11 & 40.34 \\
& F0 Std. & 80.59 & 26.50 & 65.19 & 25.37 & 84.15 & 32.39 & 66.20 & 21.43 \\
& F0 Min. & 50.56 & 1.18 & 52.64 & 9.51 & 52.11 & 11.02 & 54.62 & 12.12 \\
& F0 Max. & 396.24 & 92.27 & 348.54 & 97.74 & 407.94 & 106.69 & 352.86 & 91.03 \\
\addlinespace
& E Mean & 19.54 & 10.29 & 11.77 & 9.19 & 12.48 & 13.18 & 23.37 & 19.92 \\
& E Med. & 10.41 & 9.60 & 9.36 & 7.98 & 8.92 & 11.72 & 20.32 & 20.19 \\
& E Std. & 22.50 & 11.23 & 11.11 & 8.57 & 12.52 & 12.45 & 20.29 & 16.29 \\
& E Min. & 0.07 & 0.01 & 0.08 & 0.06 & 0.08 & 0.11 & 0.09 & 0.06 \\
& E Max. & 81.18 & 41.83 & 41.23 & 32.42 & 46.72 & 45.45 & 73.13 & 57.26 \\
\bottomrule
\end{tabular}
}
\end{table}

\begin{table}[ht]
\centering
\caption{Mean and Standard Deviation by Severity Level (rhythm)}
\label{tab:features_by_severity_rhythm}
\resizebox{\textwidth}{!}{%
\begin{tabular}{@{}clcccccccccc@{}}
\toprule
\multirow{2}{*}{Lang} & \multirow{2}{*}{Feature} & \multicolumn{2}{c}{Healthy} & \multicolumn{2}{c}{Mild} & \multicolumn{2}{c}{Moderate} & \multicolumn{2}{c}{Severe} \\
\cmidrule(lr){3-4} \cmidrule(lr){5-6} \cmidrule(lr){7-8} \cmidrule(lr){9-10}
& & Mean & Std & Mean & Std & Mean & Std & Mean & Std \\
\midrule
\multirow{5}{*}{EN} 
& \%V & 0.34 & 0.05 & 0.35 & 0.06 & 0.43 & 0.03 & 0.43 & 0.06 \\
& VarcoV & 176.72 & 68.14 & 198.21 & 108.21 & 298.48 & 117.41 & 309.11 & 170.46 \\
& VarcoC & 292.72 & 100.55 & 309.37 & 91.07 & 458.10 & 108.93 & 438.82 & 105.46 \\
& nPVI-V & 11.70 & 5.99 & 10.57 & 5.46 & 5.73 & 2.50 & 5.78 & 3.30 \\
& nPVI-C & 7.73 & 2.72 & 7.95 & 3.79 & 8.20 & 5.01 & 9.06 & 6.07 \\
\addlinespace
\midrule
\multirow{5}{*}{KO}
& \%V & 0.40 & 0.03 & 0.41 & 0.04 & 0.42 & 0.04 & 0.46 & 0.03 \\
& VarcoV & 123.79 & 40.04 & 211.01 & 122.54 & 327.30 & 130.97 & 493.56 & 116.68 \\
& VarcoC & 366.57 & 71.32 & 398.78 & 82.19 & 407.72 & 121.48 & 422.70 & 152.77 \\
& nPVI-V & 7.87 & 3.47 & 7.56 & 3.40 & 7.97 & 4.41 & 8.53 & 5.11 \\
& nPVI-C & 12.98 & 4.26 & 10.05 & 3.70 & 7.22 & 2.68 & 5.15 & 1.49 \\
\addlinespace
\midrule
\multirow{5}{*}{TA}
& \%V & 0.38 & 0.04 & 0.43 & 0.04 & 0.43 & 0.04 & 0.46 & 0.03 \\
& VarcoV & 187.44 & 76.79 & 156.48 & 78.77 & 178.51 & 82.64 & 163.50 & 62.12 \\
& VarcoC & 343.47 & 67.16 & 397.88 & 86.95 & 379.83 & 82.51 & 362.41 & 51.68 \\
& nPVI-V & 7.05 & 3.34 & 4.72 & 2.69 & 5.52 & 3.12 & 4.50 & 2.14 \\
& nPVI-C & 9.24 & 3.52 & 13.18 & 7.12 & 12.99 & 6.63 & 17.27 & 7.36 \\
\bottomrule
\end{tabular}
}
\end{table}

\bibliographystyle{elsarticle-harv}
\bibliography{thesis}

\keywordalt{마비말장애, 마비말장애 자동 중증도 분류, 마비말장애 자동 발음 평가, 다언어분석, 다언어분류}

\begin{abstractalt}
마비말장애는 조음기관 근육의 마비로 인해 음질, 발음, 운율 등에 부정적인 영향을 미쳐 말 명료도를 저하시키고, 이는 반복되는 의사소통의 어려움으로 이어져 궁극적으로 삶의 질 저하를 야기할 수 있다. 따라서 말 명료도에 대한 신뢰성 있고 정확한 평가는 환자의 진행 상황을 모니터링하고 효과적인 치료 계획을 수립하는 데 있어 매우 중요하다. 전통적인 말 명료도 평가는 훈련된 언어치료사가 청지각적으로 수행하는 방식이 주를 이루었으나, 이는 평가자의 주관적인 판단에 의존하며 시간과 자원이 많이 소요된다는 한계가 있다. 최근 음성 언어 처리 분야의 발전과 함께, 많은 연구자들은 전문가 평가와 높은 상관관계를 가지면서 객관적이고 효율적인 자동 평가 방법론 개발에 힘쓰고 있다.

기존 마비말장애 자동 평가 연구는 주로 단일 언어 환경에 초점을 맞춰왔다. 그러나 다국어 환경에서도 효과적으로 적용될 수 있는 자동 평가 방법론 개발의 필요성이 대두되고 있다. 이는 마비말장애 연구가 부족한 언어 사용자들에게도 평가 기회를 제공함으로써, 마비말장애 평가 도구의 포괄성과 접근성을 향상시킬 수 있기 때문이다. 본 논문에서는 영어, 한국어, 타밀어 마비말장애 음성 데이터를 사용하여 다국어 환경에서의 마비말장애 자동 평가 방법론을 제안한다.

먼저, 준언어적 음향 특징과 발음의 정확성(Goodness of Pronunciation, GoP)이라는 두 가지 특징을 활용하여 다언어 마비말장애 음성 분석을 수행하였다. 분석 결과, 두 가지 특징 모두 마비말장애 평가에 유용하게 활용될 수 있음을 확인하였다. 실험을 통해, 언어 구조의 차이가 서로 다른 최적의 특징 세트와 각 특징이 말의 명료도에 미치는 영향에 차이를 초래함을 확인한다.

이러한 발견을 바탕으로, 우리는 마비말장애 음성 특성을 종합적으로 포착하기 위해 임상 지식에 기반한 특징 세트를 제안한다. 이 특징들은 음질, 발음, 운율을 포함한다. 발음 특징은 음소 정확도와 모음 왜곡으로 나누어지고, 운율 특징은 기본 주파수(F0), 에너지, 지속 시간을 포함한다. 또한 우리는 언어-보편적 특징과 언어-세부적 특징을 구분하기 위해 다국어 분석을 수행한다. 우리의 실험 결과는 언어-보편적 특징에만 의존할 경우 평가 성능이 최적이 아님을 나타내며, 언어-세부적 특성을 고려하는 것이 중요함을 강조한다.

이러한 분석 결과를 바탕으로, eXtreme Gradient Boosting (XGBoost) 모델을 활용한 다언어 마비말장애 중증도 분류 방법론을 제안하였다. 본 방법론은 언어-보편적 특징과 언어별-세부적 특징을 모두 활용하여 모델 훈련의 효율성을 극대화하면서도, 목표 언어의 특징을 반영하고자 하였다. 이러한 접근 방식은 언어 보편적인 특징만을 사용하는 기존의 다언어 분류 방법론과 비교하여 7.33\%의 상대적인 성능 향상을 달성하였다.

결론적으로, 본 연구는 다국어 마비말장애 음성 분석 및 중증도 분류를 위한 새로운 방법론을 제시하며, 마비말장애 평가 분야에 다양한 시사점을 제공한다. 먼저, 다국어 음성 분석을 통해 마비말장애 음성에는 언어-보편적인 특징과 언어-세부적 특징이 모두 존재함을 확인하였고, 이러한 특징을 활용하여 마비말장애 중증도 분류 모델을 개발하였다. 특히, 언어 보편적인 특징만으로는 최적의 성능을 얻을 수 없다는 한계를 극복하기 위해 언어별 특징을 함께 고려한 XGBoost 기반 중증도 분류 모델을 제안하여 기존 방법론 대비 유의미한 성능 향상을 이루었다.

본 연구는 다국어 마비말장애 음성 데이터를 활용하여 다양한 언어 배경을 가진 마비말장애 환자들에게 효과적인 평가 도구를 제공할 수 있는 가능성을 제시한다. 이는 마비말장애 평가의 접근성을 높이고, 언어 장벽으로 인해 적절한 평가를 받지 못하는 환자들의 어려움을 해소하는 데 기여할 수 있을 것이다. 향후 연구에서는 더욱 다양한 언어와 마비말장애 유형을 포괄하는 대규모 다국어 데이터셋 구축 및 심층 신경망 모델 기반 다국어 마비말장애 평가 방법론 개발 등을 통해 마비말장애 평가 분야의 발전에 기여하고자 한다.

\end{abstractalt}

\end{document}